\documentclass[12pt]{article}

\pdfoutput=1

\usepackage{amsmath,amsfonts}
\usepackage{graphicx}
\usepackage{graphicx}
\usepackage{multirow}
\usepackage{subcaption}
\usepackage{color}
\usepackage{url}
\usepackage{xspace}
\usepackage{float}

\newcommand{\pd}[2]{\frac{\partial #1}{\partial #2}}

\newcommand{\R}{\mathbb{R}}
\newcommand{\abs}[1]{\vert#1\vert}

\def\HR{HR\xspace}
\def\PTT{{PTT$_\textrm{ff}$}\xspace}
\def\PTTmax{{PTT$_\textrm{p}$}\xspace}
\def\Dmax{{PTT$_{D}$}\xspace}
\def\DAT{DAT\xspace}

\newcommand{\PAT}{PAT$_\textrm{ff}$\xspace}
\newcommand{\PATmax}{PAT$_\textrm{p}$\xspace}
\newcommand{\PATDmax}{PAT$_{D}$\xspace}
\newcommand{\PATDAT}{{DAT}}

\def\hl{}

\usepackage[margin=1in]{geometry}

\usepackage{url}
\makeatletter
\renewcommand\@biblabel[1]{#1.}
\makeatother

\def\supmat{the appendix}

\begin{document}

\vspace*{0.35in}

\begin{flushleft} {\Large \textbf\newline{Improving pulse transit time
      estimation of aortic pulse wave velocity and blood pressure
      using machine learning and simulated training data}}

\vspace*{0.15in}
Janne M.J. Huttunen\textsuperscript{1,*},
Leo K\"arkk\"ainen\textsuperscript{1,2},
Harri Lindholm\textsuperscript{1}
\\
\bigskip
\bf{1} Nokia Bell Laboratories, Karakaari 13, 02610 Espoo,
        Finland
\\
\bf{2} Department of Electrical Engineering and Automation, Aalto University, Espoo, Finland
\\
\bigskip
* Correspondence: janne.m.huttunen@nokia-bell-labs.com

\end{flushleft}

\begin{abstract}
  Recent developments in cardiovascular modelling allow us to simulate
blood flow in an entire human body. Such model can also be used to
create databases of virtual subjects, with sizes limited only by
computational resources. In this work, we study if it is possible to
estimate cardiovascular health indices using machine learning
approaches.  In particular, we carry out theoretical assessment of
estimating aortic pulse wave velocity, diastolic and systolic blood
pressure and stroke volume using pulse transit/arrival timings derived
from photopletyshmography signals. For predictions, we train Gaussian
process regression using a database of virtual subjects generated with
a cardiovascular simulator. Simulated results provides theoretical
assessment of accuracy for predictions of the health indices. For
instance, aortic pulse wave velocity can be estimated with a high
accuracy ($r>0.9$) when photopletyshmography is measured from left
carotid artery using a combination of foot-to-foot pulse transmit time
and peak location derived for the predictions.  Similar accuracy can
be reached for diastolic blood pressure, but predictions of systolic
blood pressure are less accurate ($r>0.75$) and the stroke volume 
predictions are mostly contributed by heart rate.
\end{abstract}

\section{Introduction}
\label{sec:introduction}

This paper considers continuous monitoring of cardiac health using
computational modelling.
Stiffening of the arterial wall, such as aorta, causes reduction in
the pulsatile properties in the vascular tree, accelerates the
vascular premature ageing and predisposes to the dysfunction of the
heart, brain and other organs \cite{lyle2017,lia19}.
Aortic stiffness can be measured by using invasive methods or medical
imaging such as ultrasound \cite{negoita118} and MRI \cite{lia19}.
Another indicator reflecting the cardiac performance is stroke volume (SV),
which is typically measured using Doppler ultrasound \cite{ihlen84}.
However, these imaging modalities typically require special expertise
and are only carried out clinically.
On the other hand, aortic stiffness is associated with the
unfavourable changes in the diastolic and systolic blood pressures
(DBP/SBP), which can have several negative consequences in cardiac
function and structure \cite{lyle2017}.
Ambulatory home measurements of DBP and SBP use the techniques based
on inflated cuffs, but continuous recording is still cumbersome.
It would be helpful to find unobtrusive methods for the long-term
monitoring of these cardiac indices during the daily activities and
sleep.

Arterial stiffness is often assessed by measuring pulse wave velocity
(PWV), which is increased in stiffer arteries.
The PWV can be estimated by measuring arrivals of pulse waves at
two arterial sites:
\begin{displaymath}
  \textrm{PWV}=\frac{\textrm{distance between the sites}}{\textrm{travel time
    between the sites}} .
\end{displaymath}
The travel time is commonly referred as pulse transit time (PTT).
Arrival of the pulse wave to distal arterial sites can be easily
measured by using a photoplethysmogram (PPG), which is an
optical non-invasive sensor that can be placed, for example, in a
wearable device \cite{moraes18}.
On the other hand, in order to predict aortic stiffness reliably, the
first arterial site should be located at the beginning of aorta (for
measurement of aortic valve opening).
However, a measurement of valve opening can require a device
such as phonocardiograph, ultrasound or MRI.

To overcome this difficulty, PTT is often approximated using pulse
arrival time (PAT) which uses the R- wave of electrocardiogram
(ECG) as a reference timing \cite{gao16}.
However, there exists controversy in the clinical accuracy of using PAT
in the predictions due to variations in pre-ejection period (PEP) from
the R-wave to aortic valve opening \cite{balmer18,kortekaas18}.
An alternative approach is to approximate the reference with a
measurement from another distal site near aorta.
For example, the gold standard for aortic PWV measurement is to
measure differences of pulse arrivals to carotid and femoral arteries.

The estimation of blood pressure from arrival of pulse waves has also
been largely studied; see e.g. \cite{mukkamala2015,gao16,payne06}.
Although promising results
have been reported, clinical use of these techniques is still limited.
Haemodynamic alterations can have significant effects on the
accuracy \cite{rader17}.

A common problem with the clinical use of the above methodologies is
that the development and validation of the methods typically require a
large set of measurements from real human subjects with sufficient
variety.
Such data collection can be a very difficult and expensive task.

A preliminary assessment of the methods without extensive
data collection can be carried out using simulators.
For example, Willemet et al \cite{willemet2015,willemet2016} proposed
approach to use cardiovascular simulator for generation of a database
of ``virtual subjects'' with sizes limited only by computational
resources.
In their study, the databases were generated using one-dimensional
(1D) model of wave propagation in a artery
network comprising of largest human arteries \cite{alastruey2012}.
Such 1D models provide computationally efficient way to simulate
blood circulation and are also used in several other applications
\cite{vosse2011}.
There are also studies validating 1D simulations against real measurement
\cite{alastruey2011,matthys2007,olufsen2000}.
The virtual database approach was used to assess accuracy of pulse
wave velocity measurements for estimation of aortic stiffness \cite{willemet2015} and
the accuracy of pulse wave analysis algorithms \cite{willemet2016}.

The aim of our study is to assess theoretical limitations for the
prediction of aortic pulse wave velocity (aPWV), blood pressures
(DBP/SBP) and SV from PTT/PAT measurements.
We apply a similar virtual database approach to find correlations
between these cardiac indices and PTT/PAT timings measured from
different locations.
In particular, we train Gaussian process regressor to predict the
cardiac indices using different combinations of PTT and PAT
measurements.
The regressor model is trained using a large set of virtual subjects
generated using 1D cardiovascular simulator, and the results are
validated using another set of virtual subjects.
The result of study can give preliminary implications for the accuracy
of such predictions in rather ideal circumstances.

Our study is based on the 1D haemodynamic model of entire adult
circulations introduced by Mynard and Smolich \cite{mynard2015}.
It includes heart functions and all larger arteries and veins
for both systemic and pulmonary circulation.
%
%
As heart is included to the model, it can also simulate variations in
PEP that are essential in the comparison of PTT and PAT timings.

This paper is organized as follows.
First, the cardiovascular model is shortly summarized in Section \ref{sec:model-entire-adult}.
The numerical model is described in Section \ref{sec:coupling-with-DG}.
Section~\ref{sec:virtual-database} describes the generation of the
database of virtual subjects and the computation of the predictions
are described in Section~\ref{sec:comp-pred}.
Numerical experiments are shown in Section~\ref{sec:results}.
Discussion is given in Section \ref{sec:discussion}.


\section{Blood circulation model}
\label{sec:model-entire-adult}

The blood circulation model is based on the 1D haemodynamic model 
described in \cite{mynard2015}, which basically extends commonly used
1D wave dynamics model (see e.g.  \cite{alastruey2012}) with heart
functions and realistic arteria and venous networks including 
pulmonary and coronary circulations.
The components of the model are shortly summarized below, see
\cite{mynard2015} for more details.

\setlength{\unitlength}{0.92px}

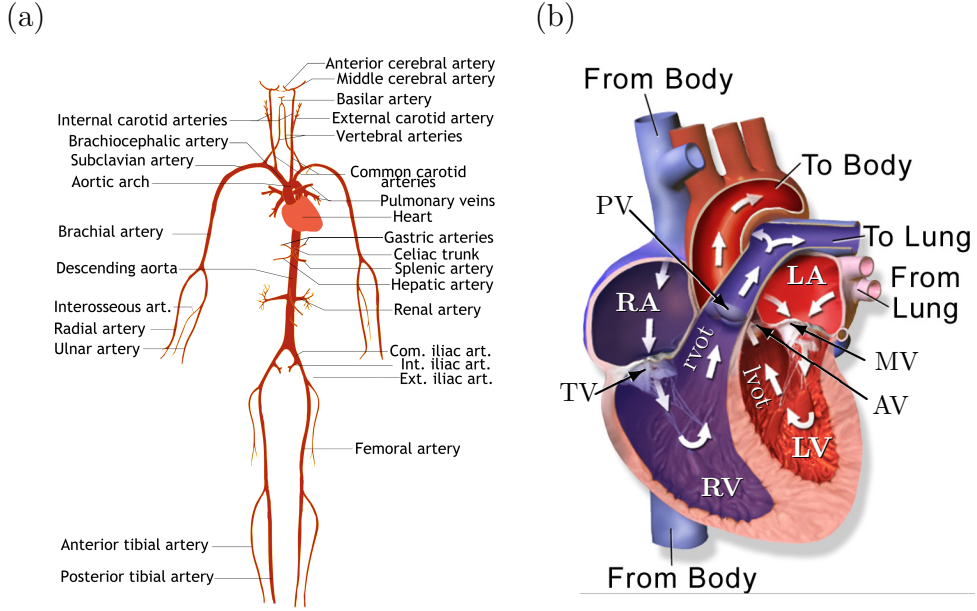
\begin{figure}[ht!]
  \centering
  \begin{tabular}{ccc}  
\vspace{-0cm}    \includegraphics[scale=0.37]{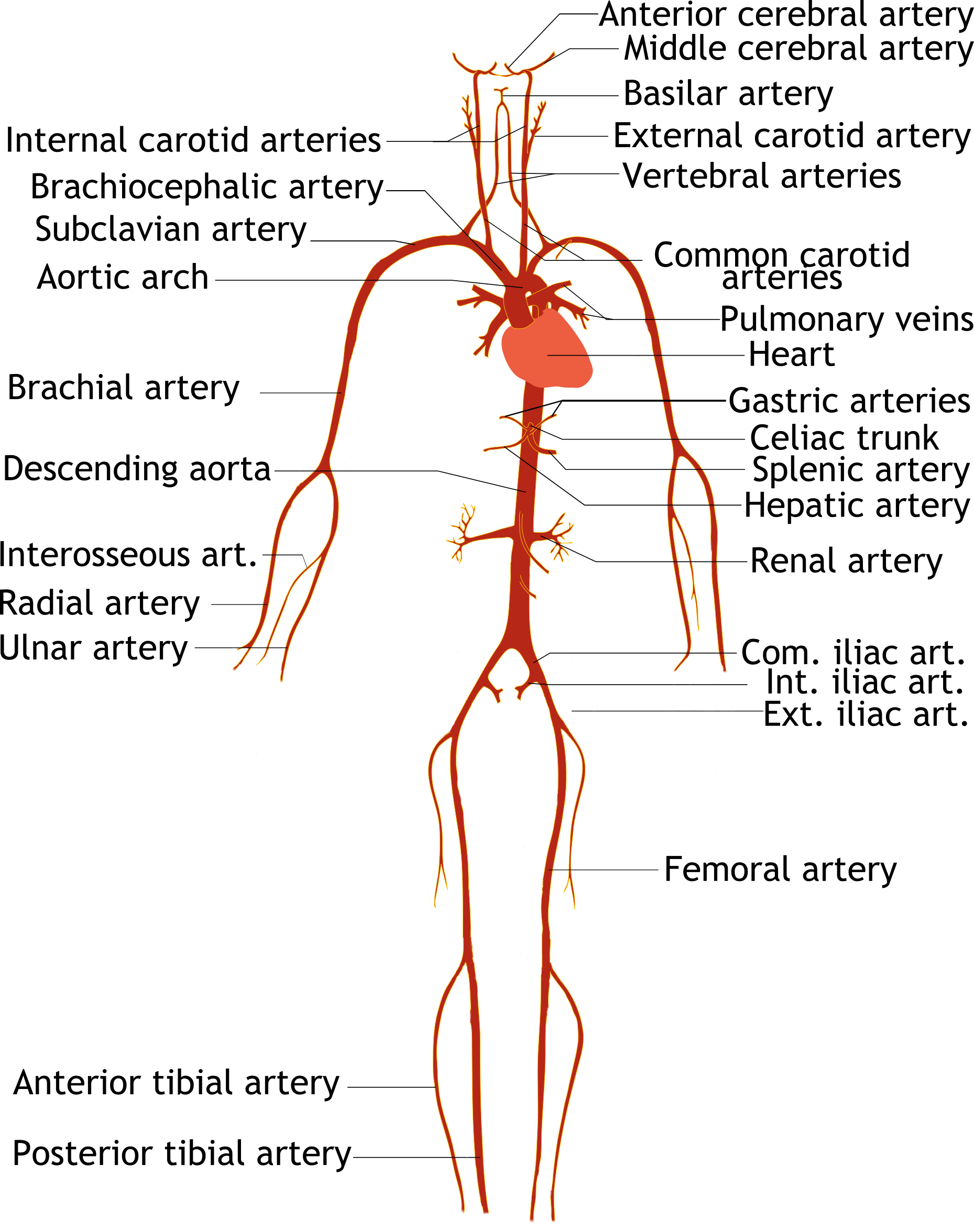}\put (-200,240){(a)}&\phantom{A}&
 \includegraphics[scale=0.24]{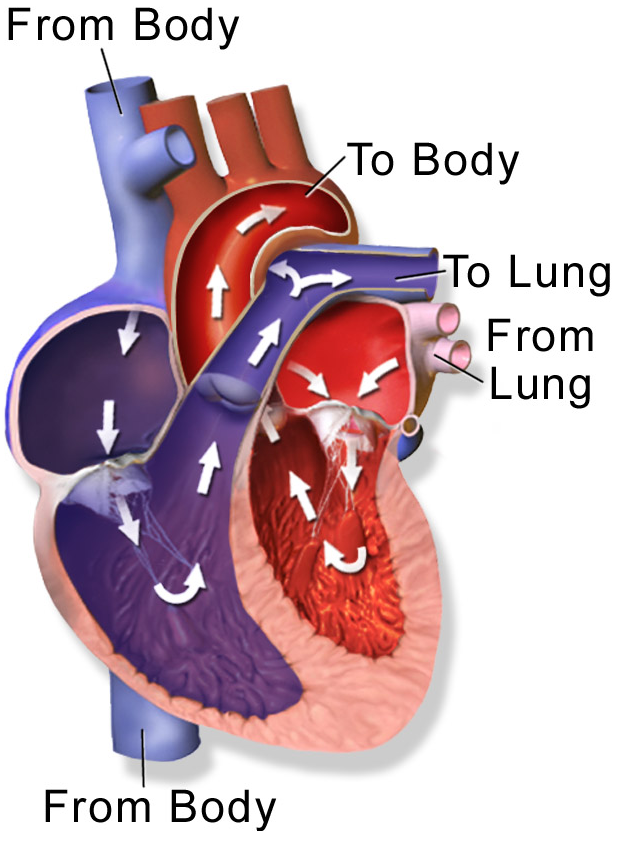}
  \put (-146.65,122.6){{\bf \footnotesize RA}}
  \put (-147,123){{\bf \color{white}\footnotesize RA}}
  \put (-111.65, 46.6){{\bf \footnotesize RV}}
  \put (-112, 47){{\bf \color{white}\footnotesize RV}}
  \put (-76.65,133.6){{\bf \footnotesize LA}}
  \put (-77,134){{\bf \color{white}\footnotesize LA}}
  \put (-74.65,62.6){{\bf \small LV}}
  \put (-75,63){{\bf \color{white}\small LV}}
   \put(-94.65,98.65){\mbox{\rotatebox{-69}{\footnotesize lvot}}}
   \put(-95,99){\mbox{\rotatebox{-69}{\color{white}\footnotesize lvot}}}
   \put(-120.65,97.65){\mbox{\rotatebox{63}{\footnotesize rvot}}}
   \put(-121,98){\mbox{\rotatebox{63}{\color{white}\footnotesize rvot}}}
  \put (-170,86){{\footnotesize TV}}
  \thicklines\put (-155.5,89){\color{white}\vector(2,1){20}}
  \thicklines\put (-155,89){\vector(2,1){20}}
  \put (-155,163){{\footnotesize PV}}
  \thicklines\put (-141.5,164){\color{white}\vector(1,-1){40}}
  \thicklines\put (-141,164){\vector(1,-1){40}}
  \put (-45,99){{ \footnotesize MV}}
  \thicklines\put (-44.75,102.75){\color{white}\vector(-2,1){30}}
  \thicklines\put (-45,103){\vector(-2,1){30}}
  \put (-46,81){{ \footnotesize AV}}
  \thicklines\put (-43.75,86.75){\color{white}\vector(-3,2){45}}
  \thicklines\put (-44,87){\vector(-3,2){45}}
%
  \put (-180,240){(b)}
  \end{tabular}
  \caption{(a) Illustration of human arterial system. The picture
    includes  only a few largest arteries; see \cite{mynard2015} for the
    complete set of arteries and veins used in the model.  (b)
    Illustration of human heart including four chambers: left atrium
    (LA), left ventricle (LV), right atrium (RA) and right ventricle
    (RV). Left and right ventricular outflow track (lvot/rvot) are
    short 1D segments before the valves. Valves: tricuspid valve (TV),
    pulmonary valve (PV), mitral valve (MV) and aortic valve (AV).
    Picture by BruceBlaus (CC BY).}
  \label{fig:Arterial_System}
\end{figure}

\subsubsection*{One-dimensional wave dynamics}

Human arterial network is illustrated in
Fig.~\ref{fig:Arterial_System}(a).
In 1D modelling, the arterial system is divided into segments
(e.g. from aortic root to the branching point of brachiochephalic
artery; see e.g. \cite{alastruey2012,mynard2015}).
Each segment is assumed to be a straight compliant tube with the
length $L$. The circular cross-sectional area $A(x,t)$ and the
velocity profile $U(x,t)$ are assumed to depend on time $t$ and a
single axial coordinate $x\in [0,L]$.
To radial direction, the velocity profile is assumed to be axisymmetric
and flat which agrees relatively well to experimental data (see
e.g. \cite{alastruey2011}).
The governing (nonlinear) equations can be written as
\cite{alastruey2012,mynard2015},
\begin{eqnarray}
  \label{eq:1Dgoverningeqs_1}
 && \pd{A}{t}+\pd{AU}{x}=0, \\
  \label{eq:1Dgoverningeqs_2}
 && \pd{U}{t}+U\pd{U}{x}+\frac 1 \rho\pd{p}{x}=\frac {f}
    {\rho A},
\end{eqnarray}
where $p$ is the pressure, $\rho$ and $\mu$ are the density and
viscosity of blood, and $f$ is the frictional force. 
With the axisymmetric and flat velocity profile, the frictional force
can be written as $f=-22\mu \pi U$  \cite{alastruey2012}.
%
%

The pressure-area relationship is written as \cite{mynard2012, mynard2015}
\begin{equation}
  \label{eq:ptoArelationship}
  p=p(A)=P_0+\frac{2\rho c_0^2}{b}\left[\left(\frac A
    {A_0}\right)^{b/2}-1\right],
\end{equation}
where $A_0$, $P_0$ and $c_0$ are the cross-sectional area, the
pressure and the wave speed at a reference state.
We have omitted the wall-viscosity in this study since the treatment of
the viscosity would result in significantly higher demands in numerical
discretization (remind that our aim is to run the model
repeatedly).
We choose $b=1$ which corresponding to the pressure law used in
Alastruey's model \cite{alastruey2011,alastruey2012}.
In Mynard et al \cite{mynard2012, mynard2015}, the constant $b$ was
specified as $b=2\rho c_0^2/(P_0-P_\textrm{collapse})$ where
$P_\textrm{collapse}$ is the collapse pressure.
However, in our experiments, this choice led to very steep raises in
pressures during systolic period due to omitted viscosity.

\subsubsection*{Heart and valves}

\begin{figure}[h!]
  \centering
  \begin{tabular}{ll}
    (a) & (b) \\
  \includegraphics[width=0.45\textwidth]{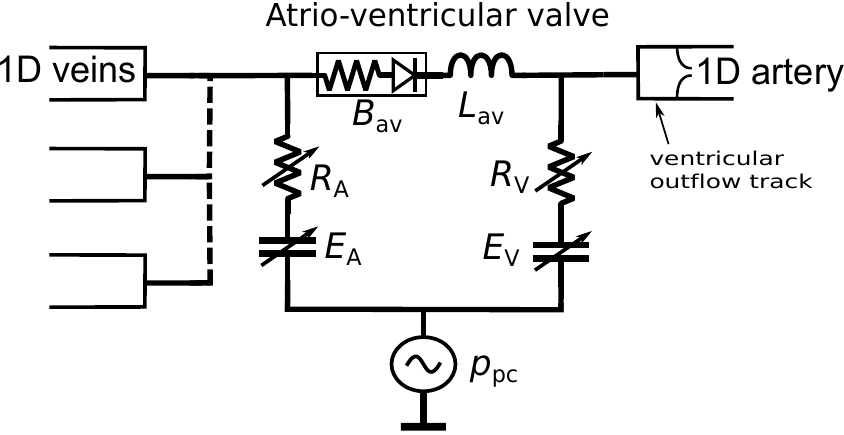}&
  \includegraphics[width=0.45\textwidth]{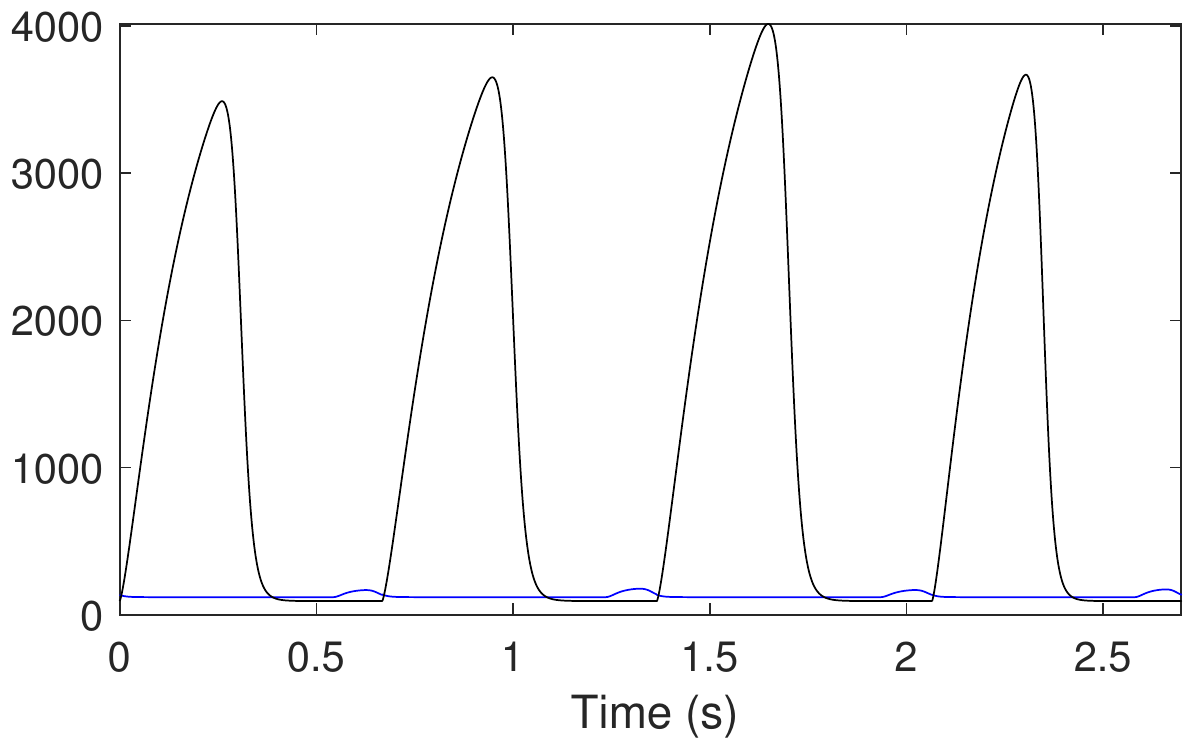}
  \end{tabular}
  \caption{(a) Schematic of atrioventriclular (av) model.  $B$ is the
    Bernoulli valve resistance, $R$ is the source resistance, $L$ is
    the blood inertance and $E$ is the elastance of the wall. The
    subscripts A and V refer to atrial and ventricular, respectively,
    and $p_\textrm{pc}$ is the pericardiac pressure. (b) Freewall
    elastance $E_\textrm{fw}$ for LA (blue) and LV (black). The figure
    includes four pulses. The duration of the pulse, the maximum
    elastance $E_\textrm{max}$ and timing parameters $\tau_1$ and
    $\tau_2$ vary between pulses.}
  \label{fig:heart_chambers}
  \label{fig:Efw}
\end{figure}

The anatomy of heart and blood circulation through heart are
illustrated in Fig.\ref{fig:Arterial_System}(b).
The blood flow through atriums (LA/RA) and ventricles (LV/RV)
is modelled using a lumped parameter model introduced
in \cite{mynard2012}, which was extended to include interactions between
heart chambers and pericardiac pressure in \cite{mynard2015}. 
The model is illustrated in Fig. \ref{fig:heart_chambers}(a).

 The relationship between the flow through valves ($q$) and the
transvalvular pressure difference
$\Delta p$ ($=p_\textrm{in}-p_\textrm{out}$) is given by the Bernoulli
equation,
\begin{equation}\label{eq:Bernoulli}
\Delta p=B_\textrm{av}q\abs{q}+L_\textrm{av}\frac{dq}{dt},
\end{equation}
where the Bernoulli resistance $B_\textrm{av}$ and the blood inertance
$L_\textrm{av}$ are
\begin{equation}
B_\textrm{av}=\frac{\rho }{2A_{\textrm{eff}}^2}\quad \textrm{and}\quad L_\textrm{av}=\frac{\rho
l_{\textrm{eff}}}{A_{\textrm{eff}}},
\end{equation}
where $A_{\textrm{eff}}$ and $l_\textrm{eff}$ are the effective valve
orifice area and length.
The valve dynamics are modelled using a state variable $\xi$ which
represents the state of the valve ($0\leq \xi\leq 1$, $\xi=0$ for
closed, $\xi=1$ for open) such that
$ A_{\textrm{eff}}(t)=\left(A_{\textrm{eff},\max
    }-A_{\textrm{eff},\min }\right)\xi (t)+A_{\textrm{eff},\min }$.
%
Valve dynamics are modelled by
\begin{equation}\label{eq:dxi}
\frac{d\xi }{dt}=K_{\text{vo}} (1-\xi )\Delta p \textrm{ when }
\Delta p\geq 0 \quad\textrm{or}\quad
\frac{d\xi }{dt}=K_{\text{vc}}\xi\Delta p \textrm{ when }
\Delta p<0 ,
\end{equation}
where $K_\textrm{vo}$ and $K_\textrm{vc}$ are rate coefficients for the
valve opening and closing, respectively.

%
%
The relationship between the pressure $p$ and the volume $V$ of a
heart chamber is given by
\begin{equation}
  \label{eq:chambers_P}
  p=p_\textrm{pc}+\frac{E_\textrm{nat}}{E_\textrm{sep}}p^*+E_\textrm{nat}(V-V_{p=0})-R_\textrm{s}q,
\end{equation}
where $p_\textrm{pc}$ is the pericardiac pressure (assumed to
depend exponentially on the total chamber volumes; see \cite{mynard2015}),
$E_\textrm{nat}$ is the native elastance of the chamber,
$E_\textrm{sep}$ is the septal elastance, $V_{p=0}$ is the volume of
the chamber in zero pressure, $R_\textrm{s}$ is the source resistance,
and $p^*$ is the pressure in the contralateral chamber.
%
The native elastance of a chamber is given by
\begin{equation}
   E_\textrm{nat}=\frac{E_\textrm{fw}E_\textrm{sep}}{E_\textrm{fw}+E_\textrm{sep}}-\mu_\textrm{AV},
   q
\end{equation}
where $E_\textrm{fw}$ is the freewall elastance of the chamber and
$\mu$ is the atrioventricular plane piston constant.
%
%
The time varying freewall elastances for each chamber are modelled by
\begin{equation}\label{eq:Efw}
  E_\textrm{fw}=k\left(\frac{g_1}{1+g_1}\right)\left(\frac{1}{1+g_2}\right)+E_\textrm{fw}^\textrm{min},\quad\textrm{where}\ 
  g_i=\left(\frac{t-t_\textrm{onset}}{\tau_i}\right)^{m_i},\quad i=1,2,
\end{equation}
and $k$ is the scaling constant chosen such that $\max(E_\textrm{fw})=E_\textrm{fw}^\textrm{max}$.
%
%
The functional properties of heart are specified via the maximum and
minimum free wall elastances ($E_\textrm{fw}^\textrm{min/max}$), the timing parameters $\tau_1$, $\tau_2$
and $t_\textrm{onset}$ and the slope parameters $m_1$ and $m_2$.
For example, increasing $E_\textrm{fw}^\textrm{max}$ increases the
contraction of the heart and the length of the pulse can be adjusted
through $\tau_1$ and $\tau_2$.
Fig. \ref{fig:Efw}(b) shows an example of the form of $E_\textrm{fw}$.
%

\subsubsection*{Vascular beds}

\begin{figure}[h!]
  \begin{tabular}{cc}
    (a) Generic vascular bed \\
\includegraphics[scale=0.8]{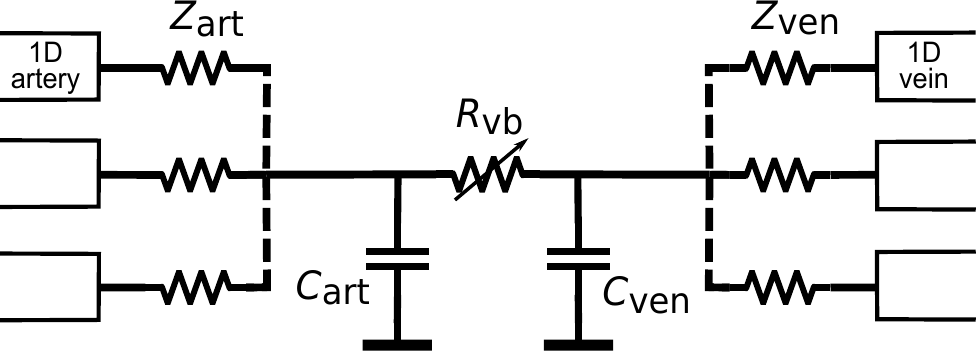}\\
    (b) Hepatic vascular bed\\
  \includegraphics[scale=0.8]{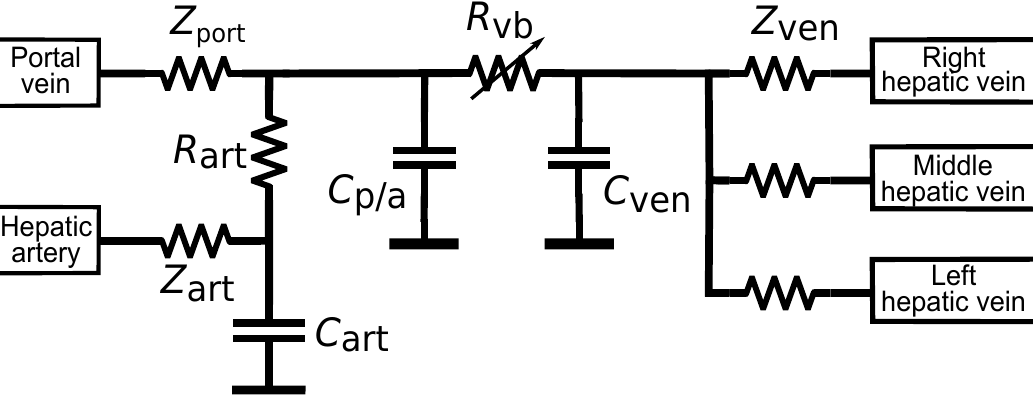}\\
    (c) Coronary vascular bed\\
    \includegraphics[scale=0.8]{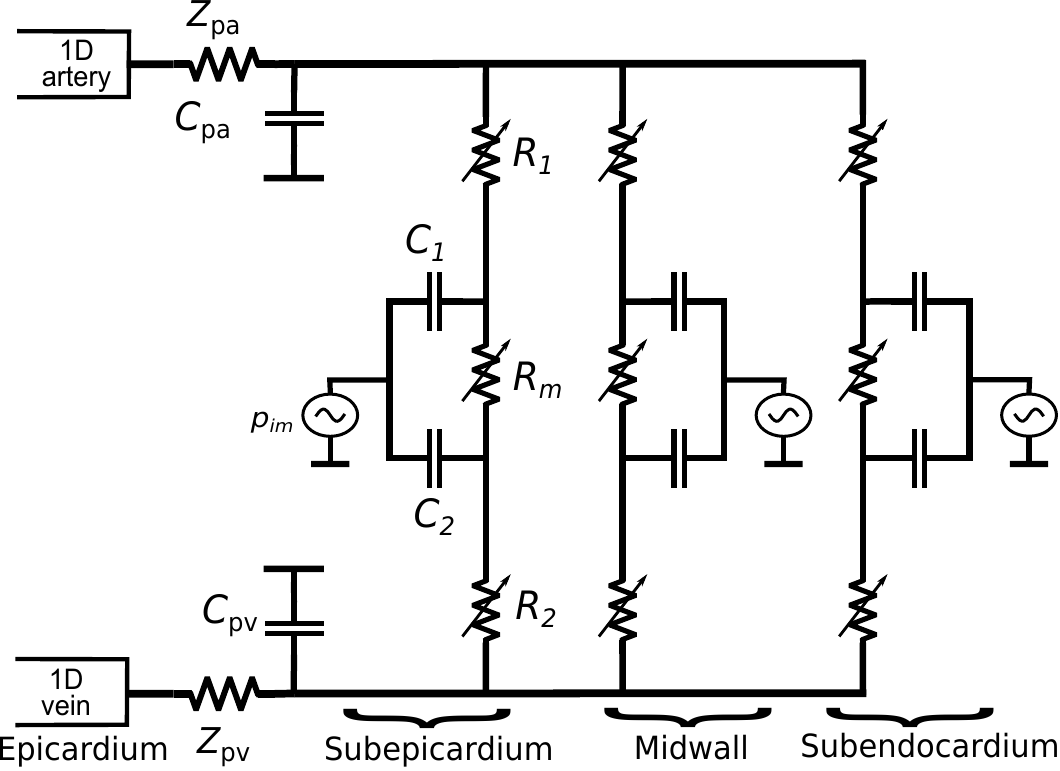}
  \end{tabular}
  \caption{(a) Generic vascular bed model; (b) Hepatic vascular bed
    model with arterial and venous inlets; (c) Coronary vascular bed
    model with compartments representing subepicardial, midwall and
    subendocardial layers.}
  \label{fig:vascular_beds}
\end{figure}

%
Mynard and Smolich \cite{mynard2015} describe models
for circulation through  three types of vascular beds (Fig. \ref{fig:vascular_beds}):
generic vascular beds, a hepatic vascular bed and coronary vascular
beds.
%
%
The generic vascular bed model (Fig. \ref{fig:vascular_beds}(a)) is used
for all microvasculature beds except the liver and myocardium.
It is based on commonly used three-element windkessel model and
consists of the characteristic impedances $Z_\textrm{art}$ and
$Z_\textrm{ven}$ (to couple the connecting 1D arteries to the vascular
bed), lumped compliances for the arterial and venous microvasculature ($C_\textrm{art}$
and $C_\textrm{ven}$) and the vascular bed resistance $R_\textrm{vb}$.
The resistance is assumed to be pressure dependent to account for the
fact that the atriovenous pressure difference remains positive even
with zero vascular bed flow:
\begin{equation}
  R_\textrm{vb}=\left\{
  \begin{array}{ll}
    R_0\left(\frac{p_\textrm{tm0}-P_\textrm{zf}}{p_\textrm{tm}-P_\textrm{zf}}\right)
    ,& p_\textrm{tm}> P_\textrm{zf},\\
    \infty,&p_\textrm{tm}\leq P_\textrm{zf},
  \end{array}\right. 
\end{equation}
where $p_\textrm{tm}=p-p_\textrm{ext}$ is the transmural pressure,
$P_\textrm{zf}$ is the zero-flow pressure and $R_0$ is the reference resistance.

The hepatic vascular bed (Fig. \ref{fig:vascular_beds}(b)) is a modification of the above to account for
both arterial and venous inlets in liver. 
It includes a compartment for the flow from hepatic artery
($R_\textrm{art}$, $C_\textrm{art}$) which connects to another
compartment ($C_\textrm{p/a}$) with common portal/arterial pressure.

The coronary vascular bed model (Fig. \ref{fig:vascular_beds}(c))
represents blood flow through intramyocardial.
%
The coronary  vessels experience a large time-varying 
myocardial pressures $p_\textrm{im}$ caused by the contracting heart
muscle.
To model depth-wise myocardial pressure, the model includes three
layers representing subendocardium, midwall and subepicardium, each
layer having three non-linear resistances $R_1$, $R_m$ and $R_2$:
\begin{equation}
  \label{eq:coronary_R12m}
  R_i(t)=R_{0,i}\frac{V_{0,i}^2}{V_i^2},   \ \  i=1,2, \quad R_m(t)=R_{0,m}\left(\frac{V_{0,1}^2}{V_1^2}+\frac{V_{0,2}^2}{V_2^2}\right),
\end{equation}
where the blood volumes $V_1$ and $V_2$ are are given by 
\begin{equation}\label{eq:coronary_Vi}
  V_i(t)=V_{0,i}+\int_0^t C_i\frac{dp_{\textrm{tm},i}(t')}{dt}dt',
  \quad p_{\textrm{tm},i}=p-p_\textrm{im},\quad i=1,2.
\end{equation}
The intramyocardial pressures $p_\textrm{im}$ is assumed to be the sum
of pressure transmitted from the ventricular cavity into the heart
muscle and pressure generated mechanically by the thickening heart
muscle.
See \cite{mynard2015} for details.


\subsection{Numerical solution of the cardiovascular model}
\label{sec:coupling-with-DG}

Our numerical solution of the wave propagation model is based on the
discontinuous Galerkin (DG) method.
The derivation of the DG solution for the 1D wave model
(\ref{eq:1Dgoverningeqs_1})-(\ref{eq:1Dgoverningeqs_2}) is described
with details e.g. in \cite{alastruey2012}, and therefore it is only
briefly summarized here.
We will give more details about the treatment of heart chambers,
valves and vascular beds as the numerical treatment differs from
\cite{mynard2015} due to the different numerical scheme.


The equations (\ref{eq:1Dgoverningeqs_1})-(\ref{eq:1Dgoverningeqs_2})
can be written in a conservative form as \cite{alastruey2012}
\begin{equation}
  \label{eq:conservative_form}
  \frac{\partial
  \pmb{U}}{\partial t}+\frac{\partial \pmb{F}}{\partial
  x}=\pmb{\hat S},\    \pmb{U}=\left(
\begin{array}{c}
 A \\
 U \\
\end{array}\right), \ \pmb{F}=\left(
\begin{array}{c}
 AU \\
 \frac{U^2}{2}+\frac{p}{\rho  } \\
\end{array}
\right)\textrm{ and } \pmb{\hat S}=\left(
\begin{array}{c}
 0 \\
 \frac{f}{\rho A} \\
\end{array}
\right) ,
\end{equation}
where $\pmb{F}$ is called the flux term.
As in the standard finite element method (FEM), each (arterial or venous) 1D
segment $[0,L]$ is divided into a non-overlapping elements
$\Omega_e$.
%
%
In addition, \eqref{eq:conservative_form} is multiplied with
a (vector valued) test function $\pmb\psi$ and integrated over the segment.
Then the integration by parts gives 
\begin{equation*}
   \sum_{e=1}^{N_\textrm{el}}\left[ \left(\frac{\partial\pmb{U}}{\partial t}, \pmb\psi\right)_{\Omega_e}
-\left(\pmb{F},  \frac{\partial\pmb\psi }{\partial x}\right)_{\Omega_e}
+\left[\pmb{F}\cdot { \pmb\psi}\right]_{x_e^l}^{x_e^r}\right]
=\sum_{e=1}^{N_\textrm{el}}\left(\pmb{\hat S}, {\pmb\psi}\right)_{\Omega_e},.
\end{equation*}
where $(\pmb u, \pmb v)_\Omega=\int_\Omega \pmb u \cdot \pmb v dx$ is
the standard $L^2(\Omega)$ inner product.
For a numerical solution, $\pmb{U}$ and $\pmb{\psi}$ are approximated which piecewise
polynomial vector functions $\pmb{U}^\delta$ and $\pmb{\psi}^\delta$.
However, contrary to the standard FEM, the approximation
$\pmb{U}^\delta$ is not enforced to be continuous across the element
boundaries.
Another application of the integration by parts gives
\begin{eqnarray}
&&\nonumber
   \sum_{e=1}^{N_\textrm{el}}\left[ \left(\frac{\partial\pmb{U}^\delta}{\partial t}, \pmb\psi ^\delta\right)_{\Omega_e}
+\left(\frac{\partial\pmb{F}({\pmb U}^\delta)}{\partial x},  {\pmb\psi}^\delta \right)_{\Omega_e}
+\left[{\pmb\psi}^\delta \cdot(\pmb{F}^*-\pmb{F}(\pmb{U}^\delta))\right]_{x_e^l}^{x_e^r}\right]\\
&&=\sum_{e=1}^{N_\textrm{el}}\left(\pmb{\hat S}(\pmb{U}^\delta),
   {\pmb\psi} ^\delta\right)_{\Omega_e},
\label{eq:DGproblem}
\end{eqnarray}
where the term $\pmb{F}^*$ is the (approximative) flux function (determined
below).
The flux $\pmb{F}^*$ is responsible of propagating information through
the elements interfaces and is also the key element in the specification of the
boundary conditions for the 1D blood vessel segments.

In order to apply a numerical integration scheme for temporal discretization, we need to find
$\mathcal F$ such that
$\frac{\partial\pmb{U}^\delta}{\partial t}=\mathcal F(\pmb{U}^\delta)$.
As in the standard FEM, this corresponds to finding the
coefficients of the approximation of
$\frac{\partial\pmb{U}^\delta}{\partial t}$ such that
(\ref{eq:DGproblem}) is satisfied for a chosen set of test functions.
However, since the approximation is discontinuous in DG, the
coefficients can be
solved separately for each element.
The problem is further simplified by using Legendre polynomials as the
basis functions of the approximation and test functions, which allows us to treat each
basis function separately due to $L^2$-orthogonality.

For the numerical integration, the second-order Adams-Bashforth
time integration scheme is used; see \cite{alastruey2012} for more
details.

\subsubsection*{Characteristic analysis and the flux $\pmb{F}^*$}

The determination of the flux $\pmb{F}^*$ and numerical boundary
conditions is based on the Riemann's method of characteristics.
The characteristic functions (or Riemann's variables) of the system
(\ref{eq:conservative_form}) can be written as (see
\cite{alastruey2012} for the derivation)
\begin{equation}
  \label{eq:charasteristicfunctions}
  W_f(A,U)=U-U_0+ \int_{A_0}^A \frac{c}{A}
  \, dA, \quad W_b(A,U)=U-U_0- \int_{A_0}^A \frac{c}{A}
  \, dA,
\end{equation}
where the subscripts $f$ and $b$ refer to information moving to
forward and backward directions, respectively, and $c$ is the wave speed (local
PWV),
\begin{equation}
  \label{eq:localPWV}
  c=\sqrt{\frac{A}{\rho}\frac{\partial p}{\partial A}}.
\end{equation}
When considering the pressure-area relationship
\eqref{eq:ptoArelationship}, 
\begin{equation}\label{eq:Psi}
c=c_0\left(\frac{A}{A_0}\right)^{b/4}\quad\textrm{and}\quad \int_{A_0}^A \frac{c}{A}
  \, dA=\frac{4c_0}{b}\left[\left(\frac A{A_0}\right)^{b/4}-1\right]
  =: \Psi(A).
\end{equation}

The fluxes $\pmb{F}^*$ at the interfaces of elements are calculated as
a solution of a Riemann problem with suitable boundary conditions, see
e.g. \cite{alastruey2012}.
The procedure involves finding a unique state $(A^*,U^*)$ such that
\begin{equation}\label{eq:Wfb_condition}
  W_f(A_L,U_L)=W_f(A^*,U^*)\quad \textrm{and}\quad   W_b(A_R,U_R)=W_f(A^*,U^*),
\end{equation}
where the subscript $L$ and $R$ refer to the value of $A$ and $U$ on
the left or the right side of the boundary of the element, respectively.
The flux is then given as $\pmb{F}^*=\pmb{F}(A^*,U^*)$.

The boundary conditions for the 1D blood vessel segments are handled
similarly by finding a state $(A^*,U^*)$ satisfying conditions similar
to \eqref{eq:Wfb_condition}.
Treatment of the boundary conditions related to splitting and merging
arteries/veins is presented in \cite{alastruey2012}.
Treatment of the boundary conditions related to the heart, valve and
vascular beds is presented below.

\subsubsection*{Numerical model for heart chambers}

We consider left heart (right heart is handled similarly).
The Trapezoidal rule applied to the net flow arriving to LV gives (see
Fig. \ref{fig:Arterial_System}(b))
\begin{equation}\label{eq:coupling_Vn}
q_{\text{LV}}=-\frac{dV_{\text{LV}}}{dt}=q_{\text{lvot},\text{in}}-q_{\text{MV}}
\quad\Rightarrow\quad V_{\text{LV}}^{n}=V_{\text{LV}}^{n-1}-\frac{\Delta t}{2}\left(q_{\text{LV}}^{n}+q_{\text{LV}}^{n-1}\right),
\end{equation}
where the superscript $n$ refers to the $n$'th temporal discretization
point and $\Delta t$ is the time step.
The above equation can be substituted to \eqref{eq:chambers_P} to give
\begin{equation}\label{eq:pn_coupling}
  p^{n}_\textrm{LV}=p^{n}_\textrm{ext}+E^{n}_\textrm{nat}\left[V_{\text{LV}}^{n-1}-\frac{\Delta t}{2}\left(q_{\text{LV}}^{n}+q_{\text{LV}}^{n-1}\right)-V_{p=0,\textrm{LV}}\right](1-K_\textrm{s,\textrm{LV}}q_\textrm{LV}^{n}),
\end{equation}
where $p^n_\textrm{ext}=p^n_\textrm{\textrm{LV},pc}+E^n_\textrm{nat}/E^n_\textrm{sep}p^n_\textrm{RV}$.

The output of LV is connected to the inlet of lvot-segment; see
Fig. \ref{fig:Arterial_System}(b).
At the inlet of lvot, we have
\begin{equation}
  \label{eq:coupling_Wb}
  W_b(A^\textrm{lvot}_\textrm{in}, U^\textrm{lvot}_\textrm{in})=W_b(A^*,U^*)=U^*-\Psi(A^*)  ,
\end{equation}
where $A^\textrm{lvot}_\textrm{in}$ and $U^\textrm{lvot}_\textrm{in}$ are the DG approximations at the inlet.
Since $q_{\text{lvot},\text{in}}^n=A^*U^*$, 
\begin{equation}\label{eq:qLV_coupling}
q_{\text{LV}}^n=A^*U^*-q^n_{\text{MV}}=A^u(\tilde W_b+\Psi
(A^*))-q_{\text{MV}}^n=A^*\tilde W_b-q_{\text{MV}}^n+A^*\Psi(A^*),
\end{equation}
where $\tilde W_b=  W_b(A^\textrm{lvot}_\textrm{in}, U^\textrm{lvot}_\textrm{in})$.
Plugging in (\ref{eq:qLV_coupling}) and
$p^n_\textrm{LV}=p^n_{\text{lvot},\text{in}}=p(A^*)$ to
\eqref{eq:pn_coupling} gives an equation from which $A^*$ can
be solved using Newton's method.
Finally, $U^*$ can be solved from \eqref{eq:coupling_Wb} and
$V_\text{LV}^n$ and $p_\text{LV}^n$ are obtained during the iteration.

The atriums have multiple vein connections; see Fig. \ref{fig:Arterial_System}(b).
Let $A^j_{\textrm{out}}$ and $U^j_{\textrm{out}}$ be the DG
approximations at the outlet of the $j$'th connecting 1D-segment
($j=1,\dots,J$).
We can write
$W_{f}(A^j_{\textrm{out}}, U^j_{\textrm{out}})=W_f\left(A_j^*,U_j^*\right)=U_j^*+\Psi(A_j^*)$,
and further
\begin{equation}\label{eq:qLA}
q_{\text{LA}}^n=q_{\text{MV}}^n-\sum _jA_j^*U_j^*=q_{\text{MV}}^n-\sum
_jA_j^*\left(\tilde W_{f}^j-\Psi_j(A_j^u)\right).
\end{equation}
where $\tilde W^j_{f}=W_{f}(A^j_{\textrm{out}}, U^j_{\textrm{out}})$
and $\Psi_j$ the function \eqref{eq:Psi} with the parameters $A_0$ and
$c_0$ corresponding to the outlet of $j$'th segment.
Then similarly as above, we can obtain a group of $J$ equations from
which $A_1^*,\ldots,A_J^*$ can be simultaneously solved using
Newton's method.
%
However, the multi-dimensional problem can be avoided by noticing that
the pressure-area relationship \eqref{eq:ptoArelationship} can be
inverted easily (i.e. we can find $A=A(p)$).
Then it is equivalent to solve $p$ from the one-dimensional problem
\begin{equation}
  p=p^n_\textrm{ext,LA}+E^n_\textrm{nat}\left[V_{\text{LA}}^{n-1}-\frac{\Delta
      t}{2}\left(q^n_\textrm{LA} (p)+q_{\text{LA}}^{n-1}\right)-V_{p=0,\textrm{LA}}\right](1-K_\textrm{s,\textrm{LA}}\tilde
  q^n_\textrm{LA} (p)).
\end{equation}
where $\tilde q^n_\textrm{LA}(p)$ is given by \eqref{eq:qLA} with
$A_j^*=A_j(p)$, where the subscript $j$ refers to the mapping in which
the parameters $A_0$, $c_0$ and $b$ in \eqref{eq:ptoArelationship} are
specified for at the outlet of the $j$'th segment.

\subsubsection*{Valves}

The application of the forward Euler method to \eqref{eq:Bernoulli}
gives
\begin{eqnarray}
\label{eq:valves_update_q}
&&q^{n+1}=q^{n}+\frac{\Delta t}{L_\textrm{av}}\left(\Delta p^{n}-B_\textrm{av} q^{n}\abs{q^{n}}\right) .
\end{eqnarray}
The equation \eqref{eq:dxi} is discretized similarly.
For MV and TV, the transvalvular pressure is the pressure difference
between artium and ventricle (e.g.
$\Delta p^{n}=p^{n}_\textrm{LA}-p^{n}_\textrm{LV}$ for MV).

PV and AV are between 1D segments (e.g. AV is between lvot and the
first segment of aorta, see Fig. \ref{fig:Arterial_System}(b)).
For the outlet of the ventricular outflow tracks, we specify the
outflow condition (e.g. $q^\textrm{out}_\textrm{lvot}=q^n_\textrm{AV}$).
For the inlet of the 1D segments behind the
valve, we specify the prescribe the inflow to be $q^n_\textrm{valve}$.
%
%
These inflow and outflow boundary conditions can be treated similarly as
above by finding the states $(A^*,U^*)$; see e.g. \cite{alastruey2012}
for details.
Then the pressures on the both sides of the valve can be computed
using the states $A^*$ and the pressure-area relationship
\eqref{eq:ptoArelationship}.

\subsubsection*{Vascular beds} 

We consider the generic vascular bed model
(Fig. \ref{fig:vascular_beds}(a)).
Arterial and venous flows $q_\textrm{art}$ and $q_\textrm{ven}$ in
the generic vascular bed model (sums of all flows from/to
1D-segments) are given by
\begin{equation}\label{sec:vascular-beds-1}
q_{\textrm{art}}=q_{\textrm{cap}}+C_{\textrm{art}}\frac{d p_{\textrm{art}}}{dt} ,\quad
q_{\textrm{ven}}=q_{\textrm{cap}}-C_{\textrm{ven}}\frac{d p_{\textrm{ven}}}{dt}.
\end{equation}
The forward Euler method gives
\begin{equation}\label{eq:partven}
p_{\textrm{art}}^{n+1}=p_{\textrm{art}}^{n}+\frac {\Delta t} {C_{\textrm{art}}}(q_{\textrm{art}}^n-q_{\textrm{cap}}^n),\quad
p_{\textrm{art}}^{n+1}=p_{\textrm{art}}^{n}+\frac {\Delta t}{C_{\textrm{ven}}} (q_{\textrm{cap}}^n-q_{\textrm{ven}}^n).
\end{equation}
The capillary flows $q_{\textrm{cap}}^n$ (flow through
$R_\textrm{vb}$) are calculated using
Ohm's law. 

Vascular beds are connected to the 1D model as the terminal resistance
boundary condition similarly as in \cite{alastruey2012}.
For example, we consider coupling of a 1D-arterial segment to the
generic vascular bed model (Fig. \ref{fig:vascular_beds}(a)).
The flow $q$ though impedance $Z_\textrm{art}$ is given by Ohm's law
$Z_\textrm{art}  q=p_{1D}-p^n_{art}$.
%
We need to
find $(A^*,U^*)$ such that 
\begin{equation}
Z_\textrm{art} A^*U^*=p(A^*)-p^n_{art}\quad\textrm{and}\quad W_f(A_L,U_L)=W_f(A^*,U^*)=U^*+\Psi(A^*)
\end{equation}
The states $A^*$ and $U^*$ can be solved by combining the 
equations as above and applying Newton's method.
Then $q_\textrm{art}^n$ is the sum of flows from all 1D-outlets
($A^*U^*$).

Portal and coronary models in Fig. \ref{fig:vascular_beds}(bc) can be
treated similarly.

\section{Virtual database}
\label{sec:virtual-database}

The database is created by running the cardiovascular model
repeatedly.
The model parameters are varied to reflect variations between
individual (virtual) subjects.

In \cite{willemet2015,willemet2016}, the seven parameters were varied:
elastic artery PWV, muscular artery PWV, the diameter of elastic arteries,
the diameter of muscular arteries, heart rate (HR), SV and peripheral
vascular resistance.
In their study, the parameters were varied by specifying a few
possible values for each parameter and the cardiovascular model was
run for all of the resulting 7776 combinations.
However, in our study, the cardiovascular model has significantly more
model parameters (e.g. parameters related to heart model and valves,
vascular beds, ...).
Such systematic variation of all essential parameters would lead to
excessively large number of combinations.

In this study, we choose ``sampling'' approach in which the model
parameters are varied randomly.
%
%
Our aim is to choose random variations that would represent 
healthy subject and, where applicable, the range of the parameters is
of similar range as in \cite{willemet2015}.
Some choices can be rather subjective due to the limited amount of
(probabilistic) information from related physiological quantities.
Our goal is to choose variations to be wide enough so that
``real world'' can be considered as a subset of the population covered
by the variations.
However, if more sufficient information about parameters becomes
available, it should be rather straightforward to carry out the analysis
with the adjusted distributions.

In the following, the superscript $(s)$ refers to a virtual subject
for which the parameters are specified.
The overbar notation (e.g. $\bar L$) refers to the values used in
\cite{mynard2015} (the baseline).
Unless otherwise mentioned, the variations are chosen to be normally distributed. 
Furthermore, the statements such as 10\% relative variation should be
understood in terms of standard deviations instead explicit
ranges of the parameter.
We use slightly unconventional notation $\mathcal N(\mu,X\%)$ to denote
the Gaussian distribution with mean $\mu$ and the standard deviation
$\sigma=X/100\mu$ (i.e. $X\%$ variation relative to the mean/baseline).
The uniform distribution is denoted as $\mathcal U(a,b)$.
%

\subsubsection*{Vascular networks}

The arterial and venous network structure is chosen to be same as in
\cite{mynard2015} (the length $L$ and $A_0$ at the inlet and outlet for each 1D segment
are given in their supplementary materials).
%
%
To include individual variations of subjects, the lengths are chosen
as
\begin{equation}
  L^{(s)}_\ell=\bar L_\ell a^{(s)} b^{(s)}_\ell,\quad a^{(s)}\sim \mathcal
  N(1,10\%),\quad b^{(s)}_\ell \sim \mathcal
  N(1,2\%),
\end{equation}
where the subscript $\ell$ refers to the $\ell$th segment.
The multiplier $a^{(s)}$ can be understood as a variation in the
height of subject total length and $b^{(s)}_\ell$ represents
individual variations of blood vessel segments.
%
With these choices, for example, distances from aortic root to the
measurement locations (see below) are $17.0\pm 1.8$ cm (left carotid
artery) and $88.9\pm 9.1$ cm (femoral artery) which are similar to the
distances reported in \cite{sugawara16,bossuyta13}.
%
The arterial diameters $D_0$($\propto\sqrt{A_0}$) are also varied
similarly, except we use separate common multipliers $a^{(s)}$ for
aorta (20\% variation) and rest of segments (10\% variation).

The elasticity $E$ of blood vessels is controlled by the
reference wave speed (PWV), which can be expressed using the empirical
formula \cite{olufsen1999}
\begin{equation}
  c_0^2=\frac 2{3\rho}\frac{Eh}{2r_0}=\frac
  2{3\rho}\left[k_1\exp(k_2r_0)+k_3\right] ,
\end{equation}
where $r_0$ is the reference radius, $h$ is the thickness of the wall
and $k_1$, $k_2$ and $k_3$ are empirical constants.
%
%
%
%
Elasticity of systemic arteries, especially aorta, have largest effect
to the condition of the cardiovascular system (increased 
significantly during ageing).
Therefore, aorta and other systemic arteries are chosen to include
largest variations:
\begin{align*}
  &  k_{1,3}^{(s)}=\bar k_{1,3} (\alpha_{\textrm{a}}^{(s)} \beta^{(s)})^2,&&\textrm{(aorta)},\\
  &  k_{1,3}^{(s)}=\bar k_{1,3} (\alpha_{\textrm{a}}^{(s)})^2 ,&&\textrm{(other systemic arteries)},\\
  &  k_{1,3}^{(s)}=\bar k_{1,3} (\gamma^{(s)})^2, &&\textrm{(all other blood vessels)},
\end{align*}
where $\alpha^{(s)}\sim\mathcal N(1,25\%)$,
$\beta^{(s)}\sim\mathcal U(1,2.5)$ and
$\gamma^{(s)}\sim\mathcal N(1,10\%)$.
The coefficient $\alpha^{(s)}$ produces 25\% variation to
the PWV of systemic arteries which, for aorta, is further
amplified with $\beta^{(s)}$ giving 65\% maximum variation.
%
%
The slope $k_2$ is also varied with 5\% variation.
To produce small variation between segments, additional 1\% variation
is added to the local PWV ($c_0$) of each segment.
%



\subsubsection*{Heart functions and valve model parameter}


The duration of the pulses $T_c$  are chosen as follows.  
For each subject, HR is drawn from
$\mathcal N(75\:\textrm{min}^{-1},35\%)$, which is rejected if HR $<$
50 min$^{-1}$ to avoid too low heart rates.
For normal sinus rhythm, pulse lengths $T_c$ are shown to follow the
distribution of a (correlated) pink noise \cite{scarsoglio14}.
Therefore, $T_c$ are chosen to be realizations of
pink noise with the mean 60/HR and the variance $\sigma^2$, which varies
among the subjects ($\sigma\sim\mathcal N(0.07,2\%)$).
%

%
To consider variations in heart pumping, we vary
$E_\textrm{fw}^\textrm{max}$ and $\tau_1$ and $\tau_2$ randomly.
For each pulse, we choose
\begin{equation}
E_\textrm{fw}^\textrm{max}\sim \mathcal N(\bar
E_\textrm{fw}^\textrm{max},P^{(s)}\%), \quad   \tau_1=\bar\tau_1 c, \quad
  \tau_2=\bar\tau_2 c, \quad c\sim\mathcal N(1,1\%), 
\end{equation}
 where
$P^{(s)}\sim\mathcal U(0,15)$ represents to level of variations in heart
muscle contraction between pulses, which is modelled to vary between subjects.
The valve model parameters $A_\textrm{eff,max}$,
$A_\textrm{eff,min}$, $\ell_\textrm{eff,min}$, $K_\textrm{vo}$,
and $K_\textrm{vc}$ are varied with 10\% variation.

\subsubsection*{Vascular beds}

%
Microvasculature compliances ($C$) and the reference capillary
resistances ($R_0$ or $R_{0,m}$) are chosen as
$C\sim\mathcal N(\bar C,5\%)$ and $R_0\sim(1.2\bar R_0,20\%)$.
The mean resistance is increased slightly to provide higher,
physiologically more relevant diastolic and systolic pressure levels.
For coronary vascular beds (see Fig.~\ref{fig:vascular_beds}), the
resistances $R_{1}$ and $R_{2}$ and the initial volumes $V_{0,1}$ and
$V_{0,2}$ are perturbed with 10\% variation.

\subsubsection*{Generation of the virtual database}

We generate two datasets: the first is used to train predictors
(training set), and another for the validation of predictions (test
set).
The generation of the training set is described first.

The model is run repeatedly for the parameter variations described
above. 
The initial state for the solution and the model parameters not
specified above are set as in \cite{mynard2015}.
The 1D-model is discretized using varying number of elements in each
segment ($N_\textrm{el}=\lceil 0.5L\rceil$ where $L$ is the length of
the segment) and the 3rd/2nd order
(arteries/veins) Legendre polynomials.
The time stepping for temporal discretization is chosen to be
$\Delta t=2\cdot 10^{-6}$ s.
The level of discretization is experimentally verified to result
sufficiently small discretization error (compared to a very dense
discretization).
We simulate 11 heart cycles to ensure that the
simulation has been converged (e.g. the dependency to the initial
condition is negligible) and the last pulse of each run is used in the
analysis.
The model is run 9986 times.
However, we noticed that similar results can also be achieved with
significantly less samples (e.g. 1000) and therefore we can assume
that the size of database is sufficient.

To ensure that simulations represent physiologically reasonable
solutions, the filtering criteria used in
\cite{willemet2015,willemet2016} are also applied here: a simulation
is accepted only if 1) DBP at the brachial arteries are higher than 40
mmHg, 2) SBP at the brachial arteries are lower than 200 mmHg, the
pulse pressures (SBP - DBP) at the brachial arteries are between
25-100 mmHg, 4), the reflection coefficient of the aortic-iliac
bifurcation satisfies $\abs{R_f}\leq 0.3$.
The reflection coefficient is calculated as
\begin{equation}
  R_f=\frac{Y_\textrm{abd}-Y_\textrm{il,left}-Y_\textrm{il,right}}{Y_\textrm{abd}+Y_\textrm{il,left}+Y_\textrm{il,right}},
\end{equation}
where the characteristic admittances $Y=A_d/(\rho c_d)$ (the subscript
$d$ refers to diastole) are for the distal abdominal aorta
($Y_\textrm{abd}$) and the proximal common iliac arteries
($Y_\textrm{il,left}$, $Y_\textrm{il,right}$).
%

Out of the 9986 cases, 5222 samples are accepted after applying the
above filtering criteria. Out of the rejected samples, 4543 have too
small or large reflection coefficient, 70 have to too small diastolic
BP, 9 have too large systolic pressure, and pulse pressure is too
large for 1115 samples.
The large portion of rejected samples due to insufficient reflection
constants can perhaps be avoided if more precise information about
spatial variations of arterial diameters and stiffness would be
available.

The test set is generated similarly, but with a denser discretization
($\Delta t=0.5\cdot 10^{-6}$ s, 4th/3rd order Legendre polynomials for
arteries/veins).
This dataset comprises of 943 virtual subjects (1792 before filtering).
The training and test set have their own unique virtual patients
without overlap.

\subsubsection*{Simulated PPG signal and calculation of PTT/PATs}

In this study, we consider predictions based on pulse transit and
arrival timings derived from simulated PPG signal.
The measurement locations ($x_\textrm{obs}$) considered in this work are listed
in Table~\ref{tab:measurement_locations}.
PPG signal can be understood as a differential measurement of blood
volume under the sensor.
If we assume that longitudinal variations in the blood veins are
negligible, the blood volume can be assumed to proportional to
$A(x_\textrm{obs},t)$.
Therefore PPG signal is simulated by removing the scale information:
\begin{equation}
  PPG(t)=\frac{A(x_\textrm{obs},t)-A_\textrm{min}}{A_\textrm{max}-A_\textrm{min}}.
\end{equation}
where $A_\textrm{min}$ and $A_\textrm{max}$ are the minimum and
maximum of $A(x_\textrm{obs},t)$ over a period of time.
We, however, note that the scale does matter when considering PTT/PAT timings.

\begin{figure}[htb!]
  \centering
  \begin{tabular}{cc}
    Carotid artery&Femoral artery\\
  \includegraphics[width=0.45\textwidth]{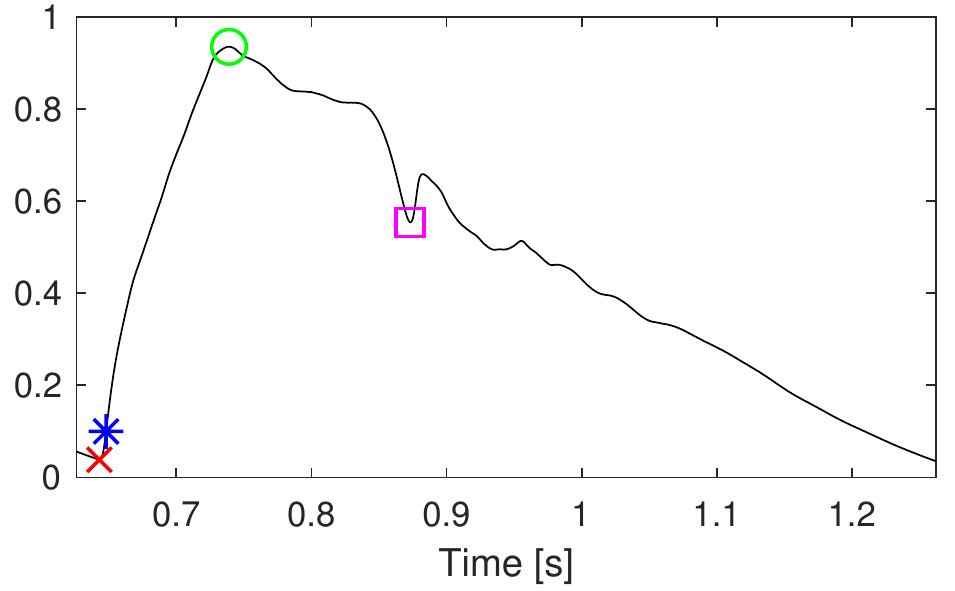}&
  \includegraphics[width=0.45\textwidth]{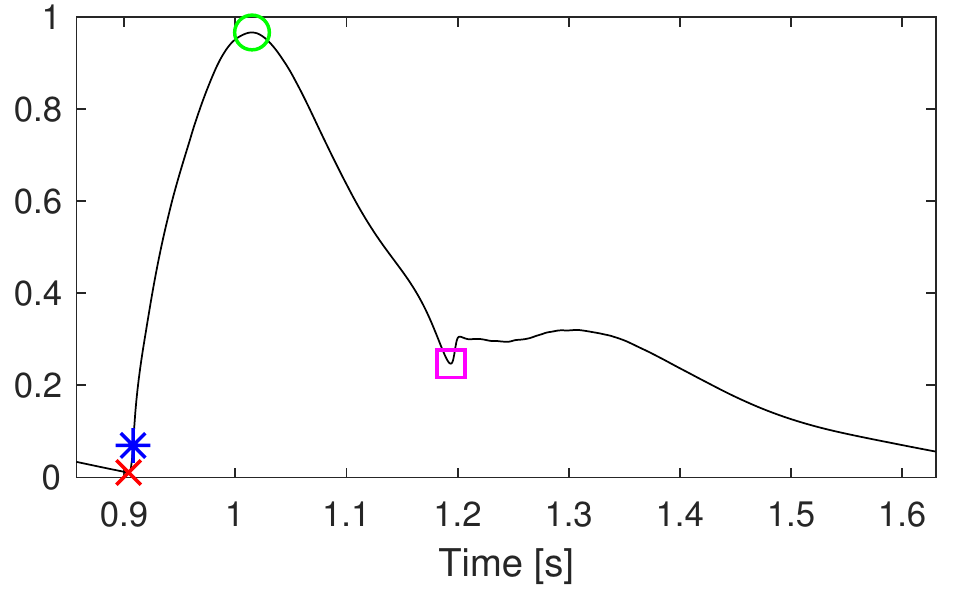}
  \end{tabular}
  \caption{Two example pulses with the considered timings marked: the
    minimum/foot (\PTT; red cross), the maximum (\PTTmax; green
    circle), the maximum of the first derivative (\Dmax; blue star),
    and dicrotic notch (\DAT; magenta square). }
  \label{fig:PTTsexpained}
\end{figure}

Arrival of the pulse can be detected as a valley at the beginning of
systolic period when pressure $p(x_\textrm{obs},t)$ starts increasing
(foot-to-toot PTT; \PTT).
Other timings can also be considered: the peak (maximum; \PTTmax), the
steepest raise (the maximum of the derivate; \Dmax), and the location
of the dicrotic notch (\DAT); see Fig. \ref{fig:PTTsexpained}.
DAT can be detected as the peak in the second derivate during the
diastolic period.
%

The pulse transit times are relative to aortic valve opening which can
be easily detected from simulations: we detect a valley in the
simulated pressure $p(x,t)$ at aortic root (the inlet of the 1D
segment connecting to aortic valve).
For pulse arrival times, simulated R-wave locations can chosen to be
the initiation of the pulse (foot) in the prescribed $E_\textrm{fw}$
for LV.

\begin{table}[!tb]
  \caption{The sensor locations considered in this
    work. $x_\textrm{obs}$ is the location of the sensor within the
    segment and $L$ is the length of the segment.}
  \label{tab:measurement_locations}
  \centering
  \begin{tabular}{llll}
      \hline
      Arteria&Abbreviation&$x_\textrm{obs}$\\
      \hline
      Left common carotid artery&LCA&$0.65L$\\
      Right common carotid artery&RCA&$0.6L$\\
      Left/right radialis artery&LRad/RRad&$0.9L$\\
      Right femoral artery&Fem&$0.5L$\\      
      \hline
    \end{tabular}
  \end{table}


We note that our simplified PPG signal model does not take into account
phenomena such as optical scattering which can induce nonlinear
effects to pulse waveform.
However, we use PPG signal only to infer timings in the pulse and
therefore possible nonlinearities do not have significant effects to
results as long as foots, peaks and notches can be estimated accurately.
Furthermore, we note that other measurement modalities measuring
volume/area of the artery (e.g. ultrasound) can also be considered.

\subsubsection*{Extraction of aPWV, DBP, SBP and SV}

Thea aim is to predict aPWV, DBP, SBP or SV using the combination of
PTT/PAT times and/or HR (input).  These can be extracted from
simulated pulses as follows.
\begin{itemize}
\item aPWV: the wave speed $c(t,x)$ given by (\ref{eq:localPWV})
  averaged over a pulse (integrated numerically). The location $x$ is
  chosen to be the center point of the segment of aorta between the branching
  points of brachiocephalic artery and LCA).
\item DBP, SBP: the minimum and maximum value of $p(x,t)$ at the
  aortic root (the inlet of the 1D segment connecting to aortic valve)
\item SV: the integral of flow $q=AU$ at the aortic root over the
  pulse (calculated numerically).
\end{itemize}
There are also other options to specify aortic PWV.
For example, we can use the foot-to-foot aortic PWV by detecting
arrivals of pulses to aortic root and the aortic-iliac bifurcation,
but this leads only to very minor differences in the results (the Pearson
correlation for between these aPWVs is $r>0.99$).
Relationships between these different options are studied in \cite{willemet2015}.

\begin{figure}[htb!]
  \centering
  \begin{tabular}{llll}
    (a)&\hspace{-.55cm} (b)&\hspace{-.55cm} (c)&\hspace{-.55cm} (d)\\
  \includegraphics[height=3cm]{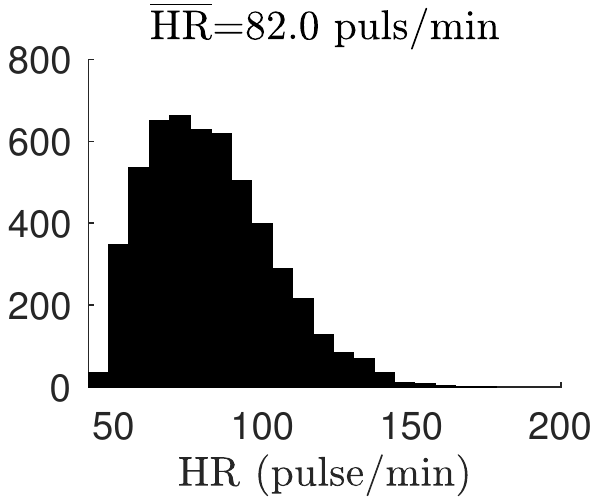}&\hspace{-.55cm}
  \includegraphics[height=3cm]{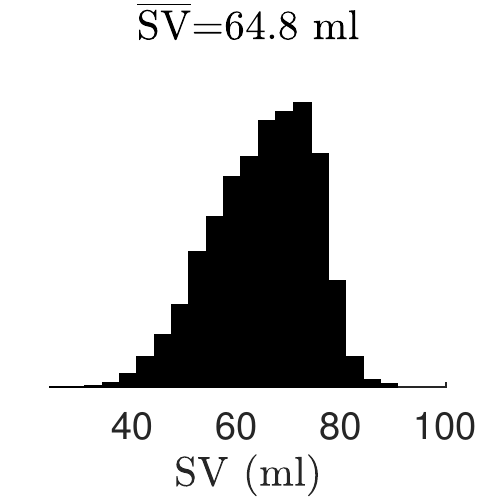}&\hspace{-.55cm}
  \includegraphics[height=3cm]{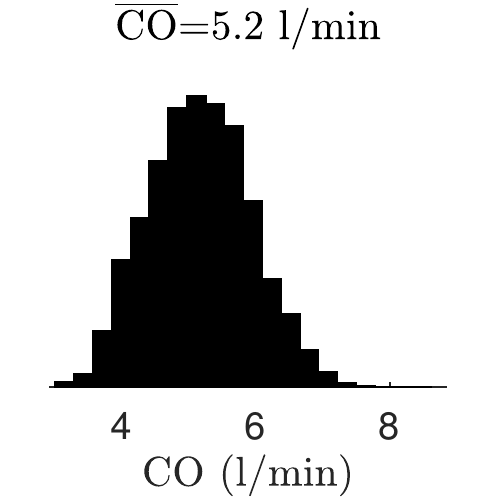}&\hspace{-.55cm}
  \includegraphics[height=3cm]{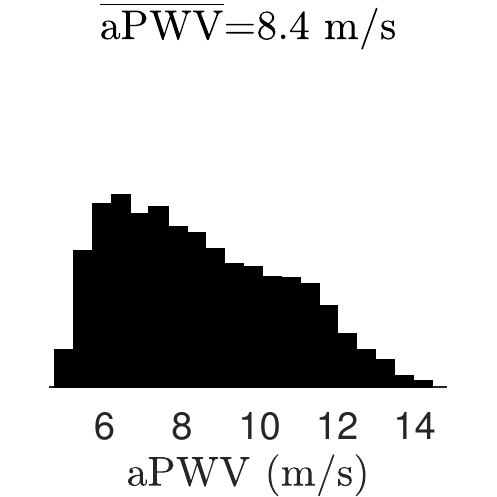}\\
    (e)&\hspace{-.55cm} (f)&\hspace{-.55cm} (g)&\hspace{-.55cm} (h)\\
  \includegraphics[height=3cm]{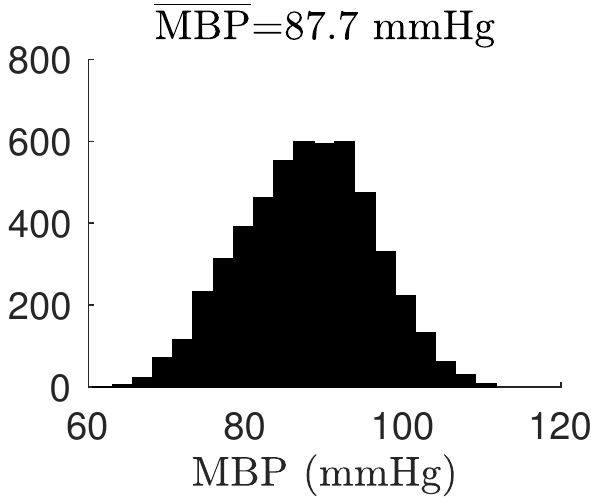}&\hspace{-.55cm}
  \includegraphics[height=3cm]{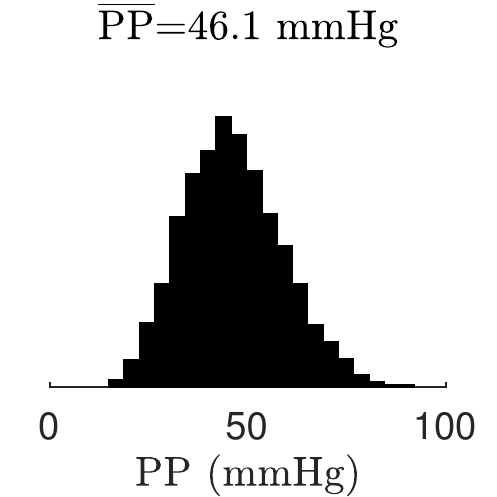}&\hspace{-.55cm}
  \includegraphics[height=3cm]{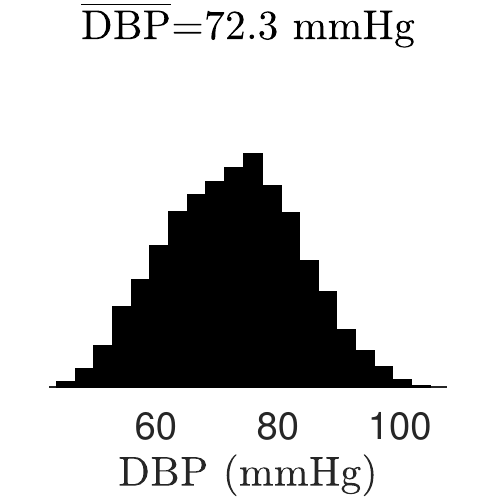}&\hspace{-.55cm}
  \includegraphics[height=3cm]{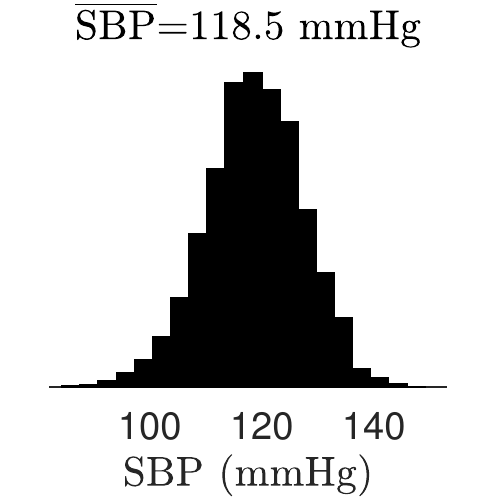}
  \end{tabular}
  \caption{Distributions of selected metrics for the virtual database
    (training set; after filtering): (a) heart rate (HR), (b) stroke
    volume (SV), (c) cardiac output (CO), (d) aortic PWV, (e) mean
    blood pressure (MBP), (f) pulse pressure (PP), (g) diastolic
    pressure (DPB), and (h) systolic blood pressure (SPB). The means
    of the metrics are shown in the title. }
  \label{fig:histograms}
\end{figure}

The distributions of selected metrics of the generated virtual database are
shown in Fig. \ref{fig:histograms}.
As a general finding, we note that there are strong correlations
between DAT and pulse length (1/HR) signals (Pearson correlation
$r=0.96-0.98$).
Due to this strong correlation, using HR and DAT as input provides
very similar predictions (which can also be seen in the results below).

\section{Gaussian process model for predictions}
\label{sec:comp-pred}

We apply Gaussian process regression for the computation of
predictors.
GPs are widely used, for example, in machine learning,
hydrogeology and analysis of computer experiments (e.g. see
\cite{rasmussen06,rubin03,kennedy01}).
GPs also provide flexible predictors that can handle non-linear
relationship between input data and the response variable as well as
uncertainties in the data.
However, we note that any other class of regressions capable of
nonlinear relationships can also be used for the analysis.
For example, similar results can be achieved with multivariate adaptive regression
splines \cite{friedman91}.

A GP is a stochastic process
$f(z)$ ($z\in \R^d$) such that $f(z_1), \ldots, f(z_n)$ is a multivariate Gaussian
random variable for all combinations of $z_1,\ldots, z_n$.
It can be described by the specifying mean function $\mu(z)=\mathbf{E} (f(z))$
and the covariance function $k(z,z')=\textrm{cov}(f(z),f(z'))$.
For more details, see e.g. \cite{rasmussen06}.

Consider a case in which the inputs $z$ are a vector of PTT or PATs and
possibly HR and $y$ is the response variable (aPWV, DBP, SBP or SV).
We model the response variables as
\begin{equation}\label{eq:GPmodel}
  y(z)=h(z)^T\beta +f(z) +\epsilon,
\end{equation}
where $h(z)$ is a vector of (deterministic) basis functions, $\beta$
is a vector of basis function coefficients, $f(z)$ is a GP with zero
mean and covariance function $k(z,z')$, and $\epsilon$ is an Gaussian
white noise.
The first term represents mean behavior of the GP model.
The GP term models non-linear relationship between input data and the
response variable as well as correlated uncertainties in the data.

Training data comprises of input-output pairs
$\{(z_i,y_i); i=1,\ldots,N\}$.
We assume that $y_i$'s are output of the above model
i.e. $y_i=y(z_1)$.
Furthermore, let $Z'=(z'_1,\ldots,z'_p)$ be inputs for which we want
to calculate predictions.
Then $Y=(y_1,\ldots,y_N)$ and $Y'=(y(z'_1),\ldots,y(z'_p))$ are both
Gaussian and the conditional distribution of $Y'$ given $Y$ is (see
e.g. \cite[Appendix A.2]{rasmussen06}),
\begin{equation}
  p(Y'|Y)=\mathcal N(\mu_{Y'}+\Sigma_{Y'Y}\Sigma_{Y}^{-1}(Y-\mu_Y), \Sigma_{Y'}+\Sigma_{Y'Y}\Sigma_{Y}^{-1}\Sigma_{YY'})
\end{equation}
where $\mu_Y$ and $\Sigma_Y$ denotes the mean and covariance
of $Y$ and $\Sigma_{YY'}$ is the cross-covariance of $Y$ and $Y'$.
The means and covariances can be calculated by pluggin in the model
\eqref{eq:GPmodel}, which gives
\begin{eqnarray}
\mu_{Y'|Y}&=&h(Z')^T\beta +k(Z',Z)(k(Z,Z)+\sigma^2_\epsilon I)^{-1}(Y-h(Z)^T\beta)\\
\Sigma_{Y'|Y}&=&k(Z',Z') -k(Z',Z)(k(Z,Z)+\sigma^2_\epsilon I)^{-1}k(Z,Z')
\end{eqnarray}
where $h(Z')$ and $k(Z',Z)$ are shorthand notations for the
vector and matrix with the components $h(z'_i)$ and $k(z'_i,z_j)$, respectively.
The above conditional mean gives us an prediction of $Y'$ with a
confidence estimate given by the conditional covariance.

In this study, the covariance function are chosen to be Matern kernel
function with $\nu=3/2$ with a separate length scales for each input
parameter.
This kernel function can be written as
\begin{equation}
  k(z,z')=\sigma^2\left(1+\frac{\sqrt 3} r\right)\exp\left(-\sqrt 3 r\right), \quad
  r=\left(\sum_m^d \frac{(z_i-z_j)^2}{\ell_m^2}\right)^{1/2}
\end{equation}
where $\sigma^2$ is the variance and $\ell_m$ are the length scales
for each input.
We note that the choice of the kernel function does not have a large
effect to the results as our sample size is large.
For example, our experiments show that use of the squared exponential
covariance function gives very similar results with differences of the
same scale as the prediction uncertainty.

The predictors are computed using fitrgp function in MATLAB
Machine Learning Toolbox which provides numerically efficient
implementation for the GP regression.
The basis functions $h(z)$ are chosen to be linear.
The fitrgp function also estimates hyperparameters $\theta$ ($\beta$,
$\sigma_\epsilon^2$, $\sigma^2$, $\ell_1$, \ldots, $\ell_d$) by
minimizing the negative loglikelihood,
\begin{equation}
  \mathcal L(\theta)=-\log p(y|Z,\theta)=\frac 1 2 y^T
  \Sigma_\theta^{-1}y+\frac 1 2 \log\mathrm{det}\Sigma_\theta +\frac n
  2 \log 2\pi
\end{equation}
where $\Sigma_\theta = k(Z,Z;\theta)+\sigma_\epsilon^2I$.
The optimization is carried out using a subset of observations to
avoid high computational load.
The parameters of fitrgp related to this hyperparameter optimization
are chosen to be the default values.


\section{Results}
\label{sec:results}
In this section, we apply GP regression to predict aPWV, DBP, SBP and
SV using combinations of different type of PTT/PAT timings and HR
as input.
We train a GP predictor separately for each considered combination as
described above.
For validation, we apply the trained predictor to the test set and calculate
Pearson correlation between the predictions and ground truth values.
Tables in \supmat{} also report 95\% confidence intervals (CI)
for the Pearson correlations (BCa bootstrapping intervals
\cite{efron87_BCa}).
Each table  also highlights selected predictions having with
Pearson correlations.
However, we note that the order of Pearson correlations should be
considered as indicate rather than a definite order of performance due
to the uncertainty especially when differences are small.



\subsection*{Predictions of aPWV}
\label{sec:pred-aort-pwv}

\begin{figure}[ht!]
  \centering
  \begin{tabular}{ccc}
    \PTT   & \PTTmax  &\Dmax \\
  \includegraphics[width=0.31\textwidth]{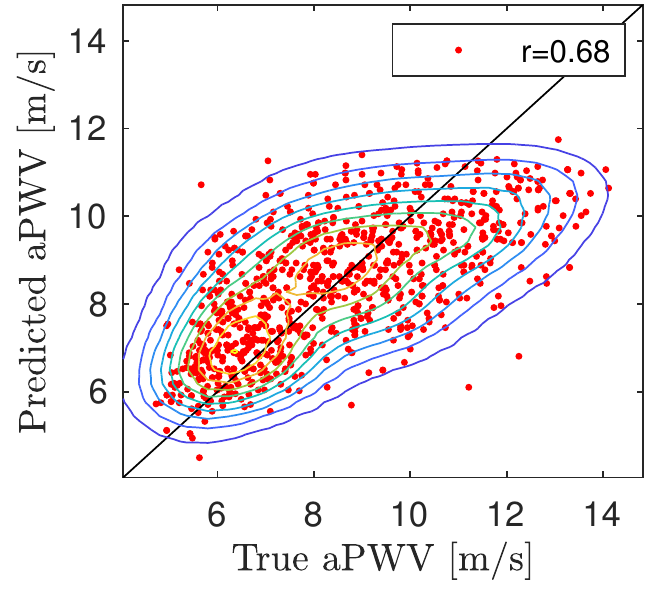}&
  \includegraphics[width=0.31\textwidth]{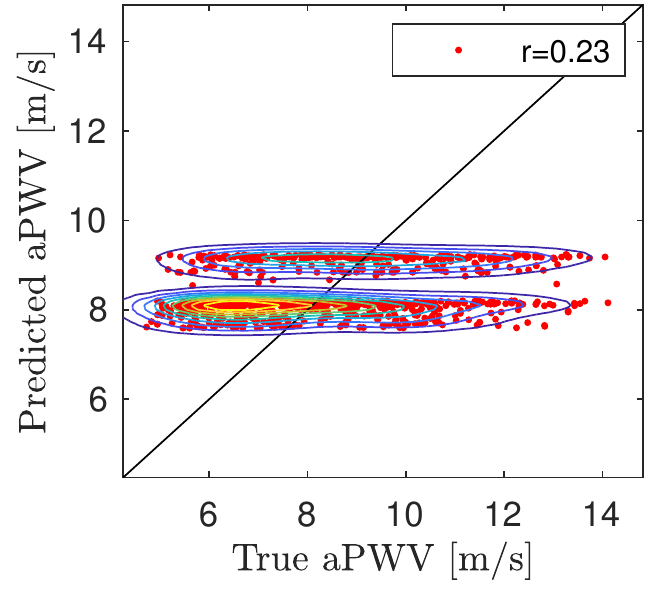}&
  \includegraphics[width=0.31\textwidth]{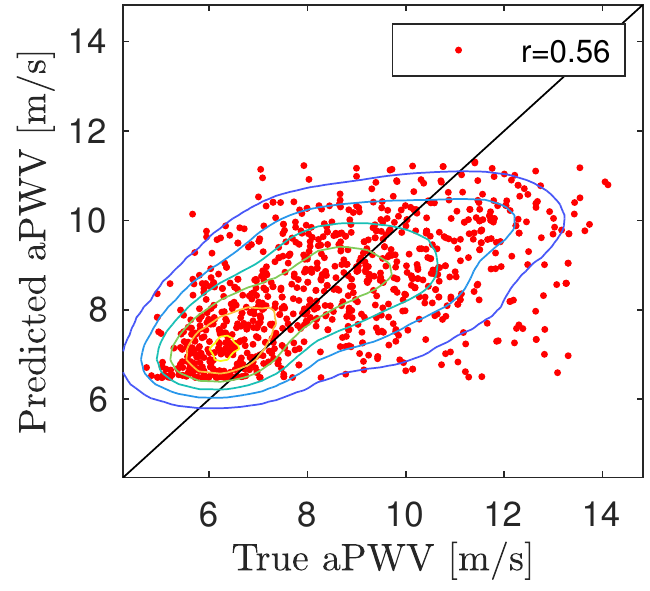}\\
   \DAT& \PTT + \PTTmax &\PTT + \PTTmax + \DAT \\
  \includegraphics[width=0.31\textwidth]{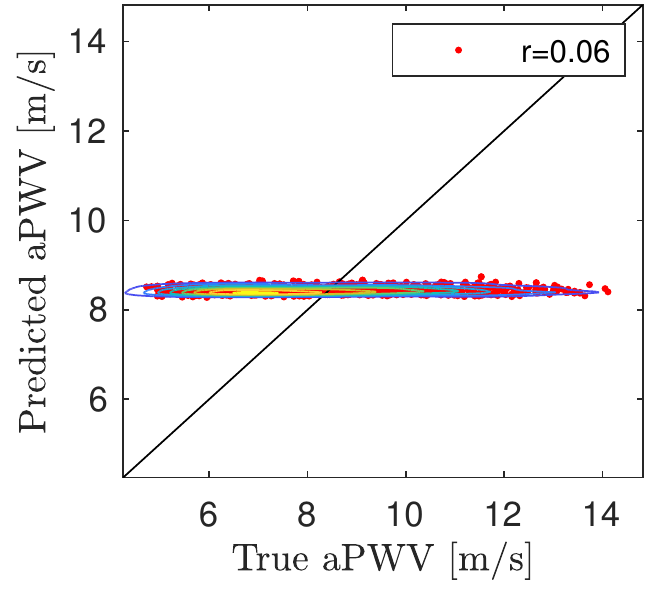}&
  \includegraphics[width=0.31\textwidth]{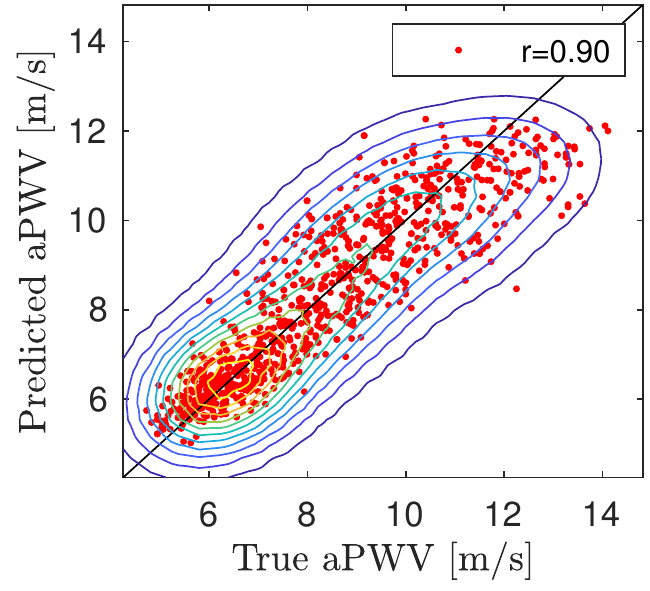}&
  \includegraphics[width=0.31\textwidth]{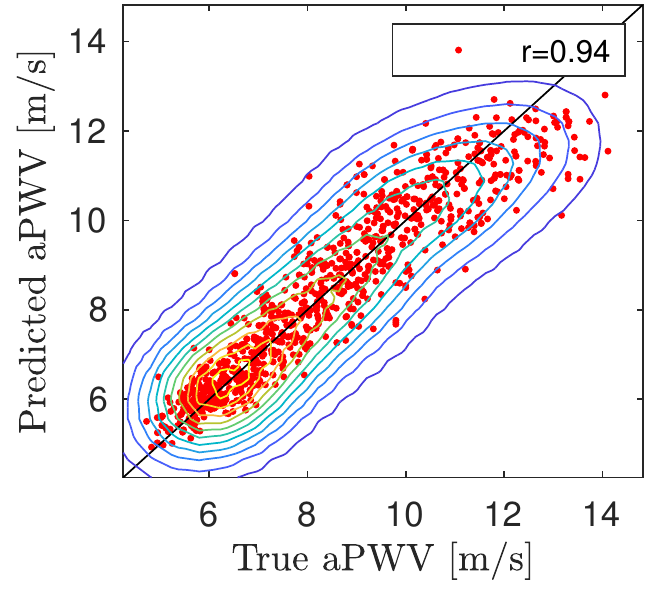}
  \end{tabular}
  \caption{ Accuracy of the aortic PWV predictions using pulse transit time (PTT)
    measurements from left carotid artery (LCA). Signals: heart rate
    (\HR) and pulse transit times to the foot of signal (\PTT), peak
    of signal (\PTTmax), the point of steepest raise (\Dmax), and the
    dicrotic notch (\DAT).  }
  \label{fig:PTT_LCA_PWV}
\end{figure}

%
%
Fig. \ref{fig:PTT_LCA_PWV} shows predictions of aPWV for a selected
set of combinations when the measurement location is LCA.
Table \ref{tab:results_PPT_LCA} in \supmat{}) summarizes the results
for the complete set of combinations.

%
%
The results show that using \PTT or \Dmax as a single input gives
moderate accuracy and predictions using either \HR, \PTTmax, or \DAT
are insufficient.
Performance can be improved by combining multiple
different timings.
For example, the accuracy is significantly improved if both \PTT and
\PTTmax are used for predictions ($r=0.90$).
Furthermore, including also \DAT provides the accuracy of $r=0.94$, and adding
other timings does not significantly improve accuracy any further.

Measurements from RCA provide less accurate predictions
(Table~\ref{tab:results_PTT_RCA}): for example, the combination of \PTT, \PTTmax,
\Dmax{} and \DAT provides one of highest accuracies for RCA ($r=0.79$),
but is still only moderate.
Such results can be expected as pulse waves travel shorter distance in
aorta and also travel through brachiocephalic artery (see
Fig.~\ref{fig:Arterial_System}) inducing additional variations to the
(average) wave speeds.

Performance of wrist measurements (LRad / RRad) are even worse (see
Table~\ref{tab:results_PTT_LRad} for LRad; results for RRad are
similar). For example, the highest accuracy ($r=0.73$) can be achieved
with the combination of \PTT, \PTTmax, \Dmax{} and \DAT.
This is also expected as relative large part of the arterial tree to
these measurement locations are comprised of brachial and radialis
arteries with their own variations to PWV.
On the other hand, measurements from lower limb could provide better
performance: for right femoral artery, we can achieve $r=0.75$ using
\PTT and $r=0.84$ using \PTT, \PTTmax, \Dmax{} and \DAT (Table
\ref{tab:results_PTT_RFem}).
In this case, pulse travels though the whole aorta to reach these
measurement locations.

As mentioned above, in practice, the R-peak location in ECG signal is
often used as a surrogate to aortic valve opening due to simpler
measurement.
However, using PATs gives only mediocre accuracy
compared to PTT due to the physiological variations in
PEP \cite{balmer18,kortekaas18}.
Our finding are similar, see for example, Fig.~\ref{fig:PAT_LCA_PWV}
and Table \ref{tab:results_PAT_LCA} for LCA.
The highest accuracy is $r=0.79$ (e.g. \PAT, \PATmax, \PATDmax{} and \HR) which is significantly worse
compared to using PTTs.

Another approach to avoid measurement of aortic valve opening is to
consider differences of pulse arrival times to two distal locations.
Such setup also allows us to avoid the influence of PEP variations.
Results for measurement between LCA and Fem can be seen in
Fig.~\ref{fig:PTT_LCA-Fem_PWV} and Table~\ref{tab:results_PTT_FemLCA}:
difference of \PTT gives $r=0.76$ which is slightly better than using
normal \PTT measurement from Fem, but not as good as normal \PTT
measurement from LCA.
The highest accuracy ($r=0.87$) can be obtained, for example, with
\PTT, \PTTmax, \Dmax{} and HR.
The predictions of PWV that use the difference between LCA and RCA or
the difference between LRad and RRad are less accurate
($r\approx 0.75-0.78$ at best); see
Tables~\ref{tab:results_PTT_LCARCA} and
\ref{tab:results_PTT_LRadRRad}.
%


\subsection*{Predictions for blood pressure}

\begin{figure}[ht!]
  \centering
  \begin{tabular}{ccc}
    \HR & \PTT   & \PTTmax  \\
  \includegraphics[width=0.3\textwidth]{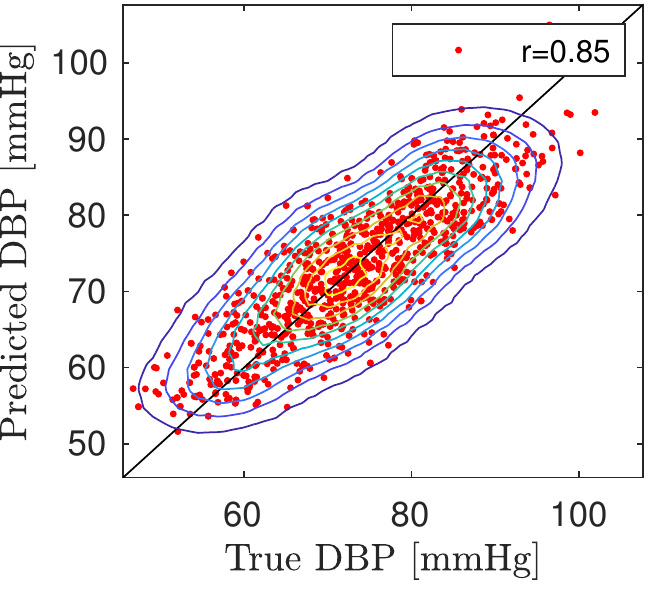}&
  \includegraphics[width=0.3\textwidth]{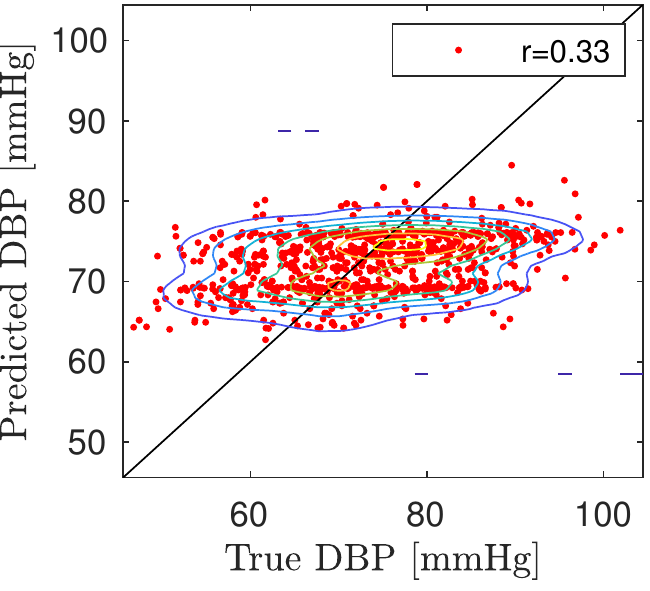}&
  \includegraphics[width=0.3\textwidth]{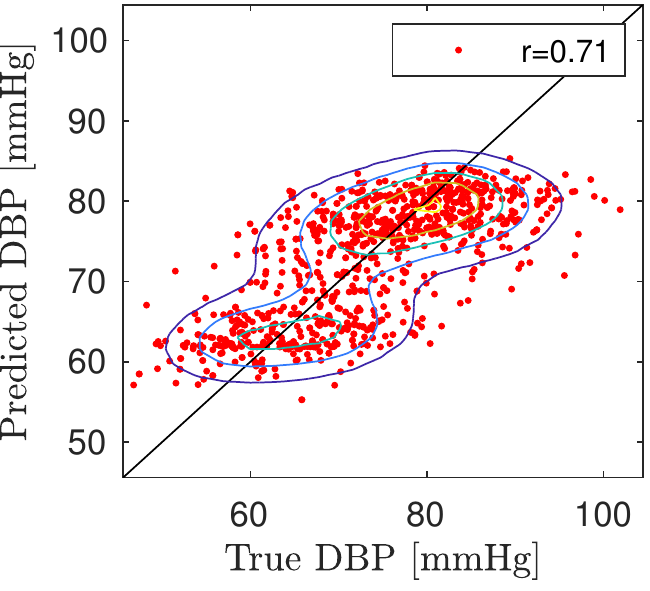}\\
    \DAT &  \PTT + \HR  & {\small \PTT + \PTTmax+ \Dmax + \DAT}\\
  \includegraphics[width=0.3\textwidth]{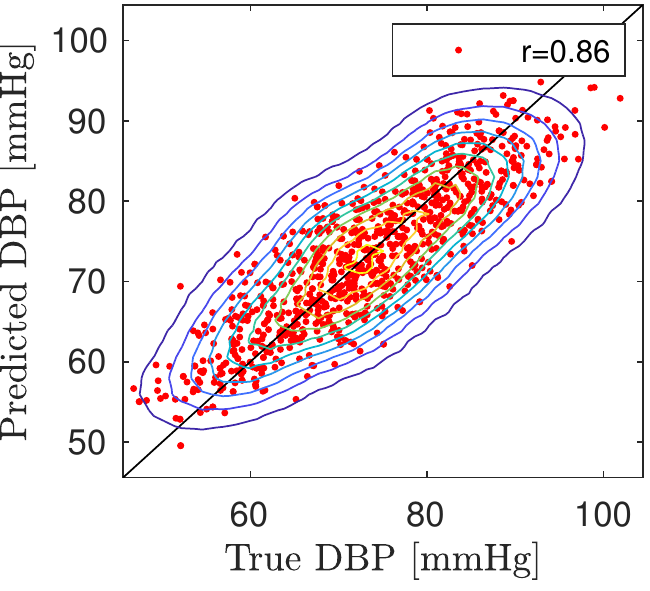}&
  \includegraphics[width=0.3\textwidth]{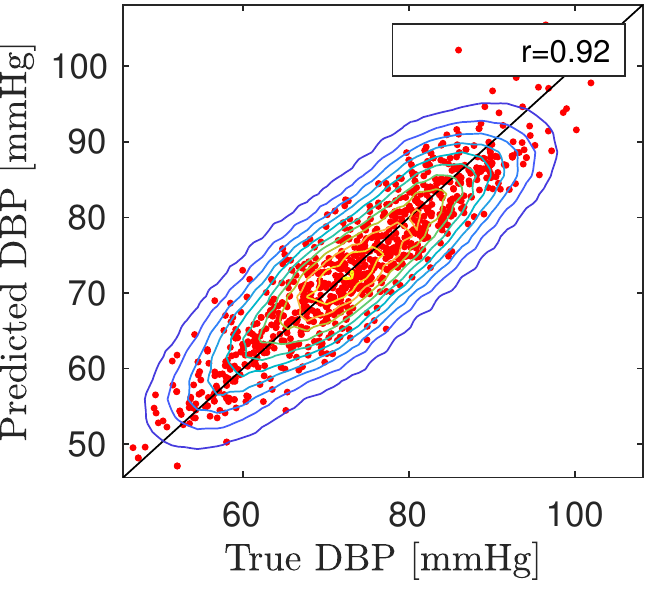}&
  \includegraphics[width=0.3\textwidth]{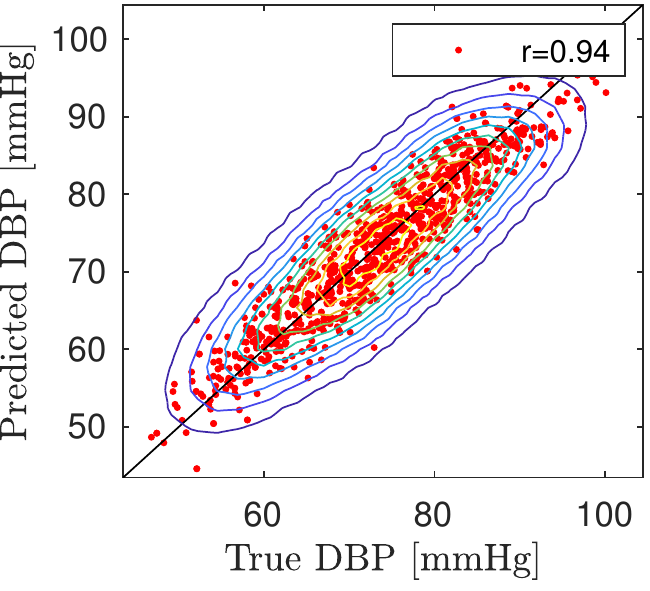}
  \end{tabular}
  \caption{ Accuracy of the DBP predictions using pulse transit time (PTT)
    measurements from left carotid artery (LCA). Signals: heart rate
    (\HR) and pulse transit times to the foot of signal (\PTT), peak
    of signal (\PTTmax), the point of steepest raise (\Dmax), and the
    dicrotic notch (\DAT).  }
  \label{fig:PTT_LCA_DBP}
\end{figure}

\begin{figure}[ht!]
  \centering
  \begin{tabular}{ccc}
    \HR & \PTT   & \PTTmax  \\
  \includegraphics[width=0.3\textwidth]{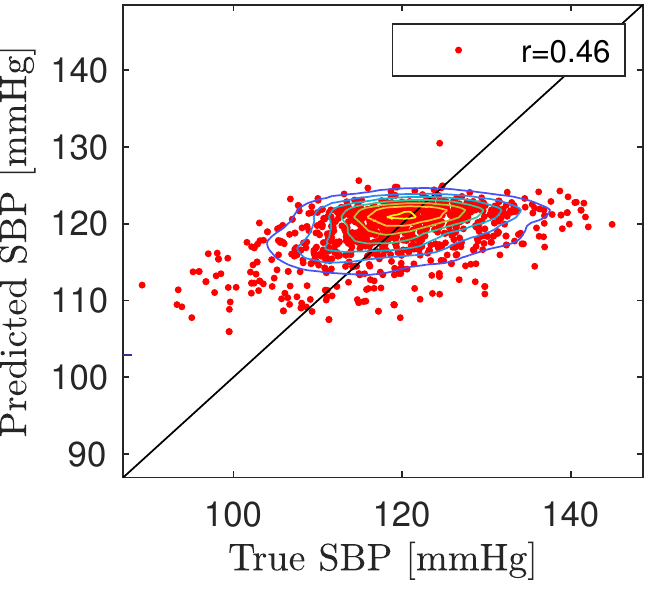}&
  \includegraphics[width=0.3\textwidth]{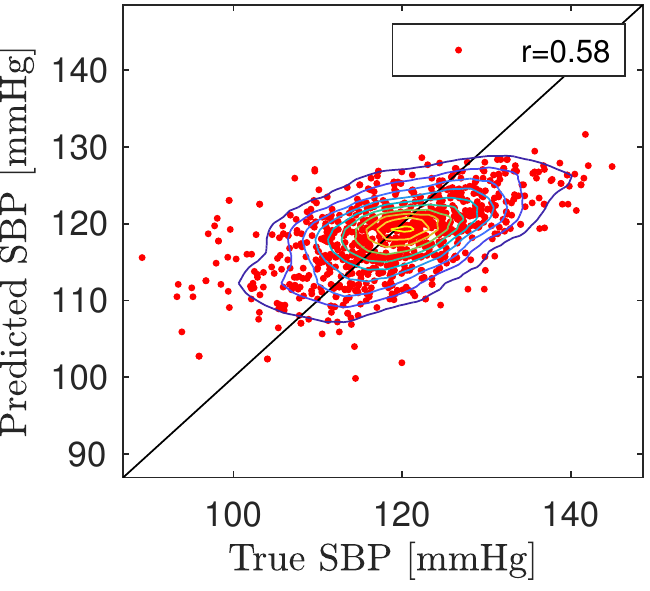}&
  \includegraphics[width=0.3\textwidth]{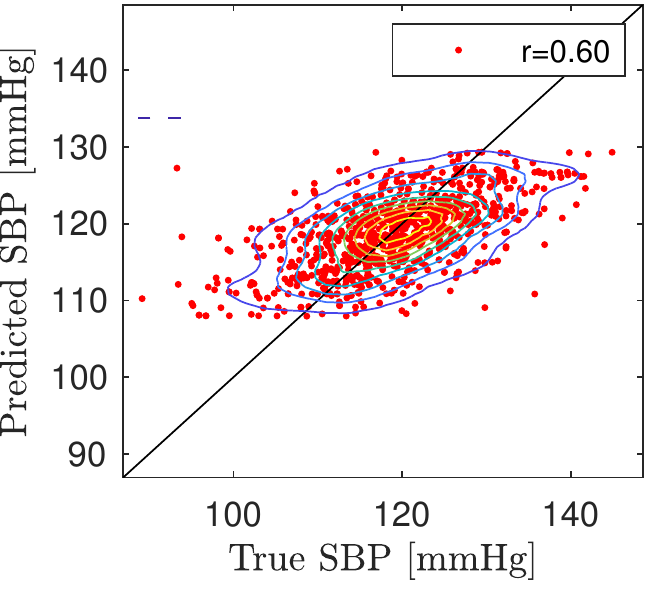}\\
    \Dmax &  \PTT + \HR &  \PTT +  \Dmax + \DAT\\
  \includegraphics[width=0.3\textwidth]{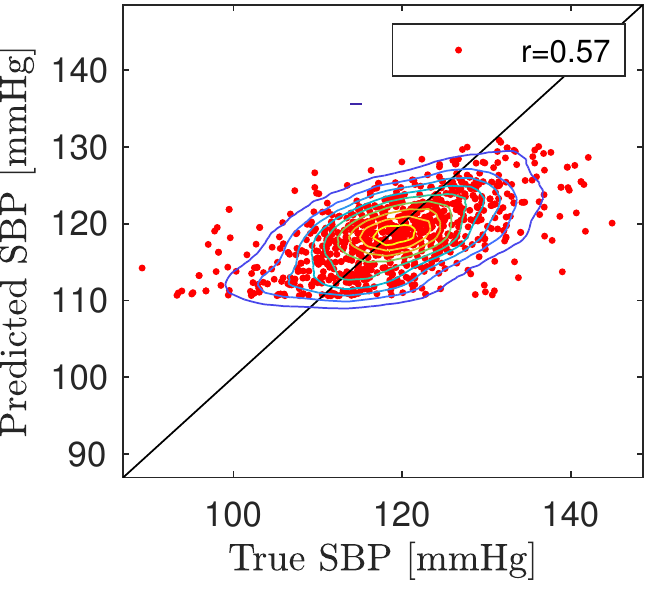}&
  \includegraphics[width=0.3\textwidth]{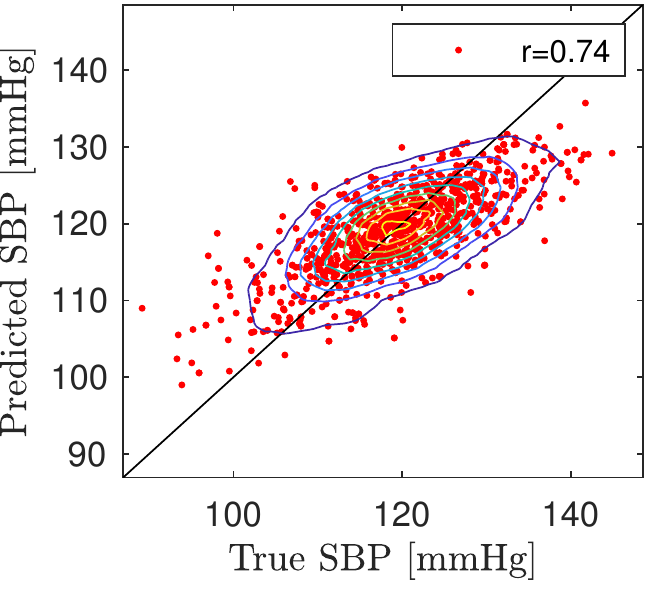}&
  \includegraphics[width=0.3\textwidth]{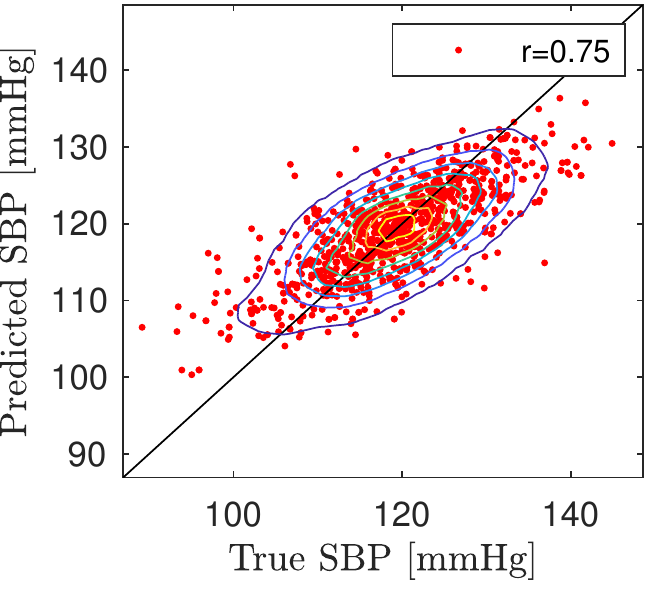}
  \end{tabular}
  \caption{ Accuracy of the SBP predictions using pulse transit time (PTT)
    measurements from left carotid artery (LCA). Signals: heart rate
    (\HR) and pulse transit times to the foot of signal (\PTT), peak
    of signal (\PTTmax), the point of steepest raise (\Dmax), and the
    dicrotic notch (\DAT).  }
  \label{fig:PTT_LCA_SBP}
\end{figure}

Fig.~\ref{fig:PTT_LCA_DBP} and \ref{fig:PTT_LCA_SBP} show predictions
for DBP and SBP for selected PTT time combinations when measurements
are taken from LCA; see also Table \ref{tab:results_PPT_LCA} for all
combinations.
For DBP, predictions using \PTT as a single input achieves very low
accuracy ($r=0.33$).
Significantly more accurate predictions can be achieved using HR
($r=0.85$) or \DAT{} ($r=0.86$).
For SBP, the performance of PTT based predictions is better but still
quite low ($r=0.58$ for \PTT{} and $r=0.60$ for \PTTmax{}).
Predictions can be improved by adding additional input timings.
For DBP, combining \PTT with \HR{} or \DAT{}  gives $r= 0.92$ and the highest
accuracy $r=0.94$ is obtained with \PTT, \PTTmax, \Dmax{} and \DAT.
Additional input timings also improves performance of SDB predictions:
\PTT and HR/\DAT{} results in $r=0.735$ and the highest accuracy is
$r=0.75$ (\PTT, \PTTmax, \Dmax{} and \DAT{}).
Findings the other measurements locations are similar; see Tables
\ref{tab:results_PTT_RCA}, \ref{tab:results_PTT_LRad} and
\ref{tab:results_PTT_RFem}.

We also consider predictions from pulse arrival times (i.e. using
R-peak as a reference timing).
Compared to PTT times, the results are of mixed accuracy; see Table
\ref{tab:results_PAT_LCA} for PAT measurements from LCA.
For DBP, using \PAT{} as single input yields insufficient predictions
($r=0.19$), but \PATmax{} gives
moderate accuracy ($r=0.67$).
Combinations of different PAT timings can even achieve higher
accuracy than using PTTs: for example, $r=0.95$ with \PAT{} and \DAT{} and
$r=0.96$ for \PAT{}, \PATmax{}, \PATDmax{} and \PATDAT{}.
For SBP, \PAT{} provides slightly better accuracy compared to \PTT{}
($r=0.62$), but otherwise results are similar.

As with aPWV, we consider differences of pulse transit/arrival times
measured with two sensor.
Measuring between LCA and Fem gives very similar performance to
PTT measurements from LCA (Table \ref{tab:results_PTT_FemLCA}).
However, other considered setups provide less accurate results: see
Table \ref{tab:results_PTT_LCARCA} for differences between LCA and RCA
measurements and Table
\ref{tab:results_PTT_LRadRRad} for differences between measurements from radialis arteries.

\subsection*{Prediction of SV}

Results show that HR has largest contribution to the predictions of
SV, meanwhile performance with pulse transit or arrival timings
(without HR information) can only provide moderate accuracy at best.
For example, Fig. \ref{fig:PTT_LCA_SV} and
Table~\ref{tab:results_PPT_LCA} show the predictions using
measurements from LCA.
Predictions with HR as a single input reaches $r=0.81$, but
predictions using \PTT or \Dmax are insufficient estimates ($r<0.25$)
and predictions with \PTTmax{} are of moderate accuracy ($r=0.60$).
SV can be predicted with good accuracy with \DAT, but this is due
to the strong correlation between HR and \DAT as mentioned above.
Furthermore, significant improvements will not be achieved by combining
several inputs.
For example, highest accuracy is $r=0.83$ which can be obtained, for
example, with \PTT, \Dmax{} and \HR).
Results are similar for all other measurement setups; see Tables
\ref{tab:results_PTT_RCA}-\ref{tab:results_PTT_LRadRRad}.

\section{Discussion} 
\label{sec:discussion}


This paper assessed theoretical limitations for the prediction of aortic
pulse wave velocity (aPWV), DBP/SBP and SV from pulse transit and
arrival time measurements.
We applied a virtual database approach proposed by Willemet et al
\cite{willemet2015,willemet2016} in which a cardiovascular simulator
is used to generate a database of virtual subjects.
In this work, we applied one-dimensional haemodynamic model by Mynard
and Smolich \cite{mynard2015} to construct a simulator
for entire adult circulation.
This simulator was used to generate a large database of synthetic
blood circulations with varied physiological model parameters.
The generated database was then used as training data for Gaussian process
regressors.
Finally, these trained regressors were applied to another synthetic
database (test set) to assess capability of regressors to predict
aPWV, SDB, DBP and SV using different combinations pulse
transit/arrival time and HR measurements.

The results indicate that aPWV and DBP can be estimated from PPG
signal with a high accuracy (Pearson correlation $r>0.9$ between true
and predicted values for measurement from left carotid artery) when,
in addition to foot-to-foot PTT time, information about the peak and
dicrotic notch location is also given as input to the predictor.
The predictions of SDB were less accurate ($r=0.75$ at best).
For SV, accurate predictions were mostly based on heart rate, with
only a very minor improvement in accuracy when PTT timings were also
included as inputs.

As this was entirely in silico study, it is not guaranteed that the
result can be applicable to the real world as is.
However, the aim of the study was to give preliminary results about
correlations between the cardiac indices and PTT/PAT timings and the
applicability of such predictions.
The hope is that the results could to be extended to real clinical
applications in future research.

The limitations to be addressed in future are the following. 
First, the cardiovascular model has its limitations.
Although previous studies have shown that similar cardiovascular
models can be used to simulate human physiology relatively well
\cite{alastruey2011,matthys2007,olufsen2000}, not all
physiological phenomena are fully covered in the Mynard's model.
One example of such phenomenon is respiration.
The effect of respiration can be important as the breathing and
cardiac cycles are in a close interaction.
Several physiological factors, such as the changes in the
intrathoracic pressure and the variation in the interbeat intervals
modulate the cardiac mechanics and blood outflow from the
heart.
Even the timing of the shorter cardiac cycles coupled with the longer
respiratory cycles has effects on the central circulation.
When we considering a healthy heart, the effects of respiration can
perhaps be managed by interpreting different virtual subjects to
represent inspiratory and expiratory phases of the breathing.
Other phenomena that are not covered by the model are, for example,
gravity and baroreceptors.
Furthermore, lumped parameter models that are used for heart and
vascular beds were relatively simple approximations.
However, new analytical methods allow us to bridge the models and
human bodily functions \cite{casas17}.

The chosen baselines and variations of the model parameters were
chosen to represent healthy subject.
The choices, however, can be subjective due to the limited amount of
(probabilistic) information.
Our attempt were to produce variations such that the virtual
population covered by the chosen parameter variations includes real
physiological variations.
We, however, emphasize that the presented approach is not limited to
the chosen parameters variations and it can be adjusted if more
precise information becomes available.

Due to the limited phenomena covered by the model, the results may not
be reliable when considering subjects with medical conditions.
For example, the simplified heart model and variations of related
model parameter may not present subjects with heart diseases.

In this study, we only considered pulse transit and arrival type of
time information as the input to the predictor.  
Predictions could potentially be improved with other kinds of
additional information.
For example, aortic PWV predictions could be improved by using
information about the distances between aorta and/or measurement
points.
Information about arterial path lengths could have been easily used in
our simulation analysis, but in practice such information would
require clinical measurements such as MRI
\cite{sugawara16,bossuyta13}.
On the other hand, the arterial path length are often estimated using
the body lengths or measuring distances of certain points in the body
\cite{sugawara16,bossuyta13}.
Such information was not used in this simulation study as precise
statistical knowledge of connection between such body measurement and
arterial length was not available.
Instead, Gaussian process regressors implicitly marginalize predictions
over different arterial lengths that are present in the virtual
database.

Ultimately it would be beneficial to develop approaches that do not
need reference measurement (aortic valve opening/R-peak).
For example, Choudhury et al \cite{choudhury2014} presented a machine
learning algorithm which uses raise times and pulse widths derived
from PPG signal to predict DBP and SBP.
Furthermore, deep learning approaches could perhaps be used to infer
optimal information from PPG waveform.
These are subject of our future research.



\section*{Acknowledgments}

This study was supported by the Seamless Patient Care project
(Business Finland - former Tekes - the Finnish Funding Agency for
Technology and Innovation grant funding grant 1939/31/2015).



\newpage 
\appendix
\section*{Appendix: Additional result}
\renewcommand\thefigure{A\arabic{figure}}    
\renewcommand\theequation{A\arabic{equation}}    
\renewcommand\thetable{A\arabic{table}}

Section \ref{sec:pred-aort-pwv-extra} presents additional predictions for
aortic PWV. 
Additional predictions for blood pressures are shown in Section
\ref{sec:addit-figur-pred} and predictions for stroke volume are shown
in \ref{sec:addit-figur-pred-1}.

All results are collected to the tables presented Section \ref{sec:tables}.
The confidence intervals (CI) that are computed using bootstrapping as
follows.
A bootstrapping dataset is formed by resampling or picking samples
from the training using sampling with replacement.
The GP regressor is then trained using this bootstrapping dataset.
The results are computed using as previously using a bootstrapped test
set (resampled with replacement from the original test set).
This procedure is repeated 500 times.
The CIs are computed as bias-corrected and accelerated (BCa) intervals
\cite{efron87_BCa,fox2002bootstrapping}.
The related acceleration parameter is commonly estimated using the
jackknife resampling which is similar to bootstrapping except 
resampled datasets are formed by excluding samples one-by-one.
However, in our study, the original jackknife would be computationally
very expensive as resampling would be repeated 5222 times.
Therefore, we used ``partial'' jackknife resampling in which the
procedure is repeated 500 times such that each time an excluded sample
is chosen randomly.
It is to be noted that in some rare cases CIs do not include the
actual estimate.
This is a property of bootstrapping intervals as bootstrapping sets
include less independent samples (random sampling with replacement
takes roughly 3000-3300 independent samples from the training set of
5222 samples) and none of GPs computed using the bootstrapping
datasets do not reach the same accuracy as the GP model trained using
the original training set.

\ifdefined\addreftosupmat
\bibliography{references}
\fi

\subsection{Additional figures for prediction of aortic PWV velocity}
\label{sec:pred-aort-pwv-extra}

  \begin{center}
  \begin{tabular}{ccc}
   \PAT   & \PATmax  &  \PATDmax\\
  \includegraphics[width=0.31\textwidth]{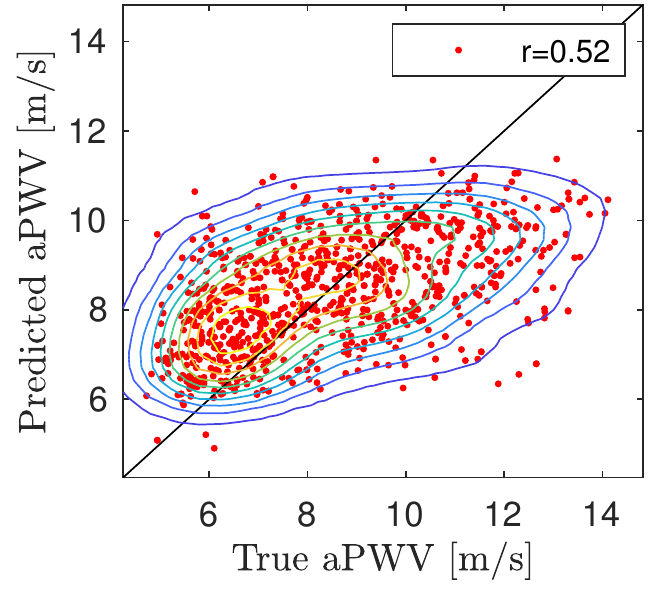}&
  \includegraphics[width=0.31\textwidth]{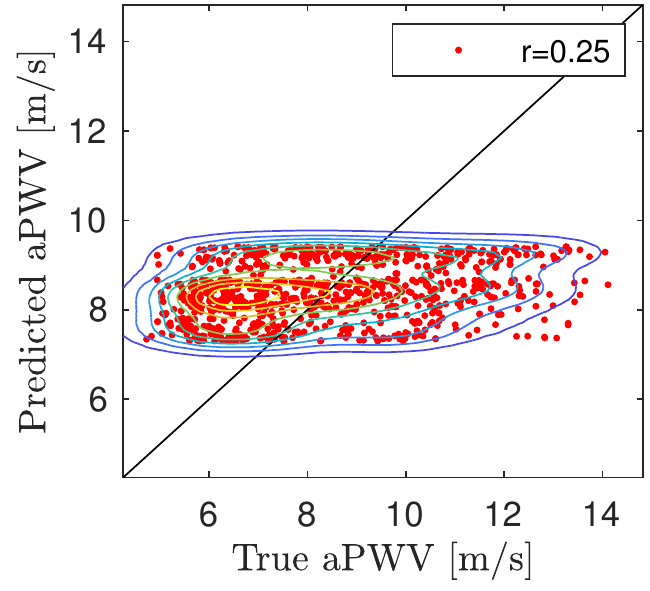}&
  \includegraphics[width=0.31\textwidth]{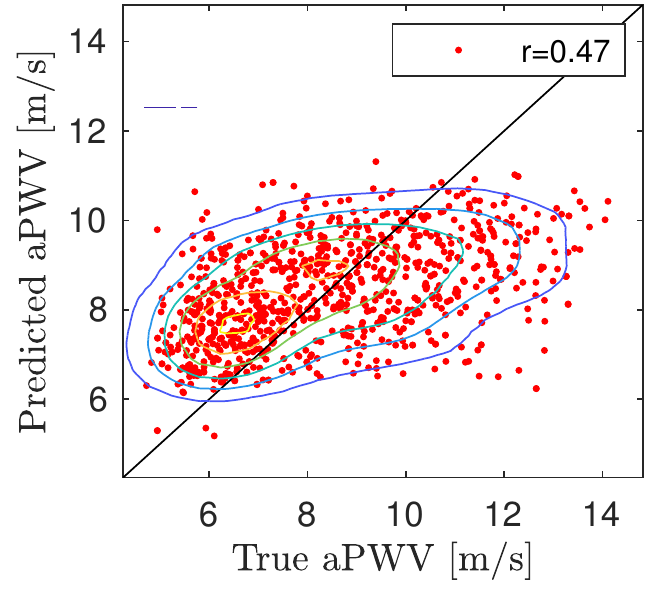}\\
   \DAT &\PAT + \PATmax &  \PAT + \PATmax+ \PATDmax + \HR\\
  \includegraphics[width=0.31\textwidth]{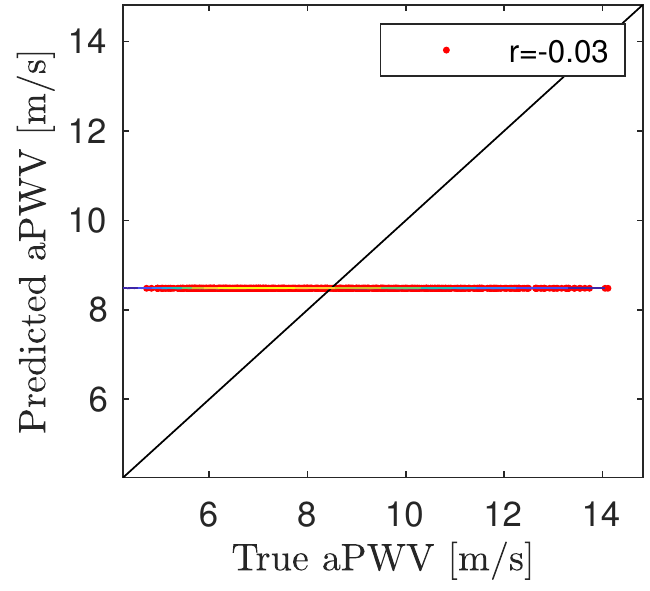}&
  \includegraphics[width=0.31\textwidth]{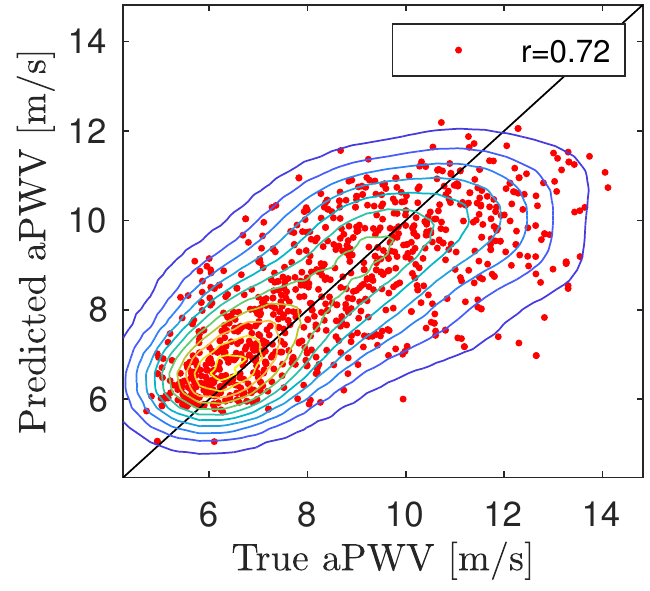}&
  \includegraphics[width=0.31\textwidth]{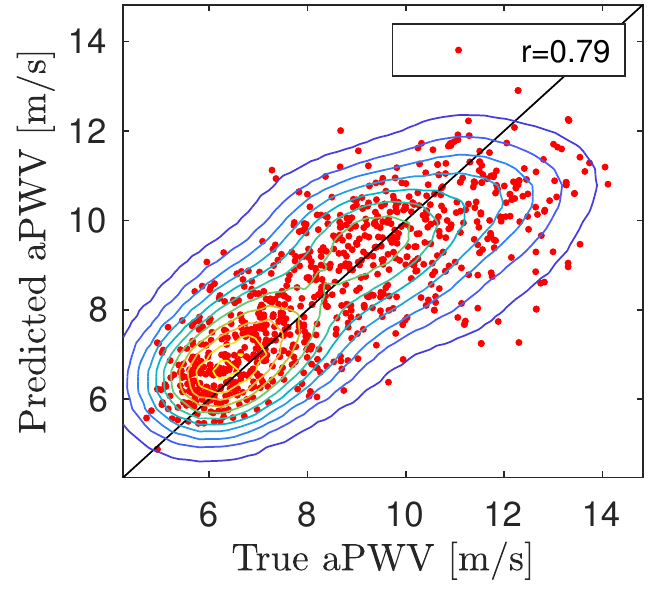}
  \end{tabular}
  \captionof{figure}{Accuracy of aortic PWV
    predictions using pulse arrival time (PAT) measurements from left
    carotid artery (LCA). Signals: heart rate (\HR) and pulse transit
    times to the foot of signal (\PAT), peak of signal (\PATmax), the
    point of steepest raise (\PATDmax), and the dicrotic notch (\PATDAT).   \label{fig:PAT_LCA_PWV}}
\end{center}
\newpage

\begin{figure}[H]
  \centering
  \begin{tabular}{ccc}
   \PTT   & \PTTmax  &  \Dmax\\
  \includegraphics[width=0.31\textwidth]{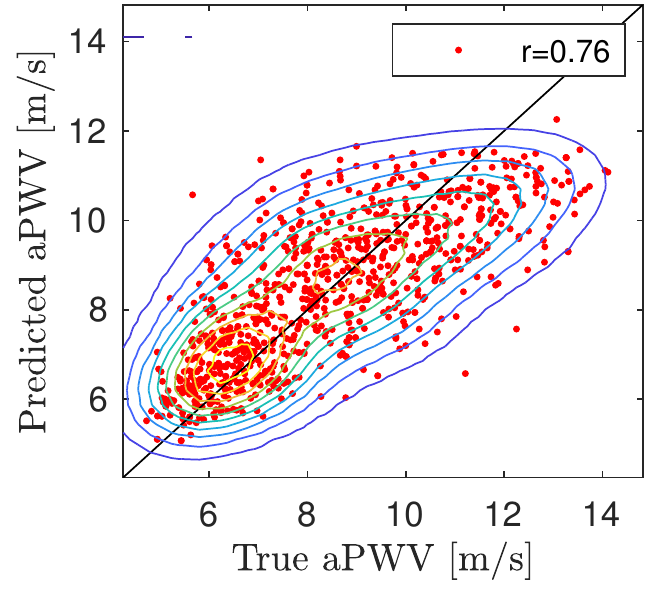}&
  \includegraphics[width=0.31\textwidth]{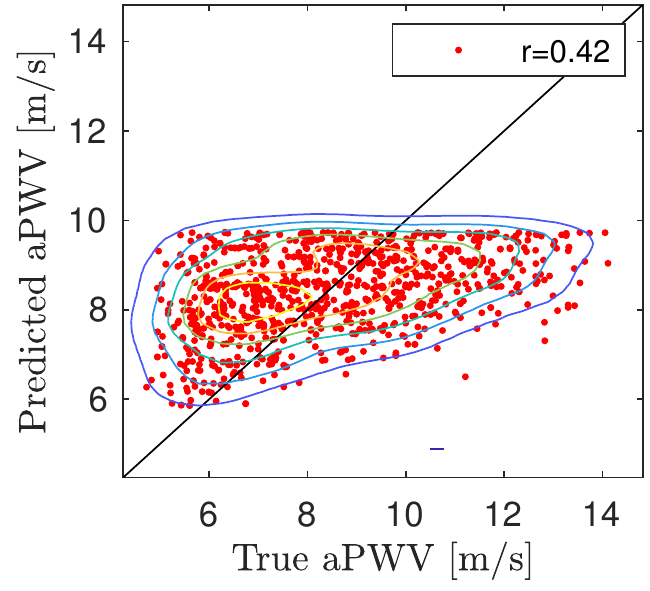}&
  \includegraphics[width=0.31\textwidth]{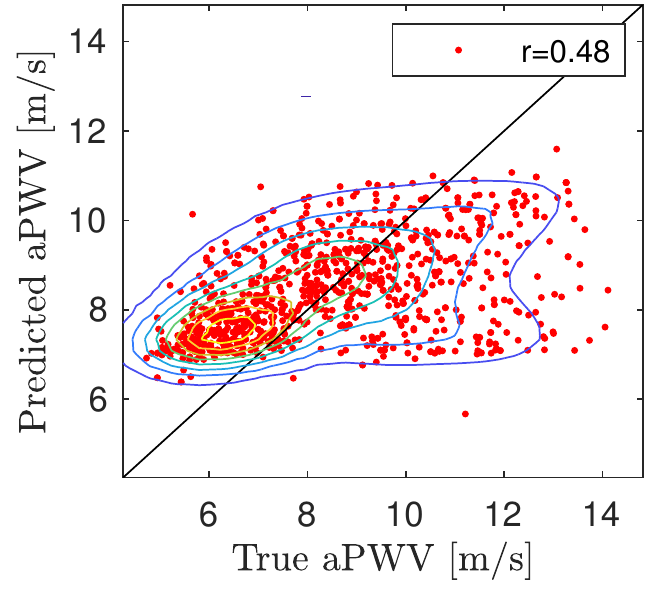}\\
   \DAT &  \PTT + \Dmax &  \PTT + \PTTmax+ \Dmax + HR\\
  \includegraphics[width=0.31\textwidth]{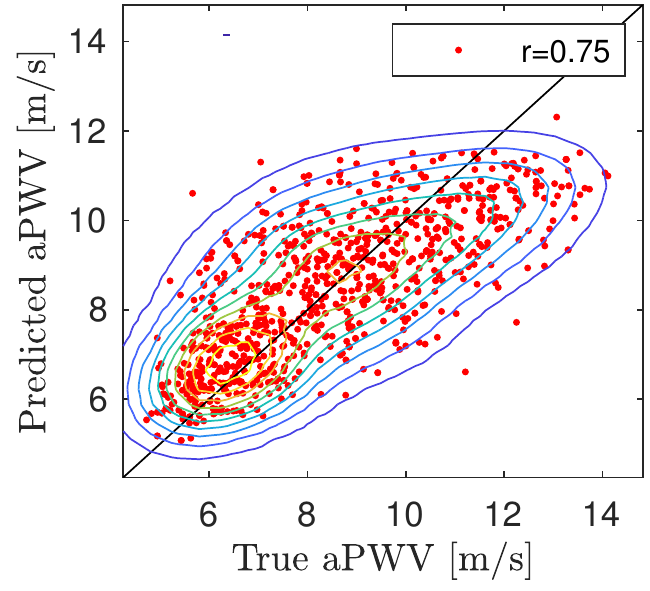}&
  \includegraphics[width=0.31\textwidth]{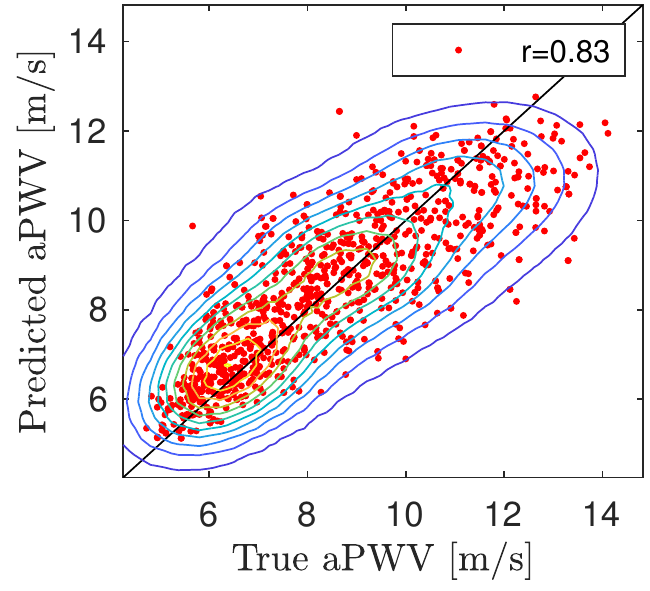}&
  \includegraphics[width=0.31\textwidth]{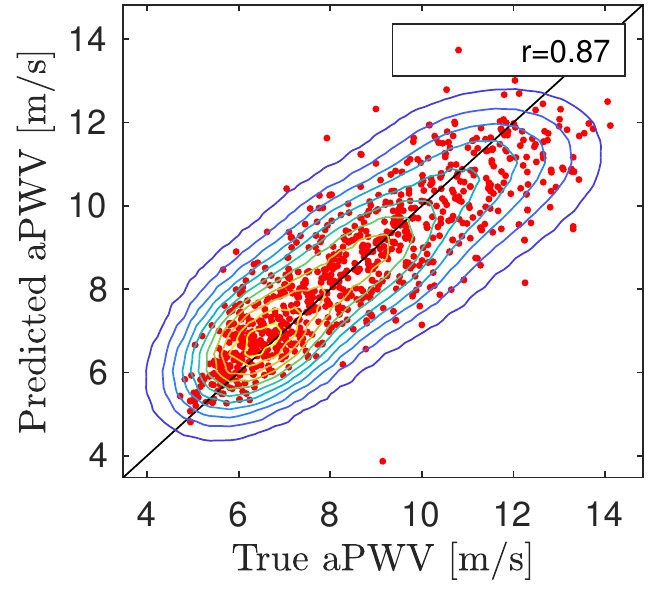}
  \end{tabular}
  \caption{ Accuracy of aortic PWV predictions using time
    difference of measurements from LCA and
    femoral arteries.    Signals: heart rate
    (\HR) and pulse arrival times to the foot of signal (\PTT), peak
    of signal (\PTTmax), the point of steepest raise (\Dmax), and the
    dicrotic notch (\DAT).  \label{fig:PTT_LCA-Fem_PWV}}
\end{figure}

\subsection{Additional figures for predictions of blood pressure levels}
\label{sec:addit-figur-pred}

\begin{figure}[H]
  \centering
  \begin{tabular}{ccc}
    \HR & \PTT   & \PTTmax  \\
  \includegraphics[width=0.31\textwidth]{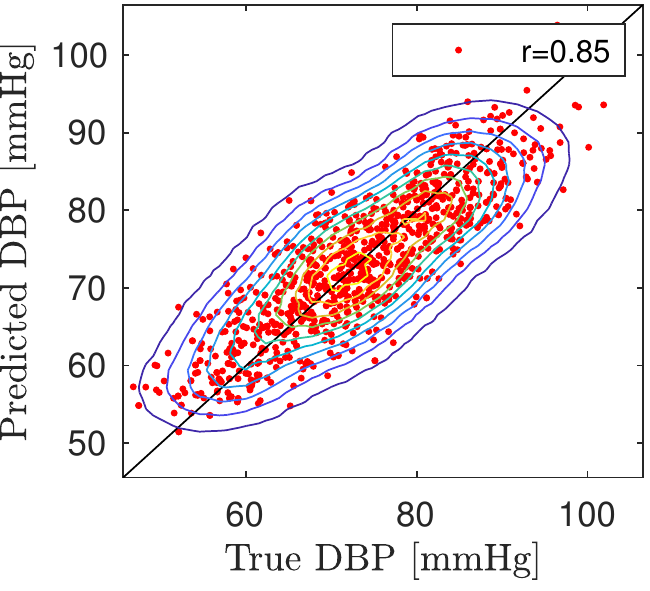}&
  \includegraphics[width=0.31\textwidth]{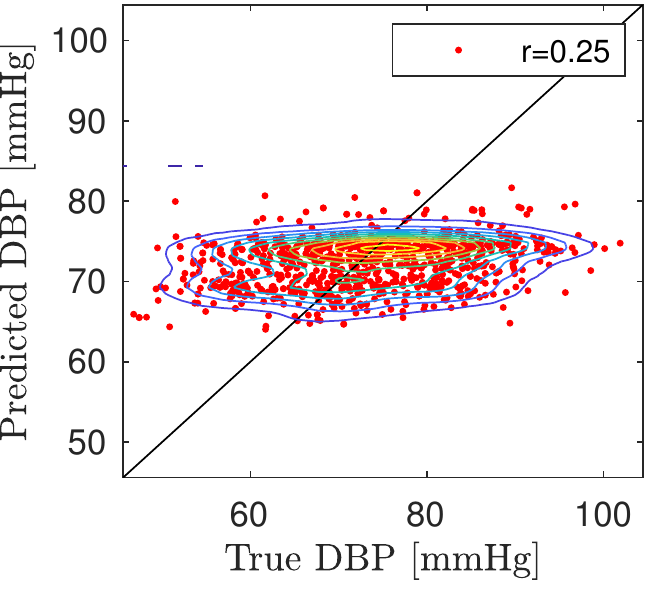}&
  \includegraphics[width=0.31\textwidth]{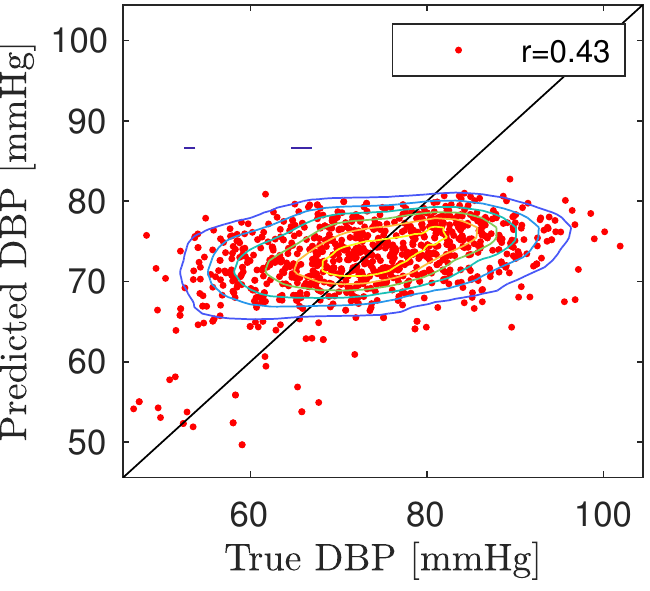}\\
    \DAT &  \PTT + HR  &  \PTT + \PTTmax+ \Dmax + \DAT\\
  \includegraphics[width=0.31\textwidth]{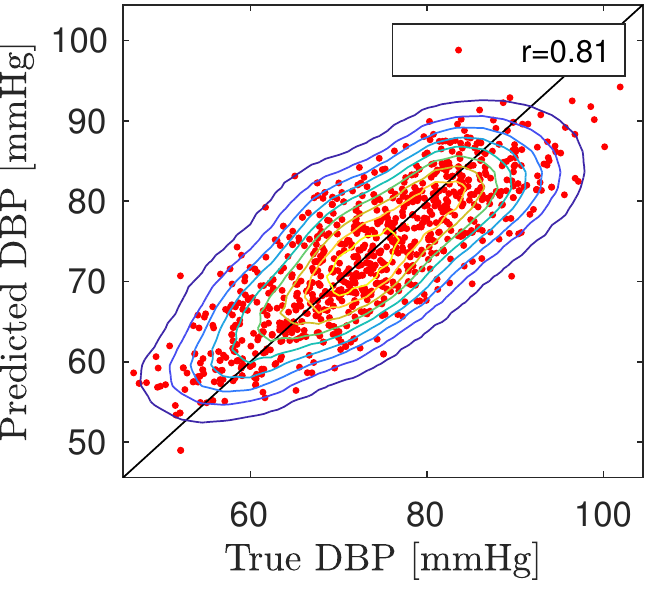}&
  \includegraphics[width=0.31\textwidth]{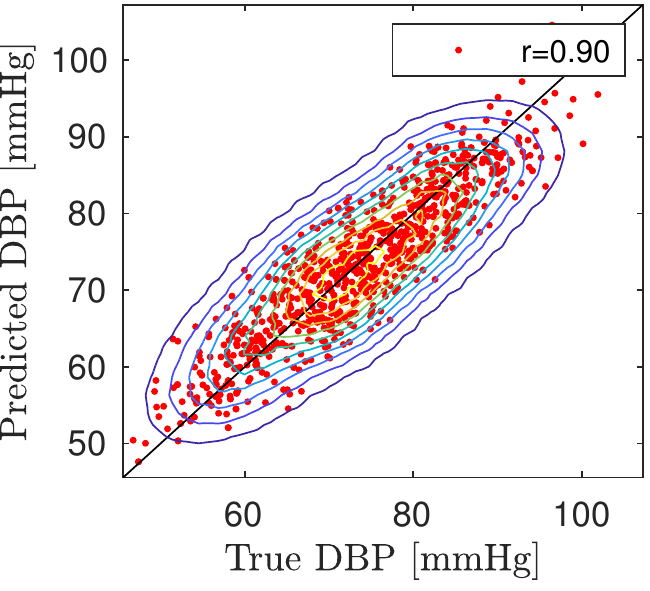}&
  \includegraphics[width=0.31\textwidth]{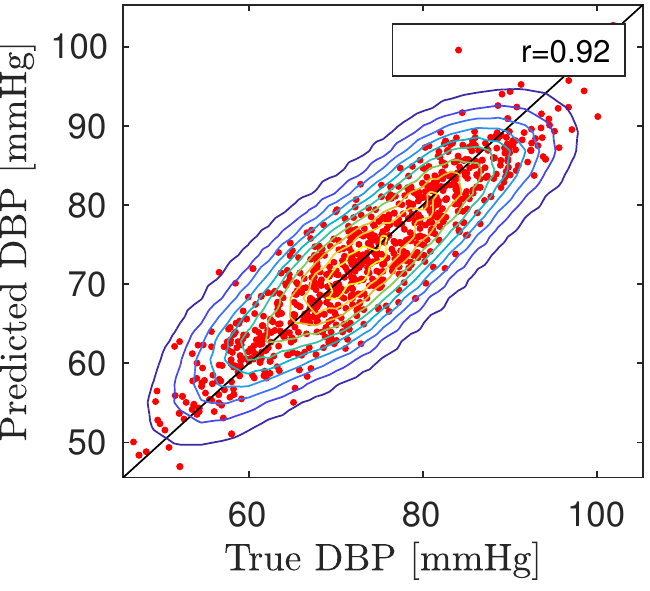}
  \end{tabular}
  \caption{Accuracy of diastolic blood pressure (DBP) predictions using pulse transit time (PTT)
    measurements from left radial artery (LRad). Signals: heart rate
    (\HR) and pulse transit times to the foot of signal (\PTT), peak
    of signal (\PTTmax), the point of steepest raise (\Dmax), and the
    dicrotic notch (\DAT).   \label{fig:PTT_LRad_DBP}}
\end{figure}

\begin{figure}[H]
  \centering
  \begin{tabular}{ccc}
    \HR & \PTT   & \PTTmax  \\
  \includegraphics[width=0.3\textwidth]{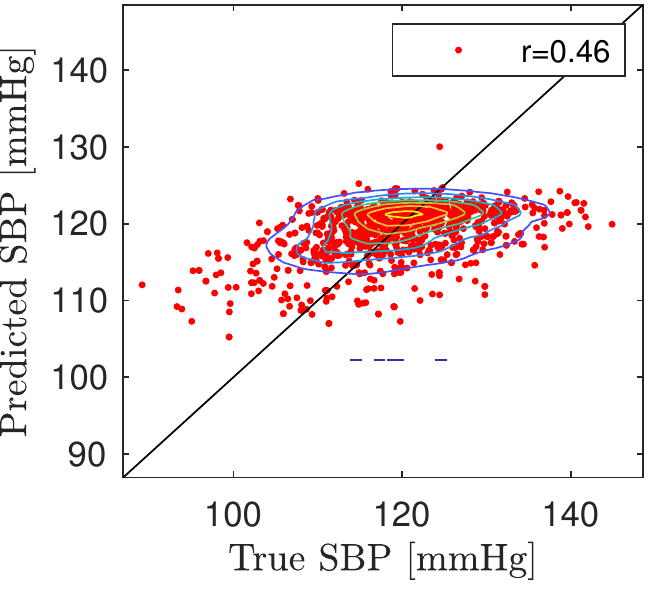}&
  \includegraphics[width=0.3\textwidth]{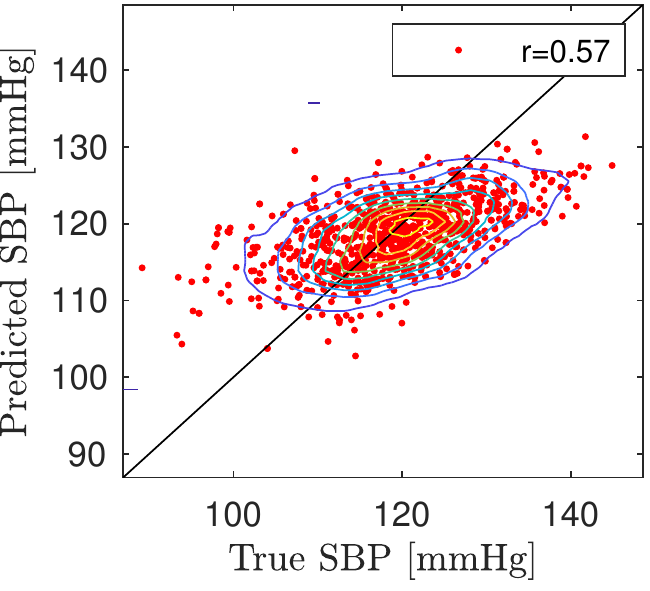}&
  \includegraphics[width=0.3\textwidth]{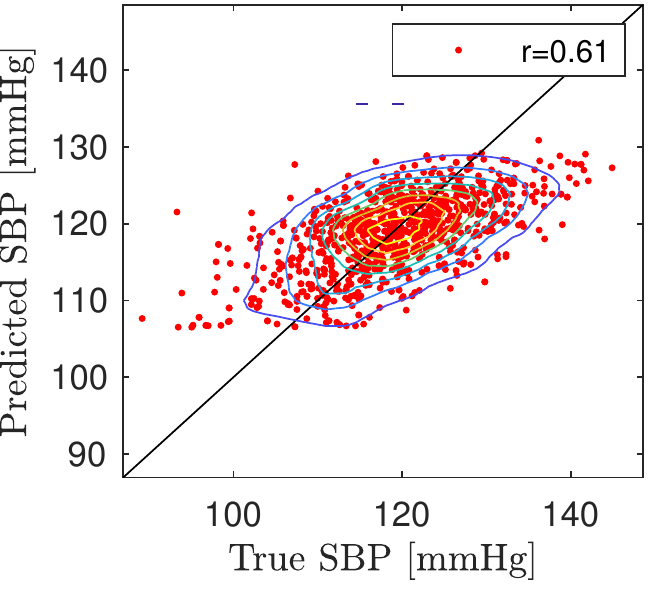}\\
    \Dmax &  \PTT + \HR &  \PTT +  \Dmax + \DAT\\
  \includegraphics[width=0.3\textwidth]{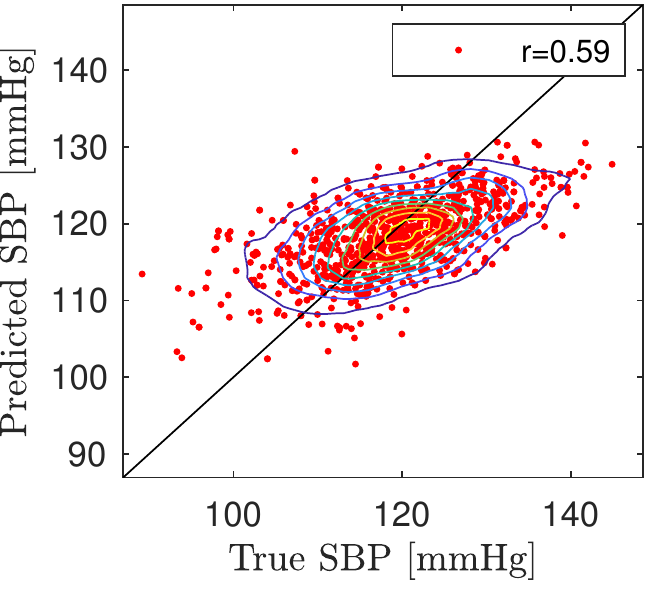}&
  \includegraphics[width=0.3\textwidth]{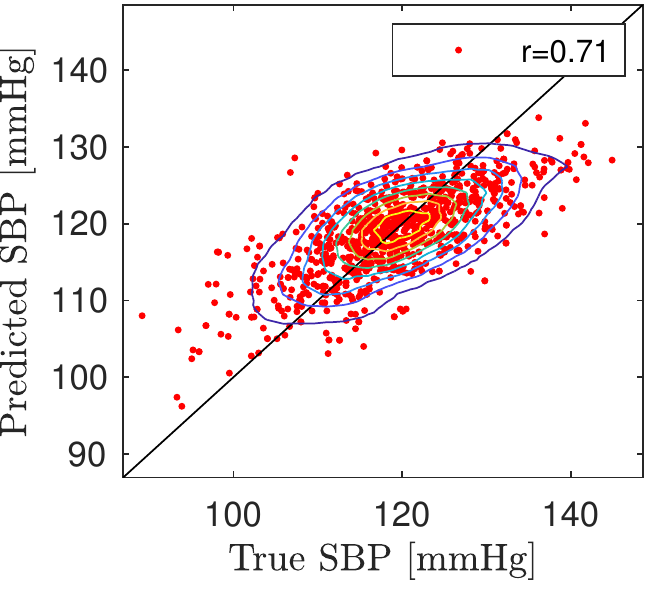}&
  \includegraphics[width=0.3\textwidth]{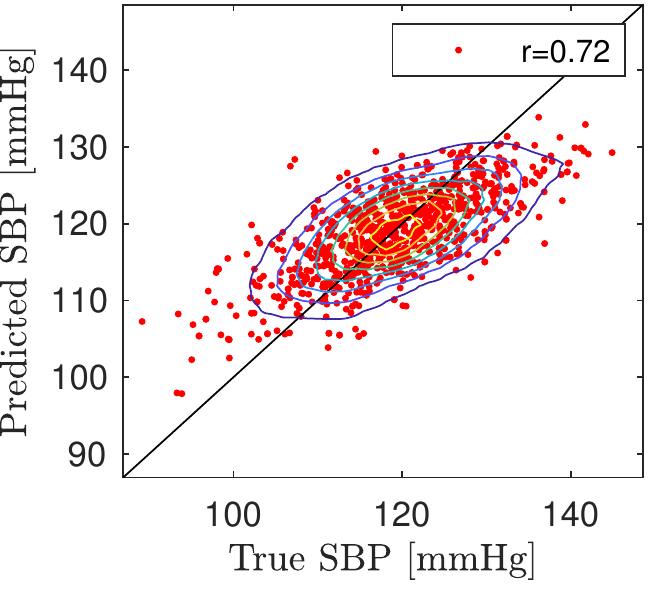}
  \end{tabular}
  \caption{ Accuracy of systolic blood pressure (SBP) predictions using pulse transit time (PTT)
    measurements from left carotid artery (LCA). Signals: heart rate
    (\HR) and pulse transit times to the foot of signal (\PTT), peak
    of signal (\PTTmax), the point of steepest raise (\Dmax), and the
    dicrotic notch (\DAT).  \label{fig:PTT_LRad_SBP}}
\end{figure}

\begin{figure}[H]
  \centering
  \begin{tabular}{ccc}
    \HR & \PAT   & \PATmax  \\
  \includegraphics[width=0.31\textwidth]{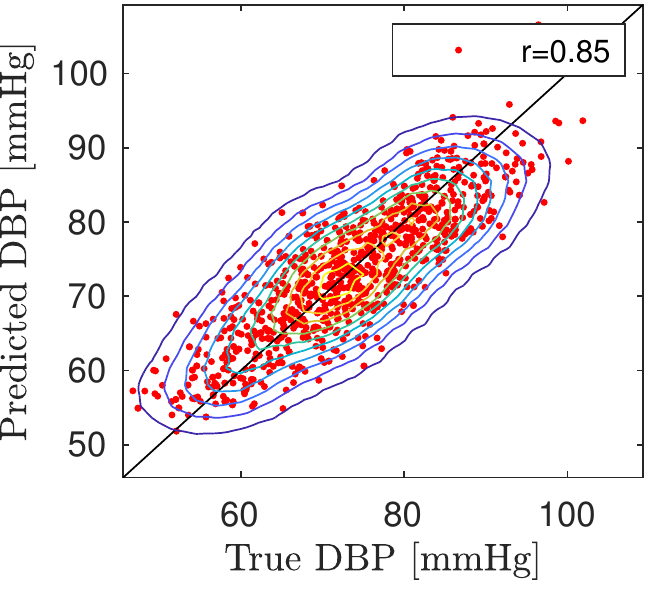}&
  \includegraphics[width=0.31\textwidth]{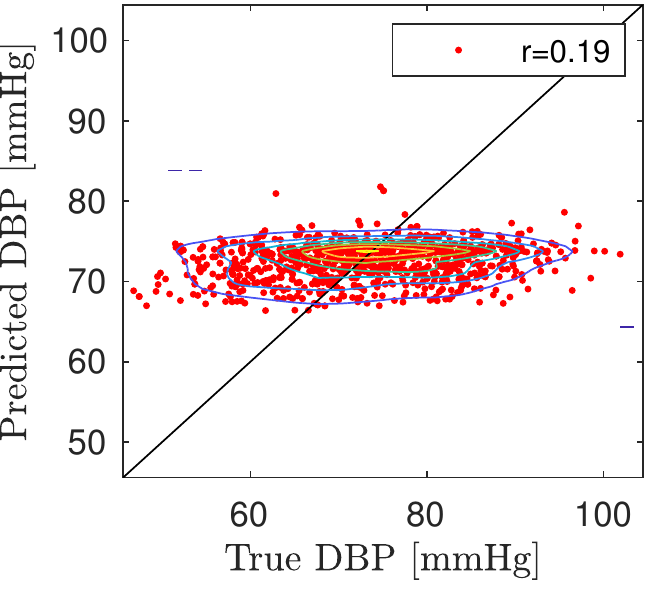}&
  \includegraphics[width=0.31\textwidth]{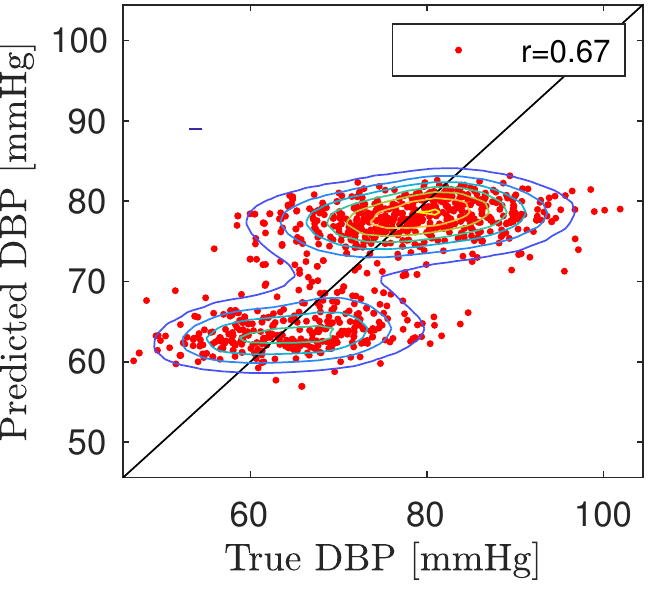}\\
    \PATDAT &  \PAT + HR  &  \PAT +  \PATDmax + \PATDAT\\
  \includegraphics[width=0.31\textwidth]{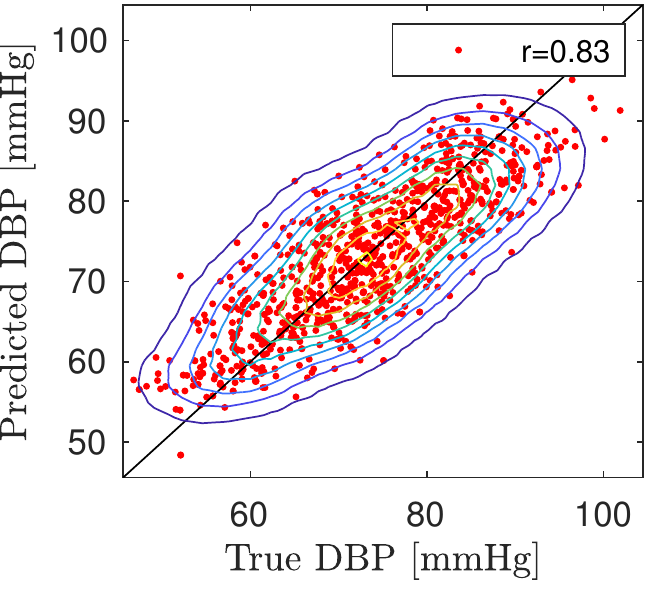}&
  \includegraphics[width=0.31\textwidth]{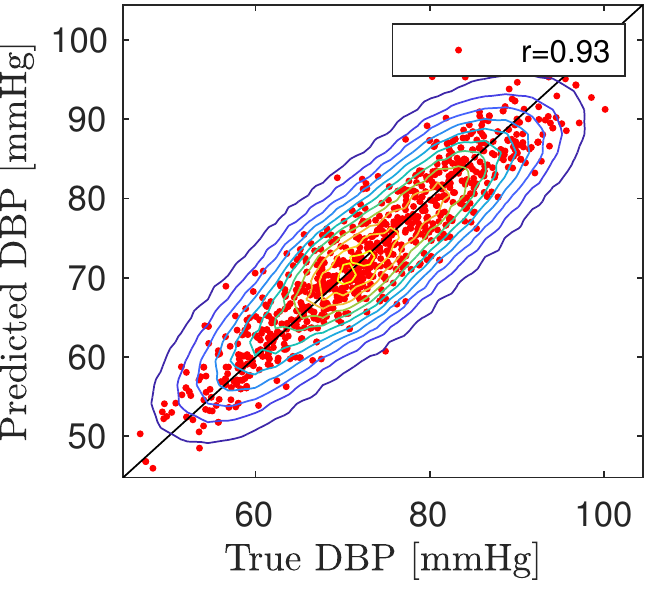}&
  \includegraphics[width=0.31\textwidth]{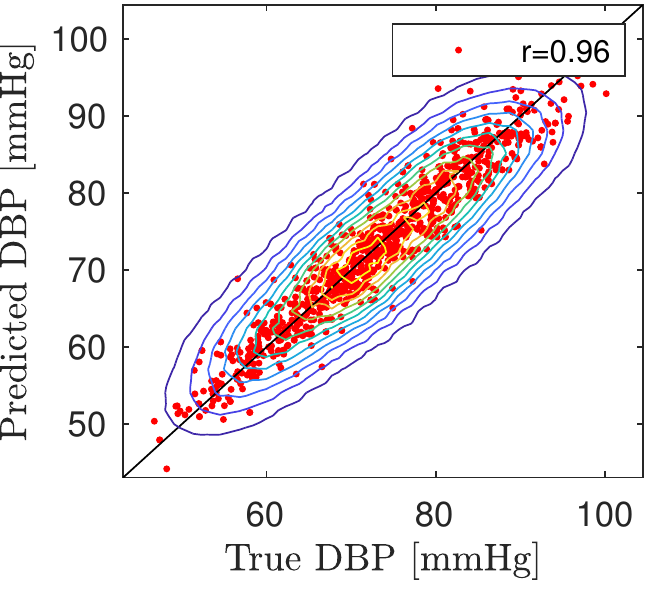}
  \end{tabular}
  \caption{ Accuracy of diastolic blood pressure (DBP) predictions using pulse arrival time (PAT)
    measurements from left carotid artery (LCA). Signals: heart rate
    (\HR) and pulse transit arrival to the foot of signal (\PAT), peak
    of signal (\PATmax), the point of steepest raise (\PATDmax), and the
    dicrotic notch (\PATDAT).  \label{fig:PAT_LCA_DBP}}
\end{figure}

\begin{figure}[H]
  \centering
  \begin{tabular}{ccc}
    \HR & \PAT   & \PATmax  \\
  \includegraphics[width=0.3\textwidth]{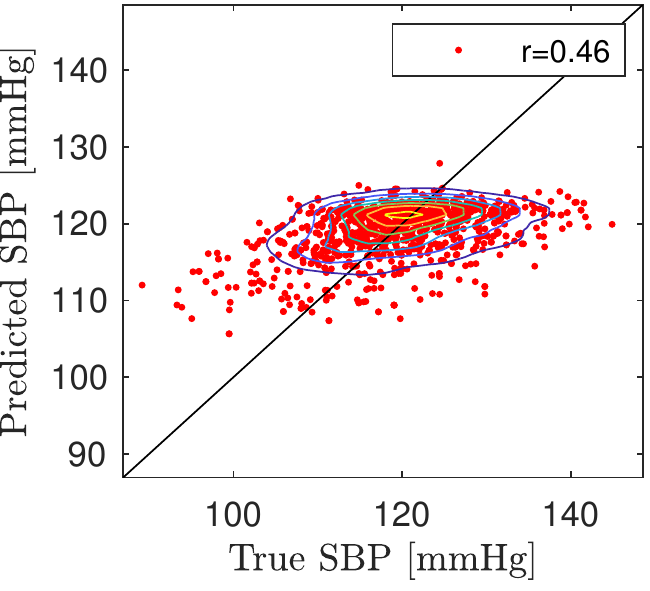}&
  \includegraphics[width=0.3\textwidth]{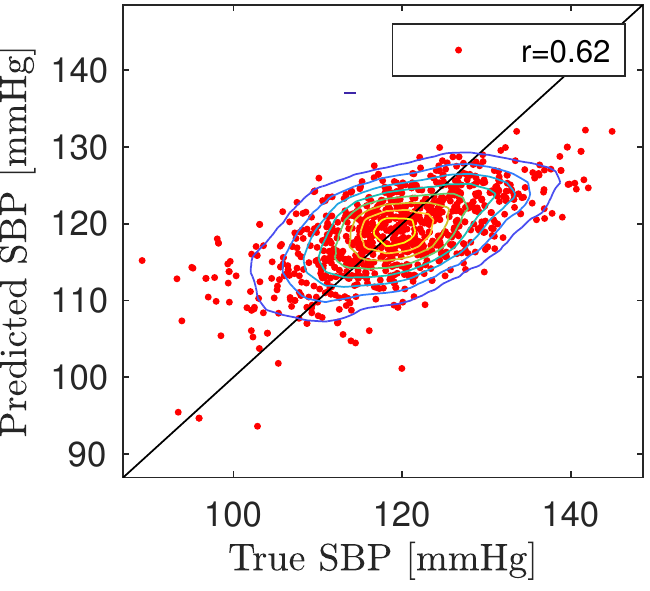}&
  \includegraphics[width=0.3\textwidth]{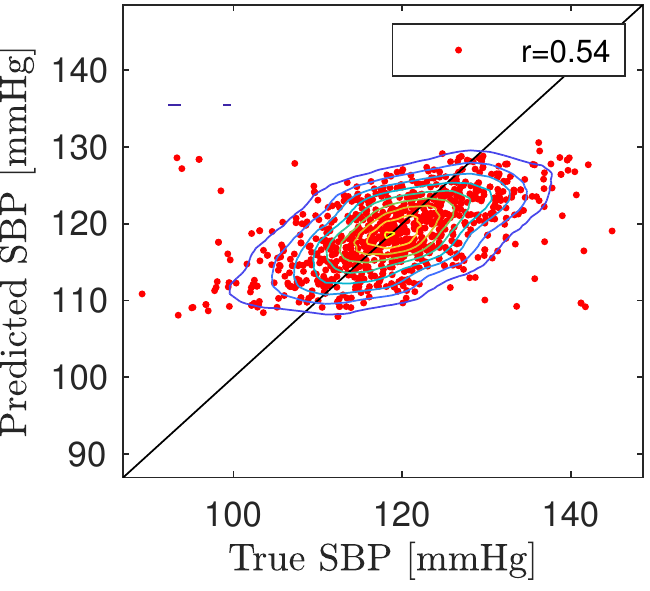}\\
    \PATDmax &  \PAT + \HR &  \PAT +  \PATmax + \PATDmax + \HR\\
  \includegraphics[width=0.3\textwidth]{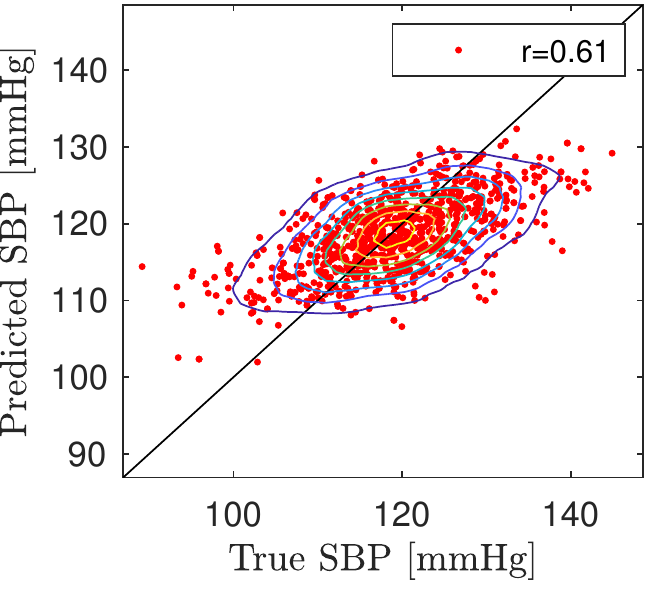}&
  \includegraphics[width=0.3\textwidth]{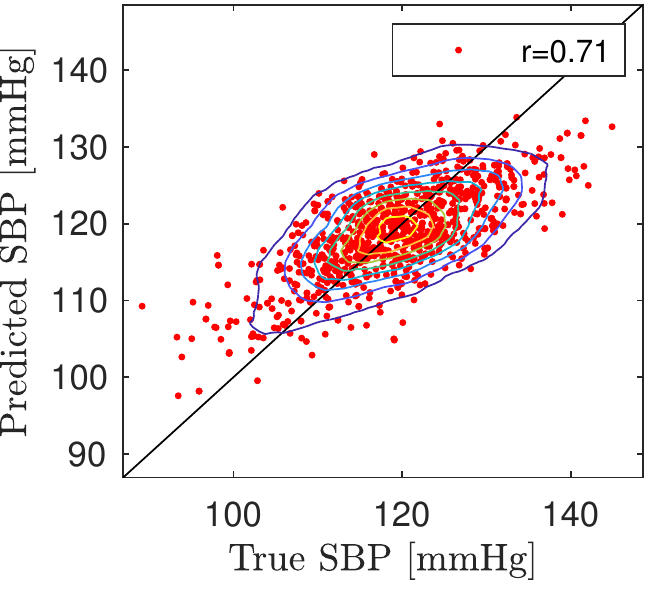}&
  \includegraphics[width=0.3\textwidth]{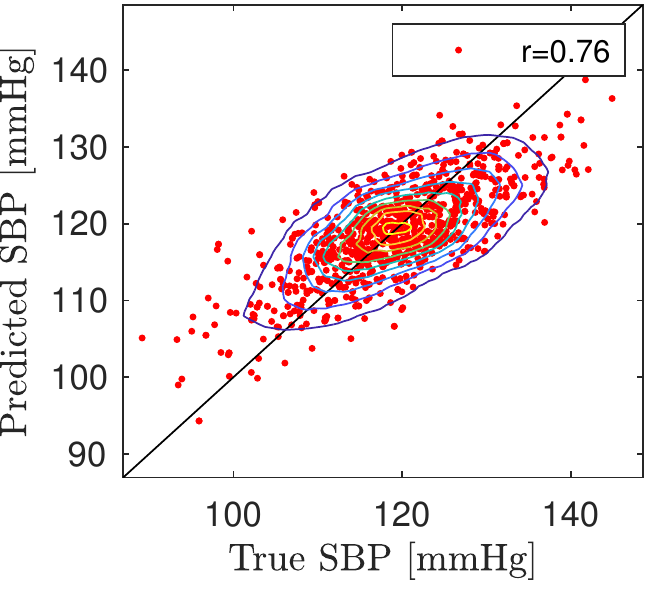}
  \end{tabular}
  \caption{ Accuracy of systolic blood pressure (DBP) predictions using pulse arrival time (PAT)
    measurements from left carotid artery (LCA). Signals: heart rate
    (\HR) and pulse transit arrival to the foot of signal (\PAT), peak
    of signal (\PATmax), the point of steepest raise (\PATDmax), and the
    dicrotic notch (\PATDAT).  \label{fig:PAT_LCA_SBP}}
\end{figure}

\subsection{Additional figures for prediction of stroke volume}
\label{sec:addit-figur-pred-1}

\begin{figure}[H]
  \centering
  \begin{tabular}{ccc}
    \HR & \PTT   & \PTTmax  \\
  \includegraphics[width=0.31\textwidth]{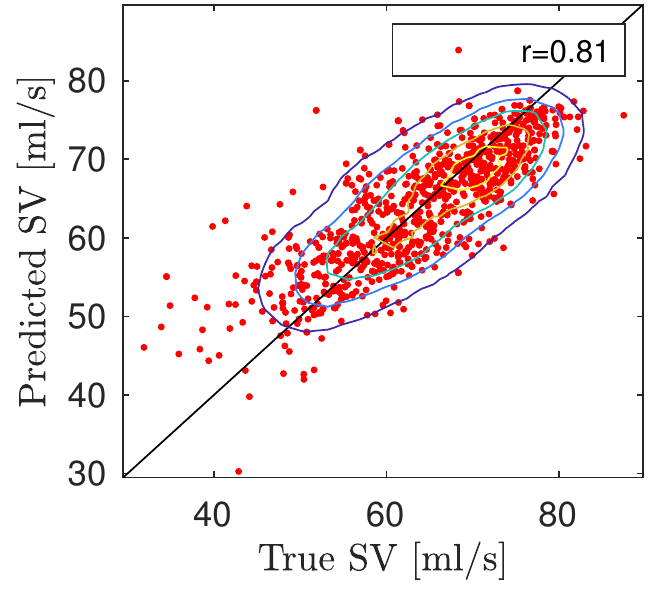}&
  \includegraphics[width=0.31\textwidth]{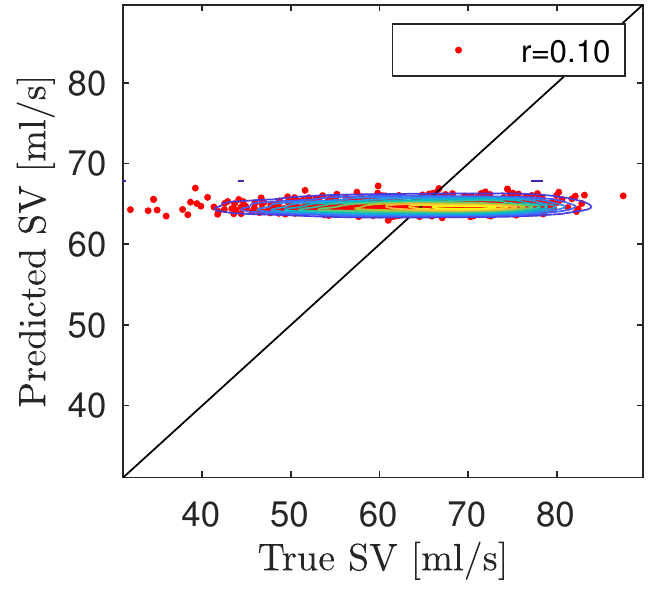}&
  \includegraphics[width=0.31\textwidth]{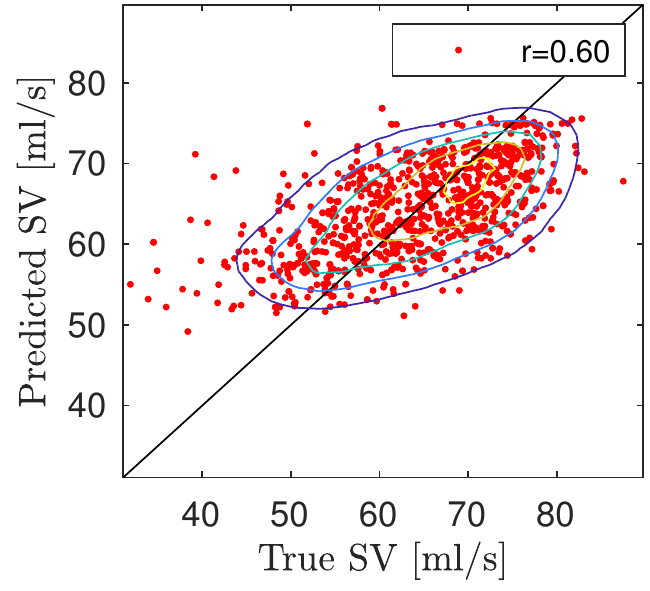}\\
    \Dmax&\DAT &  \PTT + \PTTmax+ \Dmax + \HR\\
  \includegraphics[width=0.31\textwidth]{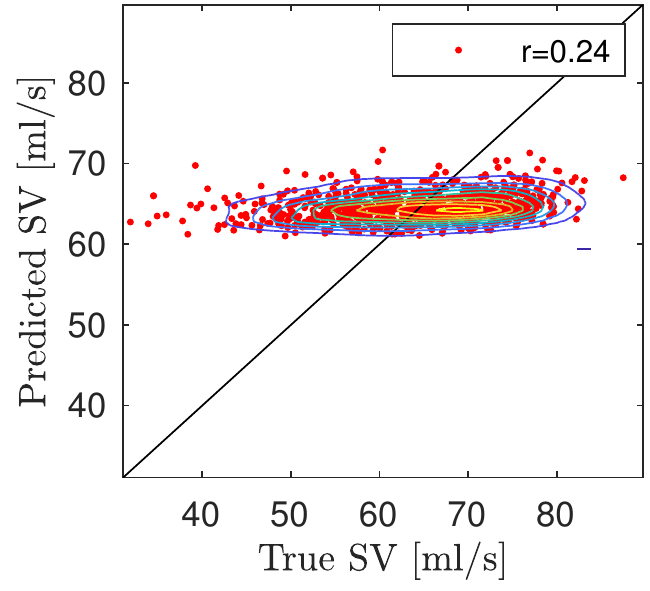}&
  \includegraphics[width=0.31\textwidth]{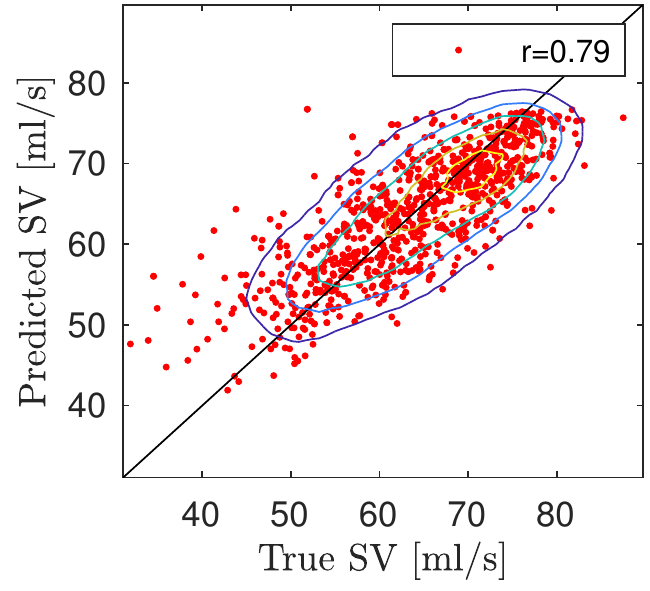}&
 \includegraphics[width=0.31\textwidth]{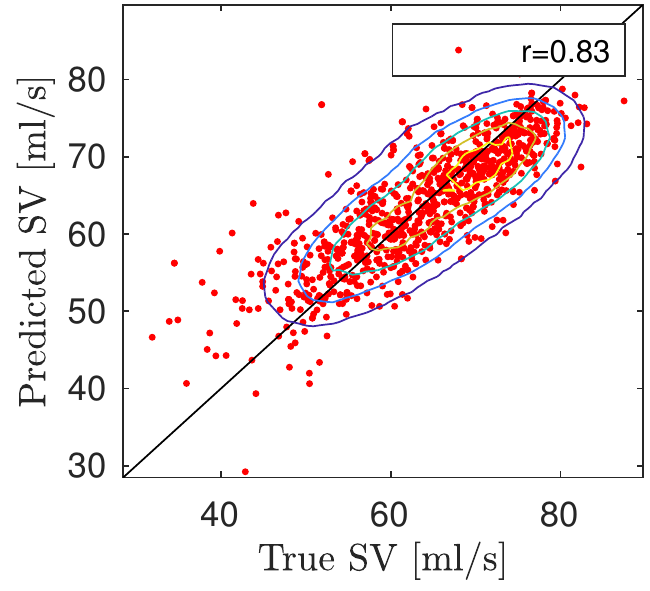}
  \end{tabular}
  \caption{ Accuracy of the stroke volumen (SV) predictions using pulse transit time (PTT)
    measurements from left carotid artery (LCA). Signals: heart rate
    (\HR) and pulse transit times to the foot of signal (\PTT), peak
    of signal (\PTTmax), the point of steepest raise (\Dmax), and the
    dicrotic notch (\DAT).  \label{fig:PTT_LCA_SV}}
\end{figure}


\newpage
\subsection{Tables}
\label{sec:tables}

\setlength{\tabcolsep}{4pt}
\begin{table}[!h]
  \caption{Accuracy of predictions using PTT measurements (the
    reference time is the aortic valve opening) from left carotid
    artery (LCA). The accuracy is expressed as Pearson correlation
    coefficients with 95\% CI (BCa intervals \cite{efron87_BCa}).  The
    predictions with largest Pearson correlations are highlighted: one
    signal (red), two signals (blue) and five most accurate
    combination (black). Signals: heart rate (\HR) and pulse transit
    times to the foot of signal (\PTT), peak of signal (\PTTmax), the
    point of steepest raise (\Dmax), and the dicrotic notch (\DAT). }
  \label{tab:results_PPT_LCA}
  \centering
    \begin{tabular}{lll@{\extracolsep{4pt}}ll@{\extracolsep{4pt}}ll@{\extracolsep{4pt}}ll}
      \hline
      &aPWV&DBP&SBP&SV\\
 \hline
{\small \HR} & 0.06 [-0.08,0.14]   & 0.85 [0.83,0.87]   & 0.46 [0.39,0.51]   & {\bf\color{red} 0.81} [0.77,0.84]  \\
{\small \PTT} & {\bf\color{red} 0.68} [0.64,0.71]   & 0.33 [0.27,0.40]   & 0.58 [0.52,0.63]   & 0.10 [0.02,0.17]  \\
{\small \PTTmax} & 0.23 [0.15,0.31]   & 0.71 [0.68,0.74]   & {\bf\color{red} 0.60} [0.54,0.65]   & 0.60 [0.55,0.65]  \\
{\small \Dmax} & 0.56 [0.51,0.61]   & 0.08 [0.00,0.17]   & 0.57 [0.52,0.62]   & 0.24 [0.07,0.30]  \\
{\small \DAT} & 0.06 [-0.08,0.13]   & {\bf\color{red} 0.86} [0.84,0.88]   & 0.46 [0.40,0.52]   & 0.79 [0.76,0.82]  \\
{\small \PTT+\HR} & 0.68 [0.64,0.72]   & 0.92 [0.90,0.93]   & 0.74 [0.69,0.77]   & 0.82 [0.79,0.84]  \\
{\small \PTTmax+\HR} & 0.31 [0.25,0.39]   & 0.89 [0.87,0.90]   & 0.67 [0.63,0.72]   & {\bf\color{blue} \hl{\bf 0.82}} [0.80,0.86]  \\
{\small \Dmax+\HR} & 0.59 [0.54,0.64]   & 0.89 [0.87,0.91]   & 0.65 [0.61,0.70]   & 0.81 [0.78,0.84]  \\
{\small \DAT+\HR} & -0.00 [-0.17,0.07]   & 0.86 [0.84,0.88]   & 0.45 [0.39,0.51]   & 0.81 [0.78,0.84]  \\
{\small \PTT+\PTTmax} & {\bf\color{blue} 0.90} [0.89,0.92]   & 0.79 [0.76,0.82]   & 0.69 [0.65,0.73]   & 0.62 [0.58,0.67]  \\
{\small \PTT+\PTTmax+\HR} & 0.93 [0.92,0.95]   & 0.92 [0.90,0.93]   & 0.74 [0.71,0.78]   & 0.82 [0.80,0.85]  \\
{\small \PTT+\Dmax} & 0.72 [0.66,0.76]   & 0.79 [0.77,0.82]   & 0.64 [0.60,0.71]   & 0.54 [0.49,0.59]  \\
{\small \PTT+\Dmax+\HR} & 0.75 [0.72,0.79]   & 0.93 [0.91,0.94]   & 0.74 [0.73,0.76]   & \hl{\bf 0.83} [0.81,0.86]  \\
{\small \PTT+\DAT} & 0.68 [0.64,0.71]   & {\bf\color{blue} 0.92} [0.91,0.94]   & {\bf\color{blue} 0.74} [0.71,0.78]   & 0.80 [0.77,0.83]  \\
{\small \PTT+\DAT+\HR} & 0.68 [0.64,0.72]   & 0.92 [0.91,0.94]   & 0.74 [0.71,0.79]   & 0.81 [0.79,0.84]  \\
{\small \PTT+\PTTmax+\Dmax} & 0.92 [0.91,0.93]   & 0.87 [0.85,0.89]   & 0.71 [0.69,0.70]   & 0.70 [0.67,0.74]  \\
{\small \PTT+\PTTmax+\Dmax+\HR} & \hl{\bf 0.93} [0.93,0.95]   & \hl{\bf 0.93} [0.92,0.94]   & 0.74 [0.73,0.76]   & \hl{\bf 0.82} [0.80,0.86]  \\
{\small \PTT+\PTTmax+\DAT} & \hl{\bf 0.94} [0.93,0.95]   & 0.92 [0.91,0.94]   & \hl{\bf 0.75} [0.73,0.79]   & 0.80 [0.77,0.84]  \\
{\small \PTT+\PTTmax+\DAT+\HR} & \hl{\bf 0.94} [0.93,0.95]   & 0.92 [0.91,0.94]   & 0.75 [0.72,0.78]   & 0.82 [0.79,0.85]  \\
{\small \PTT+\Dmax+\DAT} & 0.75 [0.72,0.79]   & \hl{\bf 0.93} [0.91,0.95]   & \hl{\bf 0.75} [0.73,0.80]   & 0.81 [0.80,0.84]  \\
{\small \PTT+\Dmax+\DAT+\HR} & 0.75 [0.72,0.78]   & \hl{\bf 0.93} [0.92,0.95]   & \hl{\bf 0.75} [0.72,0.80]   & \hl{\bf 0.83} [0.81,0.86]  \\
{\small \PTT+\PTTmax+\Dmax+\DAT} & \hl{\bf 0.94} [0.93,0.95]   & \hl{\bf 0.94} [0.92,0.95]   & \hl{\bf 0.75} [0.73,0.79]   & 0.81 [0.80,0.82]  \\
{\small \PTT+\PTTmax+\Dmax+\DAT+\HR} & \hl{\bf 0.94} [0.93,0.95]   & \hl{\bf 0.94} [0.93,0.95]   & \hl{\bf 0.75} [0.72,0.80]   & \hl{\bf 0.83} [0.82,0.86]  \\
      \hline
    \end{tabular}
\end{table}

\setlength{\tabcolsep}{4pt}
\begin{table}[!h]
  \caption{Accuracy of predictions using PTT measurements
    (the reference time is the aortic valve opening) from
    right carotid artery (RCA). Otherwise same caption as in Table \ref{tab:results_PPT_LCA}. }
  \label{tab:results_PTT_RCA}
  \centering
    \begin{tabular}{lll@{\extracolsep{4pt}}ll@{\extracolsep{4pt}}ll@{\extracolsep{4pt}}ll}
      \hline
      &aPWV&DBP&SBP&SV\\
\hline
{\small \HR} & 0.06 [-0.07,0.14]   & 0.85 [0.83,0.87]   & 0.46 [0.39,0.51]   & {\bf\color{red} 0.81} [0.77,0.84]  \\
{\small \PTT} & {\bf\color{red} 0.60} [0.54,0.64]   & 0.30 [0.24,0.37]   & {\bf\color{red} 0.57} [0.52,0.62]   & 0.11 [0.03,0.20]  \\
{\small \PTTmax} & 0.24 [0.18,0.32]   & 0.78 [0.75,0.81]   & 0.45 [0.39,0.51]   & 0.63 [0.58,0.68]  \\
{\small \Dmax} & 0.48 [0.42,0.54]   & -0.01 [-0.13,0.06]   & 0.56 [0.47,0.62]   & 0.27 [0.12,0.34]  \\
{\small \DAT} & 0.07 [-0.03,0.13]   & {\bf\color{red} 0.86} [0.84,0.88]   & 0.45 [0.37,0.51]   & 0.79 [0.76,0.82]  \\
{\small \PTT+\HR} & 0.60 [0.56,0.65]   & 0.91 [0.90,0.93]   & 0.72 [0.68,0.76]   & 0.81 [0.79,0.84]  \\
{\small \PTTmax+\HR} & 0.31 [0.23,0.40]   & 0.88 [0.86,0.90]   & 0.66 [0.62,0.70]   & {\bf\color{blue} 0.82} [0.80,0.85]  \\
{\small \Dmax+\HR} & 0.52 [0.47,0.58]   & 0.89 [0.87,0.91]   & 0.64 [0.59,0.68]   & 0.81 [0.78,0.84]  \\
{\small \DAT+\HR} & 0.06 [-0.01,0.17]   & 0.86 [0.84,0.88]   & 0.45 [0.41,0.52]   & 0.81 [0.78,0.84]  \\
{\small \PTT+\PTTmax} & {\bf\color{blue} 0.71} [0.68,0.76]   & 0.84 [0.82,0.87]   & 0.69 [0.66,0.73]   & 0.68 [0.65,0.72]  \\
{\small \PTT+\PTTmax+\HR} & \hl{\bf 0.74} [0.72,0.78]   & 0.91 [0.90,0.93]   & \hl{\bf 0.74} [0.73,0.77]   & \hl{\bf 0.82} [0.80,0.85]  \\
{\small \PTT+\Dmax} & 0.66 [0.60,0.71]   & 0.76 [0.72,0.79]   & 0.63 [0.57,0.69]   & 0.49 [0.44,0.56]  \\
{\small \PTT+\Dmax+\HR} & 0.72 [0.67,0.75]   & 0.92 [0.91,0.94]   & 0.73 [0.72,0.76]   & \hl{\bf 0.83} [0.81,0.85]  \\
{\small \PTT+\DAT} & 0.60 [0.56,0.65]   & {\bf\color{blue} 0.92} [0.90,0.93]   & {\bf\color{blue} 0.72} [0.68,0.76]   & 0.80 [0.77,0.83]  \\
{\small \PTT+\DAT+\HR} & 0.59 [0.55,0.64]   & 0.92 [0.91,0.93]   & 0.72 [0.69,0.76]   & 0.81 [0.78,0.84]  \\
{\small \PTT+\PTTmax+\Dmax} & \hl{\bf 0.77} [0.76,0.77]   & 0.87 [0.85,0.89]   & 0.70 [0.70,0.71]   & 0.71 [0.69,0.76]  \\
{\small \PTT+\PTTmax+\Dmax+\HR} & \hl{\bf 0.78} [0.75,0.76]   & \hl{\bf 0.92} [0.92,0.94]   & 0.73 [0.73,0.75]   & \hl{\bf 0.83} [0.81,0.85]  \\
{\small \PTT+\PTTmax+\DAT} & 0.74 [0.71,0.77]   & 0.92 [0.90,0.93]   & 0.73 [0.71,0.77]   & 0.80 [0.77,0.84]  \\
{\small \PTT+\PTTmax+\DAT+\HR} & 0.73 [0.73,0.76]   & 0.92 [0.91,0.93]   & 0.74 [0.71,0.77]   & 0.82 [0.79,0.85]  \\
{\small \PTT+\Dmax+\DAT} & 0.71 [0.68,0.74]   & \hl{\bf 0.93} [0.91,0.94]   & \hl{\bf 0.74} [0.72,0.77]   & 0.80 [0.77,0.83]  \\
{\small \PTT+\Dmax+\DAT+\HR} & 0.71 [0.68,0.74]   & \hl{\bf 0.93} [0.92,0.94]   & \hl{\bf 0.74} [0.71,0.77]   & \hl{\bf 0.83} [0.81,0.85]  \\
{\small \PTT+\PTTmax+\Dmax+\DAT} & \hl{\bf 0.79} [0.75,0.76]   & \hl{\bf 0.93} [0.92,0.94]   & \hl{\bf 0.74} [0.72,0.78]   & 0.80 [0.80,0.84]  \\
{\small \PTT+\PTTmax+\Dmax+\DAT+\HR} & \hl{\bf 0.79} [0.76,0.77]   & \hl{\bf 0.93} [0.92,0.94]   & \hl{\bf 0.74} [0.71,0.77]   & \hl{\bf 0.83} [0.81,0.85]  \\
      \hline
    \end{tabular}
\end{table}

\setlength{\tabcolsep}{4pt}
\begin{table}[!h]
  \caption{ Accuracy of predictions using PTT measurements
  (the reference time is the aortic valve opening) from left
  radialis artery (LRad). Otherwise same caption as in Table \ref{tab:results_PPT_LCA}. }
\label{tab:results_PTT_LRad}
\centering
    \begin{tabular}{lll@{\extracolsep{4pt}}ll@{\extracolsep{4pt}}ll@{\extracolsep{4pt}}ll}
      \hline
      &aPWV&DBP&SBP&SV\\
      \hline
{\small \HR} & 0.06 [-0.07,0.15]   & {\bf\color{red} 0.85} [0.83,0.87]   & 0.46 [0.40,0.51]   & {\bf\color{red} 0.81} [0.76,0.84]  \\
{\small \PTT} & {\bf\color{red} 0.33} [0.25,0.39]   & 0.25 [0.18,0.32]   & 0.57 [0.52,0.62]   & 0.16 [0.06,0.23]  \\
{\small \PTTmax} & 0.06 [-0.02,0.13]   & 0.43 [0.37,0.48]   & {\bf\color{red} 0.61} [0.55,0.65]   & 0.58 [0.53,0.63]  \\
{\small \Dmax} & 0.32 [0.23,0.38]   & 0.19 [0.11,0.27]   & 0.59 [0.52,0.64]   & 0.20 [0.11,0.27]  \\
{\small \DAT} & 0.04 [-0.04,0.15]   & 0.81 [0.79,0.84]   & 0.52 [0.46,0.57]   & 0.80 [0.77,0.83]  \\
{\small \PTT+\HR} & 0.33 [0.25,0.39]   & 0.90 [0.89,0.92]   & 0.71 [0.67,0.75]   & 0.82 [0.79,0.85]  \\
{\small \PTTmax+\HR} & 0.06 [-0.01,0.14]   & 0.87 [0.85,0.89]   & 0.64 [0.59,0.69]   & {\bf\color{blue} 0.82} [0.79,0.85]  \\
{\small \Dmax+\HR} & 0.34 [0.28,0.40]   & 0.90 [0.88,0.91]   & 0.71 [0.67,0.75]   & 0.82 [0.79,0.85]  \\
{\small \DAT+\HR} & 0.09 [0.01,0.17]   & 0.86 [0.83,0.87]   & 0.56 [0.51,0.61]   & 0.81 [0.78,0.84]  \\
{\small \PTT+\PTTmax} & {\bf\color{blue} 0.55} [0.51,0.60]   & 0.83 [0.80,0.86]   & 0.66 [0.61,0.71]   & 0.68 [0.64,0.72]  \\
{\small \PTT+\PTTmax+\HR} & 0.71 [0.68,0.75]   & 0.91 [0.89,0.93]   & 0.71 [0.69,0.74]   & \hl{\bf 0.82} [0.80,0.85]  \\
{\small \PTT+\Dmax} & 0.34 [0.26,0.39]   & 0.72 [0.69,0.76]   & 0.66 [0.64,0.69]   & 0.51 [0.47,0.58]  \\
{\small \PTT+\Dmax+\HR} & 0.34 [0.28,0.40]   & 0.90 [0.89,0.92]   & \hl{\bf 0.72} [0.68,0.76]   & \hl{\bf 0.83} [0.80,0.86]  \\
{\small \PTT+\DAT} & 0.34 [0.26,0.39]   & {\bf\color{blue} 0.91} [0.89,0.92]   & {\bf\color{blue} \hl{\bf 0.72}} [0.68,0.75]   & 0.80 [0.77,0.83]  \\
{\small \PTT+\DAT+\HR} & 0.35 [0.29,0.43]   & 0.91 [0.89,0.92]   & \hl{\bf 0.72} [0.68,0.76]   & 0.82 [0.79,0.85]  \\
{\small \PTT+\PTTmax+\Dmax} & 0.63 [0.60,0.66]   & 0.85 [0.83,0.88]   & 0.67 [0.66,0.71]   & 0.69 [0.64,0.74]  \\
{\small \PTT+\PTTmax+\Dmax+\HR} & \hl{\bf 0.72} [0.68,0.76]   & \hl{\bf 0.91} [0.89,0.93]   & 0.71 [0.70,0.75]   & \hl{\bf 0.83} [0.81,0.86]  \\
{\small \PTT+\PTTmax+\DAT} & \hl{\bf 0.72} [0.69,0.76]   & \hl{\bf 0.92} [0.90,0.93]   & 0.70 [0.64,0.73]   & 0.80 [0.76,0.84]  \\
{\small \PTT+\PTTmax+\DAT+\HR} & \hl{\bf 0.72} [0.69,0.76]   & \hl{\bf 0.92} [0.91,0.94]   & \hl{\bf 0.72} [0.68,0.76]   & 0.82 [0.79,0.85]  \\
{\small \PTT+\Dmax+\DAT} & 0.34 [0.27,0.40]   & 0.91 [0.90,0.93]   & \hl{\bf 0.72} [0.69,0.76]   & 0.81 [0.77,0.83]  \\
{\small \PTT+\Dmax+\DAT+\HR} & 0.35 [0.30,0.43]   & 0.91 [0.90,0.92]   & 0.71 [0.67,0.75]   & \hl{\bf 0.82} [0.80,0.86]  \\
{\small \PTT+\PTTmax+\Dmax+\DAT} & \hl{\bf 0.73} [0.69,0.77]   & \hl{\bf 0.92} [0.90,0.93]   & 0.72 [0.68,0.76]   & 0.81 [0.76,0.84]  \\
{\small \PTT+\PTTmax+\Dmax+\DAT+\HR} & \hl{\bf 0.73} [0.71,0.77]   & \hl{\bf 0.92} [0.91,0.93]   & 0.72 [0.68,0.76]   & \hl{\bf 0.83} [0.80,0.86]  \\
\hline
    \end{tabular}
\end{table}

\setlength{\tabcolsep}{4pt}
\begin{table}[!h]
  \caption{ Accuracy of predictions using PTT measurements (the
    reference time is the aortic valve opening) from right femoral
    artery (RFem). Otherwise same caption as in Table \ref{tab:results_PPT_LCA}.}
\label{tab:results_PTT_RFem}
\centering
    \begin{tabular}{lll@{\extracolsep{4pt}}ll@{\extracolsep{4pt}}ll@{\extracolsep{4pt}}ll}
      \hline
      &aPWV&DBP&SBP&SV\\
\hline
{\small \HR} & 0.06 [-0.07,0.15]   & {\bf\color{red} 0.85} [0.83,0.87]   & 0.45 [0.38,0.51]   & {\bf\color{red} 0.81} [0.76,0.84]  \\
{\small \PTT} & {\bf\color{red} 0.75} [0.71,0.78]   & 0.35 [0.29,0.42]   & 0.59 [0.51,0.64]   & 0.08 [-0.02,0.15]  \\
{\small \PTTmax} & 0.27 [0.18,0.34]   & 0.20 [0.11,0.27]   & 0.58 [0.53,0.63]   & 0.45 [0.38,0.51]  \\
{\small \Dmax} & 0.50 [0.43,0.56]   & 0.11 [0.01,0.20]   & {\bf\color{red} 0.63} [0.58,0.68]   & 0.24 [0.14,0.31]  \\
{\small \DAT} & 0.09 [0.01,0.18]   & 0.77 [0.74,0.80]   & 0.56 [0.50,0.61]   & 0.79 [0.76,0.83]  \\
{\small \PTT+\HR} & 0.75 [0.71,0.78]   & 0.92 [0.90,0.93]   & 0.75 [0.70,0.78]   & 0.82 [0.79,0.84]  \\
{\small \PTTmax+\HR} & 0.33 [0.27,0.40]   & 0.87 [0.85,0.89]   & 0.63 [0.58,0.68]   & {\bf\color{blue} \hl{\bf 0.82}} [0.79,0.85]  \\
{\small \Dmax+\HR} & 0.54 [0.48,0.59]   & 0.90 [0.88,0.92]   & 0.72 [0.68,0.76]   & 0.82 [0.79,0.84]  \\
{\small \DAT+\HR} & 0.52 [0.47,0.58]   & 0.87 [0.85,0.89]   & 0.64 [0.58,0.68]   & 0.81 [0.78,0.84]  \\
{\small \PTT+\PTTmax} & {\bf\color{blue} 0.80} [0.77,0.83]   & 0.67 [0.64,0.72]   & 0.65 [0.61,0.70]   & 0.52 [0.46,0.58]  \\
{\small \PTT+\PTTmax+\HR} & 0.82 [0.79,0.85]   & \hl{\bf 0.93} [0.91,0.94]   & 0.75 [0.71,0.78]   & \hl{\bf 0.82} [0.79,0.85]  \\
{\small \PTT+\Dmax} & 0.80 [0.76,0.83]   & 0.74 [0.71,0.78]   & 0.68 [0.67,0.71]   & 0.51 [0.49,0.56]  \\
{\small \PTT+\Dmax+\HR} & 0.81 [0.77,0.83]   & 0.92 [0.91,0.93]   & 0.76 [0.73,0.79]   & 0.82 [0.79,0.84]  \\
{\small \PTT+\DAT} & 0.75 [0.71,0.78]   & {\bf\color{blue} 0.92} [0.91,0.94]   & {\bf\color{blue} 0.75} [0.71,0.79]   & 0.80 [0.77,0.83]  \\
{\small \PTT+\DAT+\HR} & 0.75 [0.71,0.78]   & 0.93 [0.91,0.94]   & 0.75 [0.72,0.79]   & 0.81 [0.79,0.84]  \\
{\small \PTT+\PTTmax+\Dmax} & 0.82 [0.80,0.85]   & 0.77 [0.77,0.78]   & 0.68 [0.66,0.71]   & 0.56 [0.55,0.61]  \\
{\small \PTT+\PTTmax+\Dmax+\HR} & \hl{\bf 0.83} [0.81,0.86]   & \hl{\bf 0.93} [0.91,0.94]   & \hl{\bf 0.76} [0.76,0.78]   & \hl{\bf 0.82} [0.79,0.85]  \\
{\small \PTT+\PTTmax+\DAT} & \hl{\bf 0.83} [0.80,0.86]   & 0.93 [0.91,0.94]   & 0.75 [0.73,0.78]   & 0.80 [0.76,0.84]  \\
{\small \PTT+\PTTmax+\DAT+\HR} & \hl{\bf 0.83} [0.80,0.85]   & \hl{\bf 0.93} [0.91,0.94]   & 0.75 [0.73,0.78]   & \hl{\bf 0.82} [0.79,0.85]  \\
{\small \PTT+\Dmax+\DAT} & 0.81 [0.78,0.84]   & 0.92 [0.91,0.94]   & \hl{\bf 0.76} [0.74,0.79]   & 0.80 [0.76,0.83]  \\
{\small \PTT+\Dmax+\DAT+\HR} & 0.81 [0.77,0.84]   & \hl{\bf 0.93} [0.92,0.94]   & \hl{\bf 0.76} [0.75,0.79]   & 0.81 [0.79,0.85]  \\
{\small \PTT+\PTTmax+\Dmax+\DAT} & \hl{\bf 0.84} [0.82,0.88]   & 0.93 [0.91,0.94]   & \hl{\bf 0.77} [0.76,0.79]   & 0.80 [0.78,0.84]  \\
{\small \PTT+\PTTmax+\Dmax+\DAT+\HR} & \hl{\bf 0.84} [0.82,0.86]   & \hl{\bf 0.93} [0.92,0.94]   & \hl{\bf 0.77} [0.76,0.79]   & \hl{\bf 0.82} [0.80,0.85]  \\
      \hline
    \end{tabular}
\end{table}

\renewcommand{\PTT}{{PAT$_\textrm{ff}$}}
\renewcommand{\PTTmax}{{PAT$_\textrm{p}${}}}
\renewcommand{\Dmax}{{PAT$_{D}${}}}

\setlength{\tabcolsep}{4pt}
\begin{table}[!h]
  \caption{Accuracy of predictions using PAT measurements (all timings
    calculated as a difference to the ignition of the pulse in
    $E_\textrm{fw}$ for LV) from left carotid artery (LCA).  The
    predictions with largest Pearson correlations are highlighted: one
    signal (red), two signals (blue) and five most accurate
    combination (black). Signals: heart rate (\HR) and pulse transit
    arrival to the foot of signal (\PAT), peak of signal (\PATmax),
    the point of steepest raise (\PATDmax), and the dicrotic notch
    (\PATDAT). }
  \label{tab:results_PAT_LCA}
  \centering
    \begin{tabular}{lll@{\extracolsep{4pt}}ll@{\extracolsep{4pt}}ll@{\extracolsep{4pt}}ll}
      \hline
      &aPWV&DBP&SBP&SV\\
      \hline
{\small \HR} & 0.06 [-0.08,0.14]   & {\bf\color{red} 0.85} [0.83,0.87]   & 0.46 [0.41,0.52]   & {\bf\color{red} 0.81} [0.77,0.84]  \\
{\small \PTT} & {\bf\color{red} 0.52} [0.47,0.57]   & 0.19 [0.09,0.27]   & {\bf\color{red} 0.62} [0.57,0.67]   & 0.18 [0.10,0.25]  \\
{\small \PTTmax} & 0.25 [0.19,0.33]   & 0.67 [0.64,0.71]   & 0.54 [0.46,0.61]   & 0.54 [0.49,0.59]  \\
{\small \Dmax} & 0.47 [0.41,0.52]   & 0.05 [-0.03,0.13]   & 0.61 [0.52,0.66]   & 0.24 [0.09,0.31]  \\
{\small \DAT} & -0.03 [-0.15,0.04]   & 0.83 [0.80,0.85]   & 0.48 [0.43,0.55]   & 0.79 [0.75,0.82]  \\
{\small \PTT+\HR} & 0.55 [0.51,0.60]   & 0.93 [0.92,0.94]   & {\bf\color{blue} 0.71} [0.67,0.76]   & 0.81 [0.78,0.84]  \\
{\small \PTTmax+\HR} & 0.28 [0.21,0.37]   & 0.90 [0.88,0.91]   & 0.67 [0.63,0.72]   & {\bf\color{blue} \hl{\bf 0.82}} [0.79,0.85]  \\
{\small \Dmax+\HR} & 0.52 [0.47,0.57]   & 0.92 [0.90,0.93]   & 0.68 [0.64,0.73]   & 0.81 [0.78,0.84]  \\
{\small \DAT+\HR} & 0.02 [-0.06,0.05]   & 0.85 [0.83,0.87]   & 0.48 [0.30,0.52]   & 0.81 [0.78,0.84]  \\
{\small \PTT+\PTTmax} & {\bf\color{blue} 0.72} [0.69,0.76]   & 0.81 [0.79,0.83]   & 0.68 [0.64,0.73]   & 0.61 [0.57,0.67]  \\
{\small \PTT+\PTTmax+\HR} & \hl{\bf 0.78} [0.75,0.81]   & 0.94 [0.92,0.95]   & \hl{\bf 0.75} [0.71,0.79]   & \hl{\bf 0.82} [0.80,0.85]  \\
{\small \PTT+\Dmax} & 0.58 [0.56,0.63]   & 0.79 [0.76,0.83]   & 0.66 [0.65,0.68]   & 0.53 [0.49,0.59]  \\
{\small \PTT+\Dmax+\HR} & 0.63 [0.63,0.64]   & 0.95 [0.93,0.96]   & 0.73 [0.70,0.78]   & \hl{\bf 0.82} [0.79,0.85]  \\
{\small \PTT+\DAT} & 0.56 [0.52,0.62]   & {\bf\color{blue} 0.95} [0.93,0.96]   & 0.71 [0.67,0.75]   & 0.79 [0.76,0.82]  \\
{\small \PTT+\DAT+\HR} & 0.56 [0.52,0.62]   & 0.95 [0.93,0.96]   & 0.71 [0.68,0.76]   & 0.81 [0.79,0.84]  \\
{\small \PTT+\PTTmax+\Dmax} & 0.76 [0.75,0.79]   & 0.89 [0.88,0.90]   & 0.70 [0.70,0.71]   & 0.68 [0.67,0.68]  \\
{\small \PTT+\PTTmax+\Dmax+\HR} & \hl{\bf 0.79} [0.77,0.82]   & \hl{\bf 0.95} [0.95,0.96]   & \hl{\bf 0.76} [0.72,0.80]   & \hl{\bf 0.82} [0.80,0.85]  \\
{\small \PTT+\PTTmax+\DAT} & 0.76 [0.73,0.79]   & 0.95 [0.94,0.96]   & 0.75 [0.73,0.79]   & 0.80 [0.75,0.84]  \\
{\small \PTT+\PTTmax+\DAT+\HR} & \hl{\bf 0.78} [0.75,0.82]   & 0.95 [0.94,0.96]   & \hl{\bf 0.75} [0.73,0.79]   & \hl{\bf 0.82} [0.79,0.85]  \\
{\small \PTT+\Dmax+\DAT} & 0.65 [0.62,0.63]   & \hl{\bf 0.96} [0.94,0.97]   & 0.72 [0.69,0.77]   & 0.79 [0.75,0.83]  \\
{\small \PTT+\Dmax+\DAT+\HR} & 0.65 [0.63,0.64]   & \hl{\bf 0.96} [0.94,0.97]   & 0.73 [0.71,0.77]   & 0.81 [0.78,0.84]  \\
{\small \PTT+\PTTmax+\Dmax+\DAT} & \hl{\bf 0.78} [0.71,0.81]   & \hl{\bf 0.96} [0.95,0.97]   & \hl{\bf 0.76} [0.74,0.80]   & 0.80 [0.76,0.85]  \\
{\small \PTT+\PTTmax+\Dmax+\DAT+\HR} & \hl{\bf 0.79} [0.76,0.82]   & \hl{\bf 0.96} [0.95,0.97]   & \hl{\bf 0.76} [0.75,0.79]   & 0.80 [0.77,0.82]  \\\hline
    \end{tabular}
\end{table}

\renewcommand{\PTT}{{PTT$_\textrm{ff}$}}
\renewcommand{\PTTmax}{{PTT$_\textrm{p}${}}}
\renewcommand{\Dmax}{{PTT$_{D}${}}}

\setlength{\tabcolsep}{4pt}
\begin{table}[!h]
  \caption{Accuracy of predictions using difference of PTT
    measurements from right femoral artery and left carotid artery
    (LCA - Fem). Otherwise same caption as in Table \ref{tab:results_PPT_LCA}. }
  \label{tab:results_PTT_FemLCA}
  \centering
    \begin{tabular}{lll@{\extracolsep{4pt}}ll@{\extracolsep{4pt}}ll@{\extracolsep{4pt}}ll}
      \hline
      &aPWV&DBP&SBP&SV\\
      \hline
{\small \HR} & 0.06 [-0.07,0.13]   & {\bf\color{red} 0.85} [0.83,0.87]   & 0.46 [0.40,0.51]   & {\bf\color{red} 0.81} [0.77,0.84]  \\
{\small \PTT} & {\bf\color{red} 0.76} [0.72,0.79]   & 0.35 [0.28,0.42]   & 0.59 [0.53,0.64]   & 0.10 [0.02,0.17]  \\
{\small \PTTmax} & 0.42 [0.36,0.48]   & 0.67 [0.63,0.71]   & 0.32 [0.26,0.38]   & 0.49 [0.43,0.55]  \\
{\small \Dmax} & 0.48 [0.39,0.54]   & 0.17 [0.09,0.24]   & {\bf\color{red} 0.60} [0.51,0.66]   & 0.21 [0.10,0.30]  \\
{\small \DAT} & 0.75 [0.72,0.78]   & 0.36 [0.28,0.44]   & 0.59 [0.52,0.64]   & 0.10 [0.00,0.19]  \\
{\small \PTT+\HR} & 0.76 [0.73,0.79]   & 0.92 [0.91,0.93]   & 0.75 [0.70,0.78]   & 0.82 [0.79,0.84]  \\
{\small \PTTmax+\HR} & 0.45 [0.39,0.51]   & 0.88 [0.86,0.90]   & 0.62 [0.58,0.67]   & {\bf\color{blue} \hl{\bf 0.83}} [0.80,0.85]  \\
{\small \Dmax+\HR} & 0.52 [0.46,0.58]   & 0.90 [0.88,0.91]   & 0.70 [0.66,0.75]   & 0.82 [0.79,0.85]  \\
{\small \DAT+\HR} & 0.76 [0.72,0.79]   & {\bf\color{blue} \hl{\bf 0.92}} [0.91,0.94]   & {\bf\color{blue} 0.75} [0.70,0.78]   & 0.82 [0.79,0.84]  \\
{\small \PTT+\PTTmax} & 0.80 [0.77,0.83]   & 0.71 [0.67,0.74]   & 0.63 [0.58,0.68]   & 0.52 [0.46,0.57]  \\
{\small \PTT+\PTTmax+\HR} & 0.82 [0.79,0.85]   & \hl{\bf 0.93} [0.91,0.94]   & 0.75 [0.72,0.79]   & \hl{\bf 0.82} [0.80,0.86]  \\
{\small \PTT+\Dmax} & {\bf\color{blue} 0.83} [0.81,0.87]   & 0.44 [0.40,0.50]   & 0.64 [0.61,0.68]   & 0.24 [0.16,0.31]  \\
{\small \PTT+\Dmax+\HR} & 0.83 [0.81,0.86]   & 0.92 [0.90,0.93]   & \hl{\bf 0.76} [0.73,0.79]   & 0.82 [0.79,0.84]  \\
{\small \PTT+\DAT} & 0.76 [0.72,0.79]   & 0.36 [0.30,0.44]   & 0.59 [0.55,0.64]   & 0.10 [0.02,0.18]  \\
{\small \PTT+\DAT+\HR} & 0.76 [0.73,0.81]   & 0.92 [0.91,0.94]   & 0.75 [0.71,0.78]   & 0.82 [0.79,0.84]  \\
{\small \PTT+\PTTmax+\Dmax} & \hl{\bf 0.86} [0.84,0.88]   & 0.73 [0.69,0.76]   & 0.68 [0.65,0.73]   & 0.54 [0.48,0.60]  \\
{\small \PTT+\PTTmax+\Dmax+\HR} & \hl{\bf 0.87} [0.83,0.89]   & \hl{\bf 0.93} [0.90,0.94]   & \hl{\bf 0.77} [0.77,0.78]   & \hl{\bf 0.82} [0.79,0.85]  \\
{\small \PTT+\PTTmax+\DAT} & 0.80 [0.76,0.83]   & 0.74 [0.71,0.78]   & 0.65 [0.62,0.70]   & 0.61 [0.54,0.66]  \\
{\small \PTT+\PTTmax+\DAT+\HR} & 0.83 [0.78,0.86]   & \hl{\bf 0.93} [0.91,0.94]   & \hl{\bf 0.75} [0.72,0.79]   & \hl{\bf 0.82} [0.80,0.85]  \\
{\small \PTT+\Dmax+\DAT} & 0.83 [0.81,0.87]   & 0.44 [0.40,0.53]   & 0.64 [0.61,0.68]   & 0.24 [0.18,0.32]  \\
{\small \PTT+\Dmax+\DAT+\HR} & \hl{\bf 0.84} [0.82,0.85]   & 0.92 [0.90,0.94]   & \hl{\bf 0.76} [0.74,0.79]   & 0.81 [0.78,0.84]  \\
{\small \PTT+\PTTmax+\Dmax+\DAT} & \hl{\bf 0.86} [0.84,0.88]   & 0.76 [0.71,0.79]   & 0.70 [0.71,0.73]   & 0.63 [0.46,0.69]  \\
{\small \PTT+\PTTmax+\Dmax+\DAT+\HR} & \hl{\bf 0.87} [0.85,0.89]   & \hl{\bf 0.93} [0.91,0.94]   & \hl{\bf 0.76} [0.77,0.79]   & \hl{\bf 0.82} [0.79,0.85]  \\
\hline
    \end{tabular}
\end{table}

\setlength{\tabcolsep}{4pt}
\begin{table}[!h]
  \caption{Accuracy of predictions using difference of PTT
    measurements from left and right carotid artery (LCA-RCA). Otherwise same caption as in Table \ref{tab:results_PPT_LCA}. }
  \label{tab:results_PTT_LCARCA}
  \centering
    \begin{tabular}{lll@{\extracolsep{4pt}}ll@{\extracolsep{4pt}}ll@{\extracolsep{4pt}}ll}
      \hline
      &aPWV&DBP&SBP&SV\\
\hline
{\small \HR} & 0.06 [-0.08,0.14]   & {\bf\color{red} 0.85} [0.83,0.87]   & {\bf\color{red} 0.46} [0.39,0.51]   & {\bf\color{red} 0.81} [0.76,0.84]  \\
{\small \PTT} & {\bf\color{red} 0.75} [0.72,0.78]   & 0.35 [0.30,0.43]   & 0.40 [0.34,0.45]   & 0.01 [-0.06,0.08]  \\
{\small \PTTmax} & 0.24 [0.18,0.31]   & 0.46 [0.40,0.52]   & 0.22 [0.16,0.31]   & 0.29 [0.23,0.35]  \\
{\small \Dmax} & 0.59 [0.54,0.64]   & 0.31 [0.24,0.37]   & 0.35 [0.29,0.41]   & 0.07 [0.00,0.16]  \\
{\small \DAT} & 0.66 [0.47,0.72]   & 0.17 [0.06,0.25]   & 0.21 [0.10,0.38]   & 0.12 [-0.03,0.21]  \\
{\small \PTT+\HR} & 0.75 [0.71,0.78]   & {\bf\color{blue} 0.89} [0.88,0.91]   & {\bf\color{blue} 0.63} [0.57,0.67]   & 0.81 [0.79,0.85]  \\
{\small \PTTmax+\HR} & 0.31 [0.24,0.42]   & 0.87 [0.85,0.89]   & 0.57 [0.52,0.63]   & {\bf\color{blue} \hl{\bf 0.82}} [0.80,0.85]  \\
{\small \Dmax+\HR} & 0.60 [0.55,0.64]   & 0.88 [0.86,0.90]   & 0.60 [0.54,0.65]   & 0.82 [0.79,0.85]  \\
{\small \DAT+\HR} & 0.61 [0.31,0.69]   & 0.88 [0.85,0.90]   & 0.46 [0.28,0.61]   & 0.81 [0.77,0.84]  \\
{\small \PTT+\PTTmax} & {\bf\color{blue} 0.75} [0.72,0.79]   & 0.56 [0.50,0.61]   & 0.44 [0.38,0.50]   & 0.30 [0.24,0.37]  \\
{\small \PTT+\PTTmax+\HR} & 0.75 [0.73,0.79]   & \hl{\bf 0.90} [0.88,0.92]   & \hl{\bf 0.68} [0.66,0.72]   & \hl{\bf 0.82} [0.80,0.85]  \\
{\small \PTT+\Dmax} & 0.75 [0.72,0.78]   & 0.52 [0.48,0.58]   & 0.45 [0.40,0.51]   & 0.38 [0.33,0.45]  \\
{\small \PTT+\Dmax+\HR} & 0.75 [0.72,0.79]   & 0.89 [0.88,0.91]   & 0.63 [0.58,0.69]   & 0.82 [0.80,0.85]  \\
{\small \PTT+\DAT} & 0.74 [0.69,0.77]   & 0.35 [0.05,0.61]   & 0.44 [0.39,0.49]   & 0.49 [0.41,0.57]  \\
{\small \PTT+\DAT+\HR} & 0.73 [0.68,0.76]   & \hl{\bf 0.90} [0.88,0.91]   & 0.63 [0.59,0.69]   & 0.81 [0.78,0.84]  \\
{\small \PTT+\PTTmax+\Dmax} & \hl{\bf 0.76} [0.73,0.80]   & 0.58 [0.35,0.63]   & 0.44 [0.40,0.52]   & 0.31 [-0.13,0.42]  \\
{\small \PTT+\PTTmax+\Dmax+\HR} & \hl{\bf 0.76} [0.75,0.78]   & \hl{\bf 0.90} [0.89,0.92]   & \hl{\bf 0.69} [0.69,0.70]   & \hl{\bf 0.82} [0.80,0.85]  \\
{\small \PTT+\PTTmax+\DAT} & \hl{\bf 0.75} [0.73,0.79]   & 0.65 [0.51,0.69]   & 0.44 [0.38,0.52]   & 0.39 [-0.01,0.53]  \\
{\small \PTT+\PTTmax+\DAT+\HR} & 0.74 [0.71,0.78]   & \hl{\bf 0.90} [0.89,0.92]   & \hl{\bf 0.68} [0.67,0.72]   & \hl{\bf 0.82} [0.80,0.85]  \\
{\small \PTT+\Dmax+\DAT} & 0.75 [0.70,0.78]   & 0.66 [0.46,0.73]   & 0.35 [0.25,0.43]   & 0.38 [0.07,0.55]  \\
{\small \PTT+\Dmax+\DAT+\HR} & 0.74 [0.70,0.78]   & 0.89 [0.88,0.91]   & \hl{\bf 0.63} [0.60,0.69]   & 0.82 [0.79,0.85]  \\
{\small \PTT+\PTTmax+\Dmax+\DAT} & \hl{\bf 0.76} [0.74,0.80]   & 0.68 [0.63,0.73]   & 0.43 [0.42,0.50]   & 0.48 [0.25,0.55]  \\
{\small \PTT+\PTTmax+\Dmax+\DAT+\HR} & \hl{\bf 0.76} [0.75,0.77]   & \hl{\bf 0.90} [0.90,0.91]   & \hl{\bf 0.69} [0.69,0.70]   & \hl{\bf 0.83} [0.82,0.85]  \\
      \hline
    \end{tabular}
\end{table}

\setlength{\tabcolsep}{4pt}
\begin{table}[!h]
  \caption{Accuracy of predictions using difference of PTT
    measurements from left and right radialis artery (LRad-RRad). Otherwise same caption as in Table \ref{tab:results_PPT_LCA}. }
  \label{tab:results_PTT_LRadRRad}
  \centering
  \begin{tabular}{lll@{\extracolsep{4pt}}ll@{\extracolsep{4pt}}ll@{\extracolsep{4pt}}ll}
      \hline
      &aPWV&DBP&SBP&SV\\
\hline
{\small \HR} & 0.06 [-0.08,0.13]   & {\bf\color{red} 0.85} [0.83,0.87]   & {\bf\color{red} 0.46} [0.40,0.52]   & {\bf\color{red} 0.81} [0.77,0.84]  \\
{\small \PTT} & {\bf\color{red} 0.67} [0.63,0.71]   & 0.19 [0.13,0.26]   & 0.22 [0.16,0.30]   & 0.03 [-0.04,0.14]  \\
{\small \PTTmax} & 0.46 [0.41,0.51]   & 0.59 [0.53,0.63]   & 0.29 [0.23,0.36]   & 0.45 [0.40,0.50]  \\
{\small \Dmax} & 0.65 [0.60,0.69]   & 0.19 [0.14,0.29]   & 0.22 [0.16,0.29]   & 0.02 [-0.05,0.10]  \\
{\small \DAT} & 0.62 [0.58,0.67]   & 0.12 [0.01,0.18]   & 0.25 [0.18,0.32]   & 0.17 [0.12,0.26]  \\
{\small \PTT+\HR} & 0.67 [0.63,0.71]   & 0.87 [0.85,0.89]   & 0.51 [0.46,0.57]   & 0.81 [0.78,0.84]  \\
{\small \PTTmax+\HR} & 0.49 [0.44,0.55]   & {\bf\color{blue} \hl{\bf 0.88}} [0.86,0.90]   & {\bf\color{blue} 0.53} [0.47,0.59]   & 0.81 [0.78,0.84]  \\
{\small \Dmax+\HR} & 0.65 [0.60,0.69]   & 0.87 [0.85,0.89]   & 0.51 [0.46,0.57]   & 0.81 [0.78,0.84]  \\
{\small \DAT+\HR} & 0.64 [0.60,0.69]   & 0.87 [0.85,0.89]   & 0.52 [0.47,0.57]   & {\bf\color{blue} \hl{\bf 0.81}} [0.79,0.84]  \\
{\small \PTT+\PTTmax} & {\bf\color{blue} 0.76} [0.73,0.79]   & 0.62 [0.57,0.68]   & 0.34 [0.28,0.41]   & 0.47 [0.42,0.53]  \\
{\small \PTT+\PTTmax+\HR} & \hl{\bf 0.77} [0.74,0.80]   & \hl{\bf 0.89} [0.86,0.91]   & \hl{\bf 0.55} [0.49,0.61]   & 0.81 [0.79,0.85]  \\
{\small \PTT+\Dmax} & 0.68 [0.62,0.72]   & 0.38 [0.34,0.46]   & 0.30 [0.25,0.38]   & 0.02 [-0.05,0.02]  \\
{\small \PTT+\Dmax+\HR} & 0.68 [0.64,0.74]   & 0.87 [0.85,0.89]   & 0.51 [0.46,0.58]   & 0.81 [0.78,0.84]  \\
{\small \PTT+\DAT} & 0.67 [0.63,0.71]   & 0.44 [0.43,0.48]   & 0.24 [0.18,0.34]   & 0.36 [0.27,0.46]  \\
{\small \PTT+\DAT+\HR} & 0.67 [0.65,0.71]   & 0.87 [0.86,0.89]   & 0.55 [0.54,0.58]   & \hl{\bf 0.81} [0.79,0.84]  \\
{\small \PTT+\PTTmax+\Dmax} & 0.77 [0.71,0.80]   & 0.63 [0.61,0.67]   & 0.34 [0.23,0.42]   & 0.47 [0.30,0.53]  \\
{\small \PTT+\PTTmax+\Dmax+\HR} & \hl{\bf 0.78} [0.68,0.81]   & \hl{\bf 0.89} [0.87,0.91]   & \hl{\bf 0.55} [0.51,0.61]   & \hl{\bf 0.81} [0.79,0.84]  \\
{\small \PTT+\PTTmax+\DAT} & 0.76 [0.73,0.80]   & 0.69 [0.68,0.73]   & 0.37 [0.28,0.44]   & 0.52 [0.27,0.60]  \\
{\small \PTT+\PTTmax+\DAT+\HR} & \hl{\bf 0.77} [0.70,0.81]   & \hl{\bf 0.89} [0.88,0.91]   & \hl{\bf 0.58} [0.59,0.62]   & \hl{\bf 0.81} [0.79,0.85]  \\
{\small \PTT+\Dmax+\DAT} & 0.68 [0.63,0.74]   & 0.47 [0.44,0.53]   & 0.26 [0.21,0.35]   & 0.37 [0.32,0.45]  \\
{\small \PTT+\Dmax+\DAT+\HR} & 0.68 [0.68,0.70]   & 0.88 [0.87,0.89]   & \hl{\bf 0.56} [0.58,0.59]   & 0.81 [0.78,0.84]  \\
{\small \PTT+\PTTmax+\Dmax+\DAT} & \hl{\bf 0.77} [0.66,0.80]   & 0.67 [0.66,0.71]   & 0.36 [0.23,0.43]   & 0.57 [0.48,0.61]  \\
{\small \PTT+\PTTmax+\Dmax+\DAT+\HR} & \hl{\bf 0.78} [0.75,0.81]   & \hl{\bf 0.89} [0.88,0.90]   & \hl{\bf 0.58} [0.60,0.61]   & \hl{\bf 0.81} [0.79,0.85]  \\
    \hline
  \end{tabular}
\end{table}



\end{document}